\documentclass[12pt,a4paper]{article}

\usepackage{subfiles}
\usepackage{amsmath,amssymb,graphicx}
\usepackage{xspace}
\usepackage{multirow}
\usepackage{cite}

\newlength{\abstwidth}
\newcommand{\forcenewcommand}[1]{\providecommand{#1}{}\renewcommand{#1}}
\newcommand{\defparticle}[1]{
  \expandafter\forcenewcommand\csname #1\endcsname{{\mathrm{#1}}}
  \expandafter\forcenewcommand\csname #1bar\endcsname{{\bar{\mathrm{#1}}}}
}

\defparticle{g}
\defparticle{q}
\defparticle{u}
\defparticle{d}
\defparticle{s}
\defparticle{c}
\defparticle{b}
\defparticle{e}
\defparticle{p}
\defparticle{n}
\defparticle{A}
\defparticle{B}
\defparticle{D}
\defparticle{K}
\defparticle{N}
\defparticle{Pb}
\newcommand{\Jpsi}{\mathrm{J}/\psi}
\newcommand{\Pom}{\mathbb{P}}

\renewcommand{\eqref}[1]{eq.~(\ref{#1})}
\newcommand{\citeref}[1]{Ref.~\cite{#1}}
\newcommand{\figref}[1]{Figure~\ref{#1}}
\newcommand{\tabref}[1]{Table~\ref{#1}}
\newcommand{\secref}[1]{Sect.~\ref{#1}}

\newcommand{\sea}{\mathrm{sea}}
\newcommand{\Pythia}{\textsc{Pythia}\xspace}
\newcommand{\Angantyr}{\textsc{Angantyr}\xspace}
\newcommand{\QCDNUM}{\textsc{QCDNUM}\xspace}

\newcommand{\CORSIKA}{\textsc{CORSIKA}\xspace}
\newcommand{\Sibyll}{\textsc{Sibyll}\xspace}

\newcommand{\eg}{\textit{eg.}\xspace}
\newcommand{\ie}{\textit{ie.}\xspace}
\newcommand{\cf}{\textit{cf.}\xspace}

\begin{document}

\sloppy
 
\pagestyle{empty}
 
\begin{flushright}
LU TP 21--32\\
MCnet--21--14\\
August 2021
\end{flushright}

\vspace{\fill}

\begin{center}
{\Huge\bf Hadron Interactions}\\[4mm]
{\Huge\bf for Arbitrary Energies and Species,}\\[3mm]
{\Huge\bf with Applications to Cosmic Rays}\\[10mm]
{\Large Torbj\"orn Sj\"ostrand and Marius Utheim} \\[3mm]
{\it Theoretical Particle Physics,}\\[1mm]
{\it Department of Astronomy and Theoretical Physics,}\\[1mm]
{\it Lund University,}\\[1mm]
{\it S\"olvegatan 14A,}\\[1mm]
{\it SE-223 62 Lund, Sweden}
\end{center}

\vspace{\fill}

\begin{center}
\begin{minipage}{\abstwidth}
{\bf Abstract}\\[2mm]
The \textsc{Pythia} event generator is used in several contexts
to study hadron and lepton interactions, notably $\p\p$ and
$\p\pbar$ collisions. In this article we extend the hadronic
modelling to encompass the collision of a wide range of hadrons $h$
with either a proton or a neutron, or with a simplified model
of nuclear matter. To this end we model $h\p$ total and partial
cross sections as a function of energy, and introduce new parton
distribution functions for a wide range of hadrons, as required
for a proper modelling of multiparton interactions. The potential
usefulness of the framework is illustrated by a simple study of
the evolution of cosmic rays in the atmosphere, and by an even
simpler one of shower evolution in a solid detector material.
The new code will be made available for future applications.

\end{minipage}
\end{center}

\vspace{\fill}

\phantom{dummy}

\clearpage

\pagestyle{plain}
\setcounter{page}{1}

\section{Introduction}

Throughout the history of high energy particle physics, one of the most
studied processes is proton--proton collisions. Originally the \Pythia
(+ \textsc{Jetset}) event generator
\cite{Sjostrand:2006za,Sjostrand:2014zea} was designed to simulate
$\e^+\e^-/\p\p$/$\p\pbar$ collisions. Later it was extended partly to
$\e\p$ and some photon physics, while the coverage of other hadron and
lepton collision types has remained limited. For QCD studies, as well as
other Standard-Model and Beyond-the-Standard-Model ones,
$\e^+\e^-/\p\p$/$\p\pbar/\e\p$ has provided the bulk of data, and so
there has been little incentive to consider other beam combinations.

In recent years, however, there has been an increasing interest to
extend the repertoire of beams. Prompted by the ongoing heavy-ion
experiments at RHIC and the LHC, the most significant addition to \Pythia is the
\Angantyr framework for heavy-ion interactions
\cite{Bierlich:2016smv,Bierlich:2018xfw}, which implements $\p\A$
and $\A\A$ collisions building on \Pythia's existing framework for
nucleon--nucleon interactions.

A second new addition is low-energy interactions, which was developed
as part of a framework for hadronic rescattering in \Pythia
\cite{Sjostrand:2020gyg,Bierlich:2021poz}. In this framework, common
collisions (\ie mainly those involving nucleons or pions) are modelled
in detail, including low-energy versions of standard high-energy 
processes like diffractive and non-diffractive interactions,
as well as low-energy-only non-perturbative processes like resonance
formation and baryon number annihilation. Less common collisions
(involving \eg excited baryons or charm/bottom hadrons) use simplified
descriptions, the most general being the Additive Quark Model (AQM)
\cite{Levin:1965mi,Lipkin:1973nt}, which gives a cross section that
depends only on the quark content of the involved hadrons. This way,
the low-energy framework supports interactions for all possible
hadron--hadron combinations.

These non-perturbative models are accurate only for low energies, however,
up until around 10~GeV. This means that, at perturbative energies, 
still mainly nucleon--nucleon interactions are supported. While other 
hadron species seldom are used directly as beams in experiments, their
collisions still have relevance, in particular for hadronic cascades
in a medium. One such example is cosmic rays entering the atmosphere,
with collision center-of-mass (CM) energies that stretch to and above
LHC energies, and thus give copious particle production. Secondary hadrons
can be of rare species, and may interact with the atmosphere at
perturbative energies. The objective of this article is to implement
general perturbative hadron--nucleon interactions in \Pythia,
using cosmic rays as a test case for the resulting framework.

Two significant extensions are introduced to this end. One is a modelling
of total, elastic, diffractive and nondiffractive cross sections for the
various beam combinations, as needed to describe collision rates also
at energies above 10~GeV. The other is parton distribution functions
(PDFs) for a wide selection of mesons and baryons, as needed to describe
the particle production in high-energy collisions. Important is also a
recent technical improvement, namely the support for selecting beam
energies on an event-by-event basis for the main QCD processes, made
possible by initializing relevant quantities on an interpolation grid
of CM energies. At the time of writing, this is supported for
hadron--hadron beams, but not yet for heavy-ion collisions in \Angantyr,
which will prompt us to introduce a simplified handling of nuclear
effects in hadron--nucleus collisions. Nucleus--nucleus ones, such as
iron hitting the atmosphere, is not yet considered.

Key to the understanding of atmospheric cascades is the model for hadronic
interactions. Several different ones are used in the community, such as
SIBYLL \cite{Engel:1992vf,Fletcher:1994bd,Engel:2017wpa,Riehn:2019jet},
QGSJET \cite{Kalmykov:1993qe,Kalmykov:1994ys,Ostapchenko:2019few},
DMPJET \cite{Ranft:1994fd,Roesler:2000he},
VENUS/EPOS \cite{Werner:1993uh,Pierog:2013ria,Pierog:2019opp}, and
HDPM (described in \citeref{Heck:1998vt}).
It is in this category that \Pythia could offer an alternative model,
constructed completely independently of either of the other ones, and
therefore with the possibility to offer interesting cross-checks.
In some respects it is likely to be more sophisticated than some of the
models above, \eg by being able to handle a large range of beam particles
almost from the threshold to the highest possible energies, with  
semi-perturbative interactions tailored to the incoming hadron type.
In other respects it is not yet as developed, like a limited handling
of nuclear effects and a lack of tuning to relevant data.

Neither of these programs can describe the important electromagnetic
cascades, which instead typically are delegated to EGS \cite{Nelson:1985ec}.
At low energies GHEISHA \cite{Fesefeldt:1985yw} is often used for
nuclear effects, with ISOBAR (described in \citeref{Heck:1998vt}) and
UrQMD \cite{Bass:1998ca} as alternatives. Generally a typical full
simulation requires many components to be combined, under the control 
of a framework that does the propagation of particles through the
atmosphere, taking into account \eg the atmospheric density variation
and the bending of charged particles by the earth magnetic field.
Two well-known examples of such codes are \CORSIKA \cite{Heck:1998vt}
and AIRES \cite{Sciutto:1999jh, Sciutto:2001dn}. Interestingly for us,
the new \CORSIKA~8 \cite{Engel:2018akg,Dembinski:2020wrp} framework is
written in C++, like \Pythia~8 but unlike some of the other hadronic
interaction models, and \Pythia~8 is already interfaced to handle
particle decays, so a further integration is a possibility.

One should also mention that an alternative to Monte Carlo simulation
of cascades is to construct a numerical simulation from the cascade
evolution equations, examples being SENECA \cite{Nelson:1985ec} and
MCeq \cite{Nelson:1985ec}. Also in these cases the hadronic interaction
models can provide valuable input. \Angantyr has in fact already been
used to this end, to describe $\p/\pi/\K$ interactions with nuclei
\cite{Storehaug:2019}.

Another application of hadronic cascades is in detector simulations
with programs such as FLUKA \cite{Ferrari:2005zk} and GEANT
\cite{Brun:1994aa,Agostinelli:2002hh,Allison:2006ve,Allison:2016lfl},
which have also been used for cascades in the atmosphere, 
see \eg \cite{Battistoni:2010vf,Paschalis:2014csa,Sarkar:2020wrq}. 
GEANT4 depends on external frameworks for simulating collisions, 
like \CORSIKA~8, and has been explicitly designed with an
object-oriented architecture that allows users to insert their own
physics implementations, one of the current possibilities being 
\Pythia~6 \cite{Sjostrand:2006za}. One central difference is that 
the medium is much denser in a detector, so particles propagate shorter 
distances before interacting. Hence, some particles that are too 
short-lived to interact in the atmosphere can do so in detector 
simulations, \eg $\D$, $\B$, $\Lambda_\c^+$ and $\Lambda_\b^0$. 
For the rest of this article we will focus on the atmospheric case, 
but we still implement all hadronic interactions relevant 
for either medium.

To describe hadron--hadron interactions, we need to set up the relevant
cross sections and event characteristics. In particular, the latter
includes modelling the parton distribution functions for the
incoming hadrons. These aspects are developed in Section~2.
Some simple resulting event properties for $h\p$ collisions are shown 
in Section~3. In practice, mediums consist of nuclei such as nitrogen 
or lead, rather than of free nucleons. Since \Angantyr does not yet 
efficiently support nuclear collisions with variable energy, we also 
introduce and test a simplified handling of nuclear effects in Section~3. 
The main intended application of this framework is to cascades in a 
medium, so we implement a simple atmospheric model in Section~4, and 
give some examples of resulting distributions. There is also a quick
look at passage through a denser medium. Either setup is much simplified 
relative to \CORSIKA or GEANT4, so has no scientific value except to
to test and explore features of our new hadronic interactions. 
The atmospheric toy-model code will be included in a future release of 
\Pythia as an example of how to interface a cascade simulation with \Pythia. 
Finally we present some conclusions and an outlook in Section~5.

\section{Cross sections and parton distributions}
\label{sec:CSaPD}

The first step in modelling the evolution of a cascade in a medium
is to have access to the total cross sections for all relevant
collisions. Crucially, this relates to how far a particle can
travel before it interacts. Once an interaction occurs, the second
step is to split the total cross section into partial ones,
each with a somewhat different character of the resulting
events. Each event class therefore needs to be described separately.
At high energies a crucial component in shaping event properties
is multiparton interactions (MPIs). To model these, parton
distribution functions (PDFs) have to be made available for all
relevant hadrons. Special attention also has to be given
to particles produced in the forward direction, that take most
of the incoming energy and therefore will produce the most energetic
subsequent interactions. These topics will be discussed in the
following.

\subsection{New total cross sections}

The description of cross sections depends on the collision
energy. At low energies various kinds of threshold phenomena
and resonance contributions play a key role, and these can
differ appreciably depending on the incoming hadron species.
At high energies a more smooth behaviour is expected, where
the dominating mechanism of pomeron exchange should give
common traits in all hadronic cross sections
\cite{Collins:1977jy,Forshaw:1997dc,Donnachie:2002en,Barone:2002cv}.  

In a recent article \cite{Sjostrand:2020gyg} we implemented
low-energy cross sections for most relevant hadron--hadron
collisions, both total and partial ones. Input came from a
variety of sources. The main ones were mostly based on data or
well-established models, while others involved larger measures
of uncertainty. Extensions were also introduced to the traditional
string fragmentation framework, to better deal with constrained
kinematics at low energy.

In cases where no solid input existed, the Additive Quark Model
(AQM) \cite{Levin:1965mi,Lipkin:1973nt} was applied to rescale
other better-known cross sections. In the AQM, the total
cross section is assumed to be proportional to the product of the
number of valence quarks in the respective hadron, so that \eg a
meson--meson cross section is $4/9$ that of a baryon--baryon one. 
The contribution of a heavy quark is scaled down relative to that
of a $\u/\d$ quarks, however, presumably by mass effects giving a
narrower spatial wave function. Assuming that quarks contribute
inversely proportionally to their constituent masses, we define an
effective number of valence quarks in a hadron to be approximately
\begin{equation}
n_{\q,\mathrm{AQM}} = n_{\u} + n_{\d} + 0.6 \, n_{\s}
+ 0.2 \, n_{\c} + 0.07 \, n_{\b}~. 
\label{eq:nqAQM}    
\end{equation}
This expression will also be used as a guide for high-energy
cross sections, as we shall see.

The emphasis of the low-energy cross sections lies on the description
of collisions below 5~GeV, say, but the models used should be valid
up to 10 GeV. Many processes also have a sensible behaviour above that,
others gradually less so.

At the other extreme then lies models intended to describe
high-energy cross sections. Here $\p\p/\p\pbar$ collisions are 
central, given the access to data over a wide energy range,
and the need to interpret this data. A few such models have 
been implemented in \Pythia \cite{Rasmussen:2018dgo}, giving the
possibility of comparisons. Fewer models are available for diffractive
topologies than for the total and elastic cross sections.

For the purposes of this study we will concentrate on the SaS/DL 
option, not necessarily because it is the best one for $\p\p/\p\pbar$
but because we have the tools to extend it to the necessary range of
collision processes in a reasonably consistent manner. The starting
point is the Donnachie--Landshoff modelling of the total cross section
\cite{Donnachie:1992ny}. In it, a common ansatz
\begin{equation}
\sigma^{AB}_{\mathrm{tot}} = X^{AB} s^{\epsilon} + Y^{AB} s^{-\eta}
\label{eq:sigmaDL} 
\end{equation}
is used for the collisions between any pair of hadrons $A$ and $B$.
Here $s$ is the squared CM energy, divided by 1 GeV$^2$ to make it
dimensionless. The terms $s^{\epsilon}$ and $s^{-\eta}$ are assumed
to arise from pomeron and reggeon exchange, respectively, with
tuned universal values $\epsilon = 0.0808$ and $\eta = 0.4525$.
The $X^{AB}$ and $Y^{AB}$, finally, are process-specific.
$X^{\overline{A}B} = X^{AB}$ since the pomeron is charge-even, whereas
generally $Y^{\overline{A}B} \neq Y^{AB}$, which can be viewed as a
consequence of having one charge-even and one charge-odd reggeon.
Recent experimental studies \cite{Antchev:2017yns,Abazov:2020rus}
have shown that the high-energy picture should be complemented by
a charge-odd odderon \cite{Lukaszuk:1973nt} contribution, but as of
yet there is no evidence that such effects have a major impact on
total cross sections.

\begin{table}[tp!] \centering
\begin{tabular}{|c|r|r|r|l|}
\hline
$AB$        & $X^{AB}$ & $Y^{AB}$ & $Y^{\overline{A}B}$ & comment\\
\hline
$\p\p$      & 21.70 & 56.08 & 98.39 & \\
$\p\n$      & 21.70 & 54.77 & 92.71 & not used, see text\\
$\pi^+\p$   & 13.63 & 27.56 & 36.02 & \\
$\K^+\p$    & 11.82 &  8.15 & 26.36 & \\
\hline
$\pi^0\p$   & 13.63 & 31.79 & --  & $(\pi^+\p + \pi^-\p)/2$\\
$\phi^0\p$  & 10.01 & -1.51 & --  & $\K^+\p + \K^-\p - \pi^-\p$\\
\hline
$\K^0\p$      & 11.82 & 17.26 & -- & $(\K^+\p + \K^-\p)/2$  \\
$\eta\p$      & 12.18 & 19.68 & -- & $0.6 \, \pi^0\p + 0.4 \, \phi^0\p$  \\
$\eta'\p$     & 11.46 & 13.62 & -- & $0.4 \, \pi^0\p + 0.6 \, \phi^0\p$  \\
$\Jpsi\p$     &  3.33 & -0.50 & -- & $\phi^0\p/3$ \\ 
$\D^{0,+}\p$   &  8.48 & \multicolumn{2}{c|}{15.65} & $(\pi^0\p + \Jpsi\p)/2$ \\
$\D_{\s}^+\p$  &  6.67 & \multicolumn{2}{c|}{-1.00} & $(\phi^0\p + \Jpsi\p)/2$\\
$\Upsilon\p$  &  1.17 & -0.18 & -- &  $0.07 \, \phi^0\p / 0.6$ \\
$\B^{0,+}\p$   &  7.40 & \multicolumn{2}{c|}{15.81}
& $(\pi^0\p + \Upsilon\p)/2$ \\
$\B_{\s}^0\p$  &  5.59 & \multicolumn{2}{c|}{-0.85}
& $(\phi^0\p + \Upsilon\p)/2$\\
$\B_{\c}^+\p$  &  2.25 & \multicolumn{2}{c|}{-0.34}
& $(\Jpsi^0\p + \Upsilon\p)/2$ \\
\hline
$\Lambda\p$   & 18.81 & 37.39 & 65.59 & AQM, $2 \,\p\p /3$ \\
$\Xi\p$       & 15.91 & 18.69 & 32.80 & AQM, $\p\p /3$ \\
$\Omega\p$    & 13.02 &  0.00 &  0.00 & AQM, 0 \\
$\Lambda_{\c}\p$   & 15.91 & 37.39 & 65.59 & AQM, $2 \,\p\p /3$ \\
$\Xi_{\c}\p$       & 13.02 & 18.69 & 32.80 & AQM, $\p\p /3$ \\
$\Omega_{\c}\p$    & 10.13 &  0.00 &  0.00 & AQM, 0 \\
$\Lambda_{\b}\p$   & 14.97 & 37.39 & 65.59 & AQM, $2 \,\p\p /3$ \\
$\Xi_{\b}\p$       & 12.08 & 18.69 & 32.80 & AQM, $\p\p /3$ \\
$\Omega_{\b}\p$    &  9.19 &  0.00 &  0.00 & AQM, 0 \\
\hline
\end{tabular}
\caption{Coefficients $X^{AB}$ and $Y^{AB}$, in units of mb,
  in \eqref{eq:sigmaDL} for various beam combinations. First section
  is from DL \cite{Donnachie:1992ny}, second from SaS \cite{Schuler:1993td}
  and the rest are new for this study.}
\label{tab:XYcoef}
\end{table}

In the context of $\gamma\p$ and $\gamma\gamma$ studies, the
set of possible beam hadrons was extended by Schuler and Sj\"ostrand (SaS)
to cover vector meson collisions \cite{Schuler:1993td,Schuler:1996en}.
Now we have further extended it to cover a range of additional processes
on a $\p/\n$ target, \tabref{tab:XYcoef}. The extensions have been based
on simple considerations, notably the AQM, as outlined in the table. They
have to be taken as educated guesses, where the seeming accuracy of numbers
is not to be taken literally. For simplicity, collisions with protons and
with neutrons are assumed to give the same cross sections, which is
consistent with data, so only the former are shown. The reggeon term for
$\phi^0\p$ is essentially vanishing, consistent with the OZI rule
\cite{Okubo:1963fa,Zweig:1981pd,Iizuka:1966fk},
and we assume that this suppression of couplings between light $\u/\d$
quarks and $\s$ quarks extends to $\c$ and $\b$. Thus, for baryons, the
reggeon $Y^{AB}$ values are assumed proportional to the number of light
quarks only, while the AQM of \eqref{eq:sigmaDL} is still used for the
pomeron term. Another simplification is that $\D/\B$ and $\Dbar/\Bbar$
mesons are assigned the same cross section. Baryons with the same flavour
content, or only differing by the relative composition of $\u$ and $\d$
quarks, are taken to be equivalent, \ie
$\Lambda\p = \Sigma^+\p = \Sigma^0\p = \Sigma^-\p$.

The DL parametrizations work well down to 6~GeV, where testable. Thus
there is an overlap region where either the low-energy or the high-energy
cross sections could make sense to use. Therefore we have chosen to mix
the two in this region, to give a smooth transition. More precisely, the
transition is linear in the range between
\begin{align}
E_{\mathrm{CM}}^{\mathrm{begin}} &= E_{\mathrm{min}} + \max(0., m_A - m_{\p})
+ \max(0., m_B - m_{\p}) ~~\mathrm{and}\\
E_{\mathrm{CM}}^{\mathrm{end}} &= E_{\mathrm{CM}}^{\mathrm{begin}}
+ \Delta E ~,
\end{align}
where $E_{\mathrm{min}}$ is 6~GeV and $\Delta E$  is 8 GeV by default.

\subsection{New partial cross sections}

The total cross section can be split into different components
\begin{equation}
\sigma_{\mathrm{tot}} = \sigma_{\mathrm{ND}} + \sigma_{\mathrm{el}}
+ \sigma_{\mathrm{SD}(XB)} + \sigma_{\mathrm{SD}(AX)} + \sigma_{\mathrm{DD}}
+ \sigma_{\mathrm{CD}} + \sigma_{\mathrm{exc}} + \sigma_{\mathrm{ann}}
+ \sigma_{\mathrm{res}} + \ldots ~.
\label{eq:sigmasplit}
\end{equation}
Here ND is short for nondiffractive, el for elastic, SD$(XB)$ and
SD$(AX)$ for single diffraction where either beam is excited, DD for
double diffraction, CD for central diffraction, exc for excitation,
ann for annihilation and res for resonant.
Again slightly different approaches are applied at low and at high
energies, where the former often are based on measurements or models
for exclusive processes, whereas the latter assume smoother and more
inclusive distributions. The last three subprocesses in
\eqref{eq:sigmasplit} are only used at low energies. In the transition
region between low and high energies, the two descriptions are mixed
the same way as the total cross section.

High-energy elastic cross sections are modelled using the optical theorem.
Assuming a simple exponential fall-off
$\d\sigma_{\mathrm{el}}/\d t \propto \exp(B_{\mathrm{el}}t)$ and a vanishing
real contribution to the forward scattering amplitude ($\rho = 0$)
\begin{equation}
\sigma_{\mathrm{el}} = \frac{\sigma_{\mathrm{tot}}^2}{16 \pi B_{\mathrm{el}}}
\end{equation}
(with $c = \hbar = 1$). The slope is given by
\begin{equation}
\B_{\mathrm{el}}^{AB} = 2 b_A + 2b_B + 2 \alpha' \, \ln\left(\frac{s}{s_0}\right)
\to 2 b_A + 2b_B + 2 (2.0 \, s^{\epsilon} - 2.1) ~,
\end{equation}
where $\alpha' \approx 0.25$~GeV$^{-2}$ is the slope of the pomeron
trajectory and $s_0 = 1/\alpha'$. In the final expression the SaS
replacement is made to ensure that $\sigma_{\mathrm{el}}/\sigma_{\mathrm{tot}}$
goes to a constant below unity at large energies, while offering a
reasonable approximation to the logarithmic expression at low energies.
The hadronic form factors $b_{A,B}$ are taken to be 1.4 for mesons and
2.3 for baryons, except that mesons made only out of $\c$ and $\b$ quarks
are assumed to be more tightly bound and thus have lower values.
As a final comment, note that a simple exponential in $t$ is only a
reasonable approximation at small $|t|$, but this is where the bulk
of the elastic cross section is. For $\p\p$ and $\p\pbar$ more
sophisticated larger-$|t|$ descriptions are available
\cite{Rasmussen:2018dgo}.

Also diffractive cross sections are calculated using the SaS ansatz
\cite{Schuler:1993wr,Schuler:1996en}. The differential formulae are
integrated numerically for each relevant collision process and the result
suitably parametrized, including a special threshold-region ansatz
\cite{Sjostrand:2020gyg}. Of note is that, if the hadronic form factor
from pomeron-driven interactions is written as
$\beta_{A\Pom}(t) = \beta_{A\Pom}(0) \, \exp(b_A t)$ then, with suitable
normalization, $X^{AB} = \beta_{A\Pom}(0) \, \beta_{B\Pom}(0)$ in
\eqref{eq:sigmaDL}. Thus we can define $\beta_{\p\Pom}(0) = \sqrt{X^{\p\p}}$
and other $\beta_{A\Pom}(0) = X^{A\p}/\beta_{\p\Pom}(0)$. These numbers
enter in the prefactor of single diffractive cross sections, \eg
$\sigma_{AB \to AX} \propto \beta_{A\Pom}^2(0) \, \beta_{B\Pom}(0) =
X^{AB} \, \beta_{A\Pom}(0)$.
This relation comes about since the $A$ side scatters (semi)elastically
while the $B$ side description is an inclusive one, cf.\ the optical
theorem. In double diffraction $AB \to X_1X_2$ neither side is elastic
and the rate is directly proportional to $X^{AB}$.

\begin{figure}[t!]
\begin{minipage}[c]{0.49\linewidth}
\centering
\includegraphics[width=\linewidth]{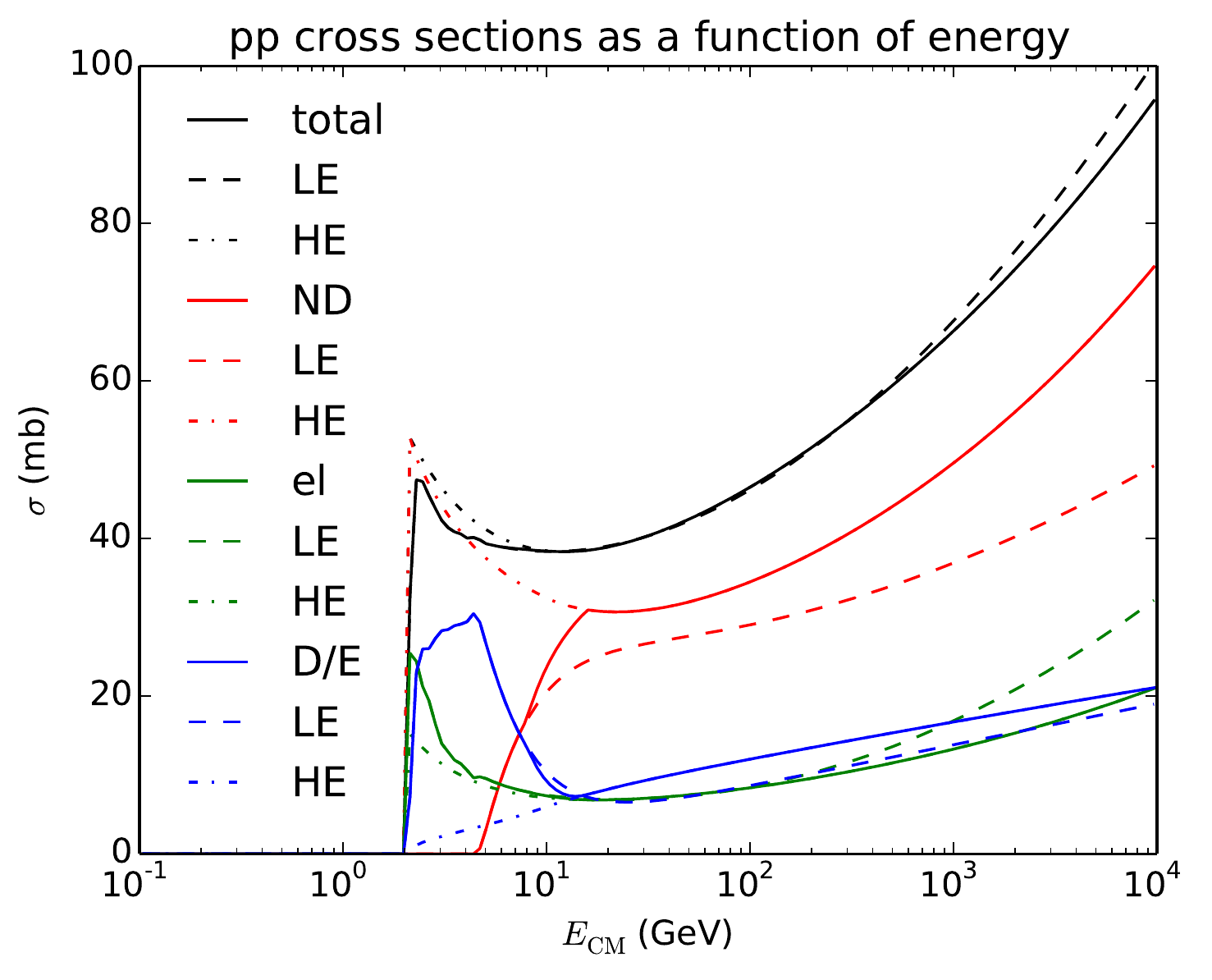}\\
(a)
\end{minipage}
\begin{minipage}[c]{0.49\linewidth}
\centering
\includegraphics[width=\linewidth]{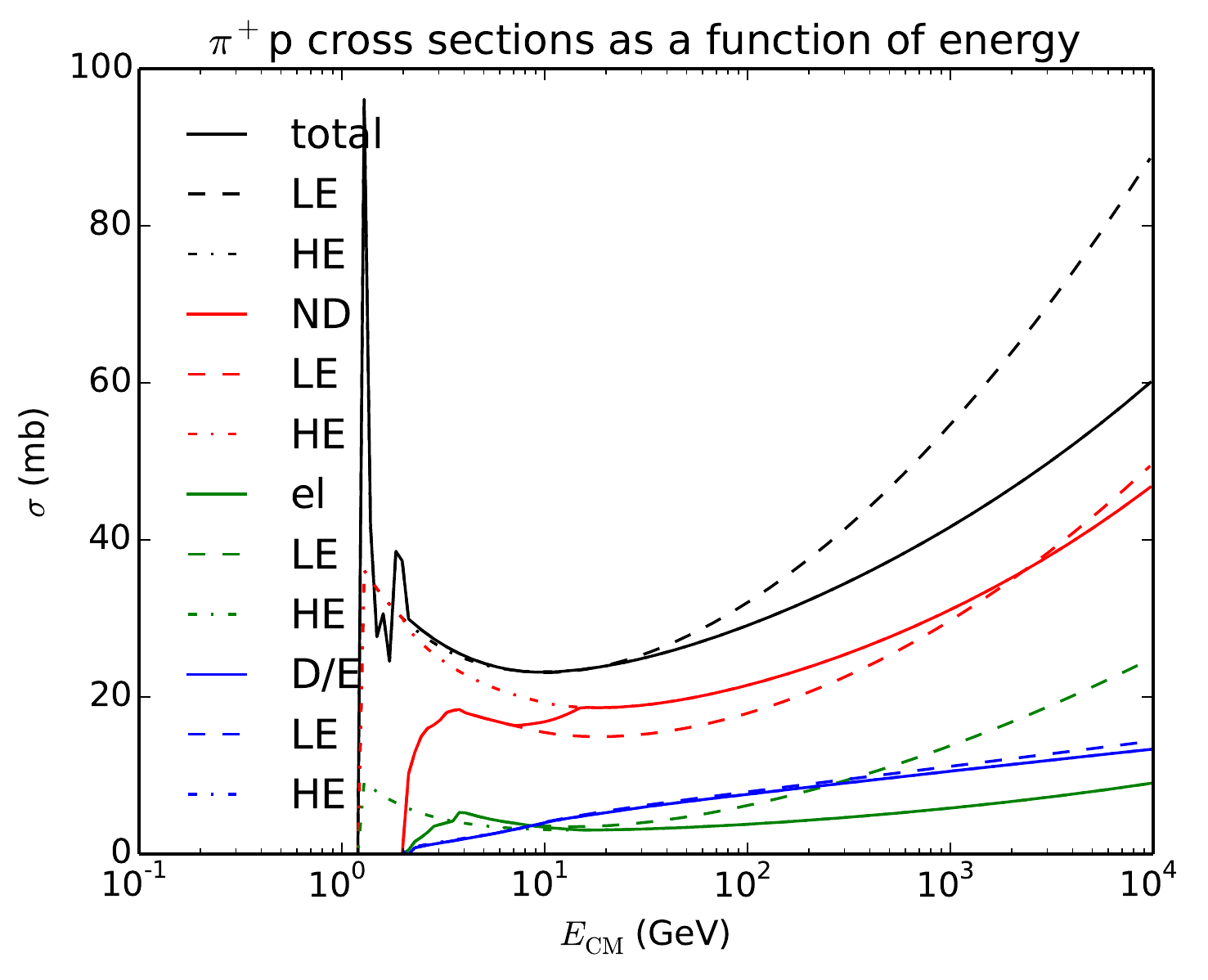}\\
(b)
\end{minipage}\\
\begin{minipage}[c]{0.49\linewidth}
\centering
\includegraphics[width=\linewidth]{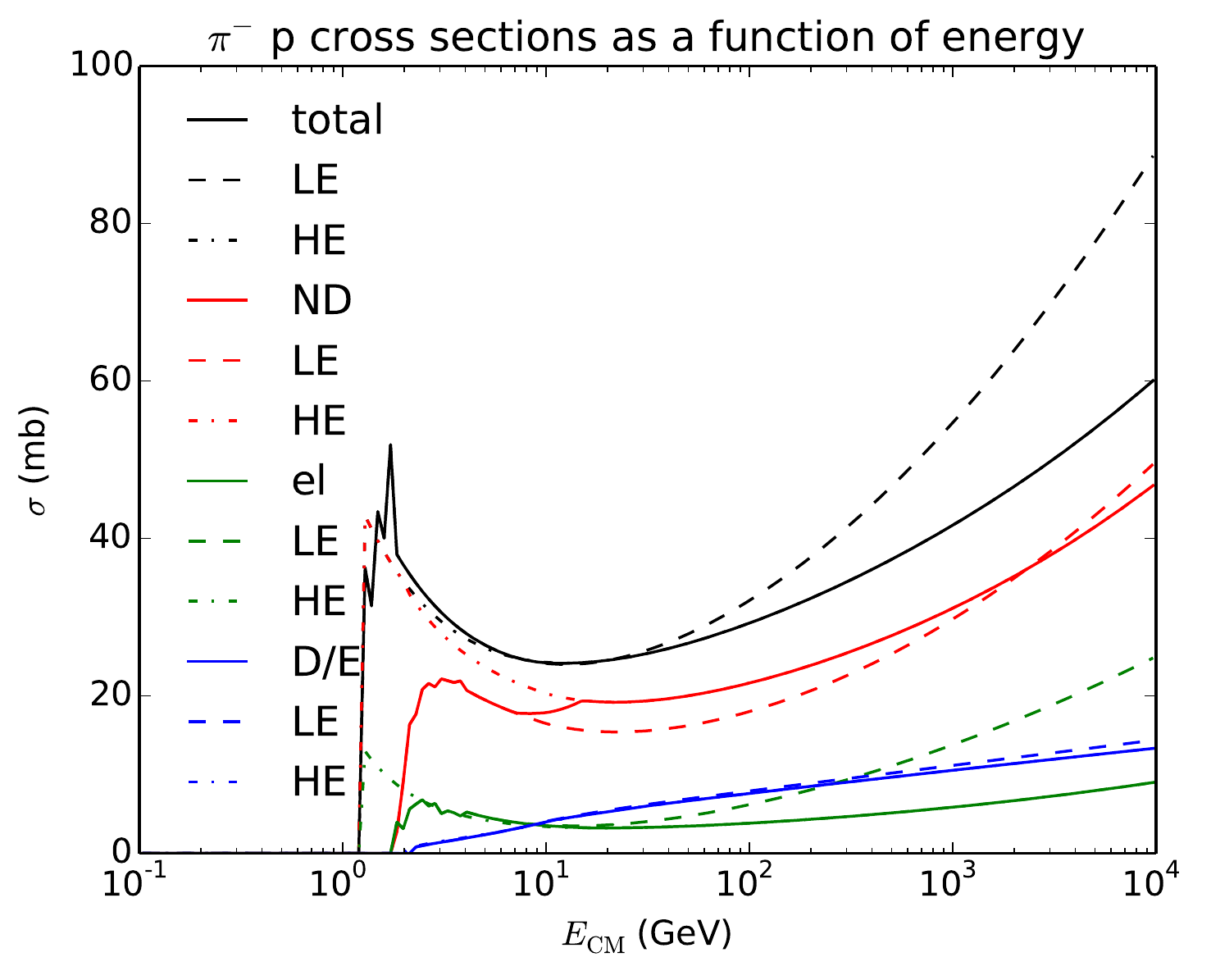}\\
(c) 
\end{minipage}
\begin{minipage}[c]{0.49\linewidth}
\centering
\includegraphics[width=\linewidth]{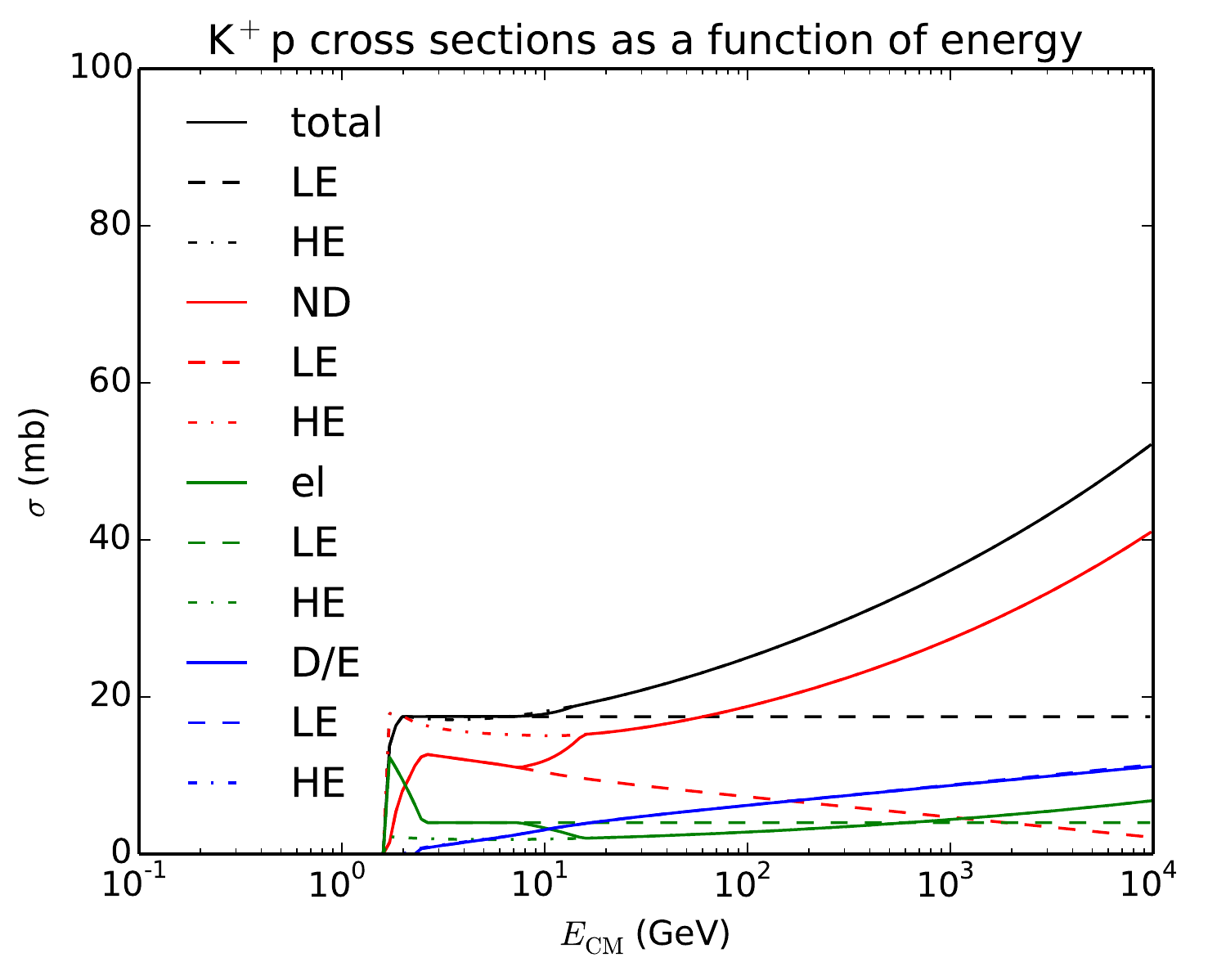}\\
(d)
\end{minipage}\\
\caption{Total, nondiffractive (ND), elastic (el) and diffractive/excitation
(D/E) cross sections for some common collision processes, (a) $\p\p$,
(b) $\pi^+\p$, (c) $\pi^-\p$ and (d) $\K^+\p$. Full lines show
the cross sections actually used, while dashed show the low-energy (LE) and
dash-dotted the high-energy (HE) separate inputs. The LE/HE curves are shown
also outside of their regions of intended validity, so should be viewed as
illustrative only.}
\label{fig:sigmaData}
\end{figure}

In addition to the approximate $\d M_X^2/ M_X^2$ mass spectrum of
diffractive systems, by default there is also a smooth low-mass
enhancement, as a simple smeared representation of exclusive resonance
states. In the low-energy description of nucleon--nucleon collisions
this is replaced by a set of explicit low-mass resonances
(\eg $AB \to AR$) \cite{Sjostrand:2020gyg}. The low-energy description
also includes single-resonance ($AB \to R$) and baryon--antibaryon
annihilation contributions that are absent in the high-energy one.

The nondiffractive cross section, which is the largest fraction at
high energies, is defined as what remains when the contributions above 
have been subtracted from the total cross section.

Some examples of total and partial cross sections are shown in
\figref{fig:sigmaData}. 

\subsection{Hadronic collisions}

At low energies the character of an event is driven entirely by
nonperturbative processes. In a nondiffractive topology, this can
be represented by the exchange of a single gluon, so soft that the
momentum transfer can be neglected. The colour exchange leads to
two colour octet hadron remnants, however. Each can be split into a
colour triplet and a colour antitriplet part, $\q$-$\qbar$ for a meson
and $\q$-$\q\q$ for a baryon. This leads to two
(Lund \cite{Andersson:1983ia}) strings being pulled out,
each between the colour of one hadron and the anticolour of the
other. In diffraction either a quark or a gluon is kicked out from
the diffracted hadron, giving either a straight string or one with
a kink at the gluon. Other processes have their own descriptions
\cite{Sjostrand:2020gyg}. 

At high energies, on the other hand, perturbative processes play a
key role. A suitable framework is that of multiparton interactions,
MPIs \cite{Sjostrand:1987su,Sjostrand:2017cdm}. In it, it is assumed
that the composite nature of the hadrons leads to several separate
parton--parton interactions, each dressed up with associated parton
showers. At first glance the interactions occur independently, but 
at closer look they are connected by energy--momentum--flavour--colour
conservation. Especially the last is nontrivial to model, and requires
a special colour reconnection step. There the total string length
is reduced relative to a first assignment where the MPIs are largely
decoupled from each other.

The probability to offer a perturbative description of a nondiffractive
event is assumed to be
\begin{equation}
P_{\mathrm{pert}} = 1 - \exp\left(
-\frac{E_{\mathrm{CM}} - E_{\mathrm{min}}}{E_{\mathrm{wid}}} \right) ~,
\end{equation}
when $E_{\mathrm{CM}} > E_{\mathrm{min}}$, and else vanishing.
Here $E_{\mathrm{CM}}$ is the collision energy in the rest frame, and
\begin{equation}
E_{\mathrm{min}} = E_{\mathrm{min,0}} + 2 \, \max(0., m_A - m_p)
+ 2 \, \max(0., m_B - m_p) ~, 
\end{equation}
while $E_{\mathrm{min,0}}$ and $E_{\mathrm{wid}}$ are two free (within reason)
parameters, both 10~GeV by default. The same transition can be used 
for the handling of diffraction, with $E_{\mathrm{CM}}$ replaced
by the mass of the diffractive system. Note that it is separate from
the transition from low- to high-energy cross section expressions.

In perturbative events the parton--parton collision rate (neglecting
quark masses) is given by
\begin{equation}
\frac{\d\sigma^{AB}}{\d p_{\perp}^2} = \sum_{i,j,k} \iiint f^A_i(x_1, Q^2) \,
f^B_j(x_2, Q^2) \, \frac{\d\hat{\sigma}_{ij}^k}{\d\hat{t}} \,
\delta \left( p_{\perp}^2 - \frac{\hat{t}\hat{u}}{\hat{s}} \right)
\, \d x_1 \, \d x_2 \, \d \hat{t}
\label{eq:dsigmadpT}
\end{equation}
differentially in transverse momentum $p_{\perp}$. Here the PDF
$f^A_i(x, Q^2)$ represents the probability to find a parton $i$ in a
hadron $A$ with momentum fraction $x$ if the hadron is probed at a
scale $Q^2 \approx p_{\perp}^2$. Different subprocesses are possible,
labelled by $k$, but the dominant one is $t$-channel gluon exchange.
It is convenient to order MPIs in falling order of $p_{\perp}$, like in
a parton shower.

A problem is that the perturbative QCD cross section in \eqref{eq:dsigmadpT}
is divergent in the $p_{\perp} \to 0$ limit. This can be addressed by
multiplying it with a factor
\begin{equation}
f_{\mathrm{damp}}(p_{\perp}) = \left( \frac{\alpha_{\mathrm{s}}(p_{\perp 0}^2
  + p_{\perp}^2)} {\alpha_{\mathrm{s}}(p_{\perp}^2)} \,
\frac{p_{\perp}^2}{p_{\perp 0}^2 + p_{\perp}^2} \right)^2 ~.
\end{equation}
which is finite in the limit $p_{\perp} \to 0$. Such a modification can
be viewed as a consequence of colour screening: in the $p_{\perp} \to 0$
limit a hypothetical exchanged gluon would not resolve individual
partons but only (attempt to) couple to the vanishing net colour charge
of the hadron. The damping could be associated only with the PDFs or
only with the $\d\hat{\sigma}/\d\hat{t}$ factor, according to taste,
but we remain agnostic on this count. The new $p_{\perp 0}$ parameter is
assumed to be varying with the collision energy, with current default
\begin{equation}
p_{\perp 0} = (2.28~\mathrm{GeV}) \, \left(
\frac{E_{\mathrm{CM}}}{7~\mathrm{TeV}} \right)^{0.215} ~,
\end{equation}
which can be related to the increase of PDFs at low $x$, leading to an    
increasing screening with energy.

Most of the MPIs occur in the nondiffractive event class. The average
number is given by
\begin{equation}
\langle n_{\mathrm{MPI}} \rangle = \frac{1}{\sigma_{\mathrm{ND}}} \,
\int_0^{E_{\mathrm{CM}}/2} f_{\mathrm{damp}}(p_{\perp}) \,
\frac{\d\sigma^{AB}}{\d p_{\perp}} \, \d p_{\perp}~.
\label{eq:nMPI}
\end{equation}
MPIs can also occur in high-mass diffraction, and is simulated in \Pythia
\cite{Rasmussen:2015qgr}, but this is a smaller fraction.

The amount of MPIs in a collision directly impacts the event activity,
\eg the average charged multiplicity. MPIs have almost exclusively
been studied in $\p\p$ and $\p\pbar$ collisions, however, so we have
no data to go on when we now want to extend it to all the different
collision types listed in \tabref{tab:XYcoef}. As a guiding principle 
we assume that $\langle n_{\mathrm{MPI}} \rangle$ should remain roughly
constant, \ie plausibly hadronic collisions at a given (large) energy
have a comparable event activity, irrespective of the hadron types.
But we already assumed that total cross sections are lower for
mesons than for baryons, and falling for hadrons with an increasing
amount of strange, charm or bottom quarks, so naively then
\eqref{eq:nMPI} would suggest a correspondingly rising
$\langle n_{\mathrm{MPI}} \rangle$. There are (at least) two ways to
reconcile this.

One is to increase the $p_{\perp 0}$ scale to make the MPI cross section
decrease. It is a not unreasonable point of view that a lower cross section
for a hadron is related to a smaller physical size, and that this implies
a larger screening. But it is only then interactions at small $p_{\perp}$
scales that are reduced, while the ones at larger scales remain. 

The alternative is to modify the PDFs and to let heavier quarks
take a larger fraction of the respective total hadron momentum, such
that there are fewer gluons and sea quarks at small $x$ values and
therefore a reduced collision rate. (A high-momentum quark will have
an enhanced high-$p_{\perp}$ collision rate, but that is only one
parton among many.) This is actually a well-established ``folklore'',
that all long-lived constituents of a hadron must travel at approximately
the same velocity for the hadron to stick together. It is a crucial
aspect of the ``intrinsic charm'' hypothesis \cite{Brodsky:1980pb},
where a long-lived $\c\cbar$ fluctuation in a proton takes a major
fraction of the total momentum. In the inverse direction it has also
been used to motivate heavy-flavour hadronization
\cite{Suzuki:1977km,Bjorken:1977md}. This is the approach we will
pursue in the following.

\subsection{New parton distribution functions}

Most PDF studies have concerned and still concern the proton, not least
given the massive influx of HERA and LHC data. Several groups regularly
produce steadily improved PDF sets
\cite{Ball:2017nwa,Hou:2019efy,Bailey:2020ooq}. The emphasis of these
sets are on physics at high $Q^2$ and (reasonably) high $x$ to NLO or
NNLO precision. In our study the emphasis instead is on inclusive events,
dominated by MPIs at scales around $p_{\perp 0}$, \ie a few GeV, and
stretching down to low $x$ values. These are regions where NLO/NNLO
calculations are notoriously unstable, and LO descriptions are better
suited.   

Moving away from protons, data is considerably more scarce. There is
some for the pion, \eg
\cite{Badier:1983mj,Betev:1985pf,Bordalo:1987cs, Conway:1989fs},
a very small amount for the Kaon \cite{Badier:1980jq}, and nothing
beyond that. There has also been some theoretical PDF analyses,
based on data and/or model input, like
\cite{Gluck:1991ey,Sutton:1991ay,Gluck:1997ww,Gluck:1999xe,Clerbaux:2000hb,
Davidson:2001cc,Bissey:2002yr,Detmold:2003tm,Aicher:2010cb,
Holt:2010vj,Han:2018wsw,Watanabe:2016lto,Barry:2018ort,
Watanabe:2019zny,Ding:2019lwe,Novikov:2020snp} for the pion and
\cite{Gluck:1997ww,Davidson:2001cc,Watanabe:2017pvl,Watanabe:2018qju,
Shi:2018mcb,Lan:2019rba,Cui:2020tdf} for the Kaon. But again nothing
for hadrons beyond that, to the best of our knowledge, which prompts
our own work on the topic.

In order to be internally consistent, we have chosen to take the
work of Gl\"uck, Reya and coworkers as a starting point. The basic
idea of their ``dynamically generated'' distributions is to start
the evolution at a very low $Q_0$ scale, where originally the input
was assumed purely valence-quark-like \cite{Gluck:1988xx}. Over the
years both gluon and $\u/\d$ sea distributions have been introduced
to allow reasonable fits to more precise data
\cite{Gluck:1991ng,Gluck:1994uf,Gluck:1998xa,Gluck:2007ck},
but still with ans\"atze for the PDF shapes at $Q_0$ that involve a
more manageable number of free parameters than modern high-precision
(N)NLO ones do. Their LO fits also work well with the \textsc{Pythia}
MPI framework. To be specific, we will use the GRS99 pion 
\cite{Gluck:1999xe} as starting point for meson PDFs, and the
GJR07 proton one \cite{Gluck:2007ck} similarly for baryons.
Also the GR97 Kaon one \cite{Gluck:1997ww} will play some role.

In the LO GRS99 $\pi^+$ PDF the up/down valence, sea, and gluon distributions
are all parameterized on the form 
\begin{equation}
f(x) = N x^a (1-x)^b (1 + A \sqrt{x} + B x)
\label{eq:basicPDF}
\end{equation}
at the starting scale $Q_0^2 = 0.26$~GeV$^2$. The sea is taken
symmetric $\u_{\mathrm{sea}} = \ubar = \d = \dbar_{\mathrm{sea}}$, while
$\s = \c = \b = 0$ at $Q_0$. The GRS97 $\K^+$ PDFs are described by assuming
the total valence distribution to be the same as for $\pi^+$ (as specified
in the same article), but the $\u$ PDF is made slightly softer by
multiplying it by a factor $(1-x)^{0.17}$. That is, the $\K^+$ valence
PDFs are given in terms of the $\pi^+$ PDFs as
\begin{equation}
\begin{gathered}
  v_\u^\K    = N_{\K/\pi} (1-x)^{0.17} v_\u^\pi, \\
  v_\sbar^\K = (v_\dbar^\pi + v_\u^\pi) - v_\u^\K.
\end{gathered}
\label{eq:KPDF}
\end{equation}
The coefficient $N_{\K/\pi}$ is a normalization constant determined by the 
flavour sum relation
\begin{equation}
\int_0^1 dx~v(x) = 1.
\label{eq:pdfValenceSum}
\end{equation}
Gluon and sea (both $\u/\d$ and $\s$) distributions are taken to be the same
as for the pion.

\begin{figure}
  \includegraphics[width=0.49\linewidth]{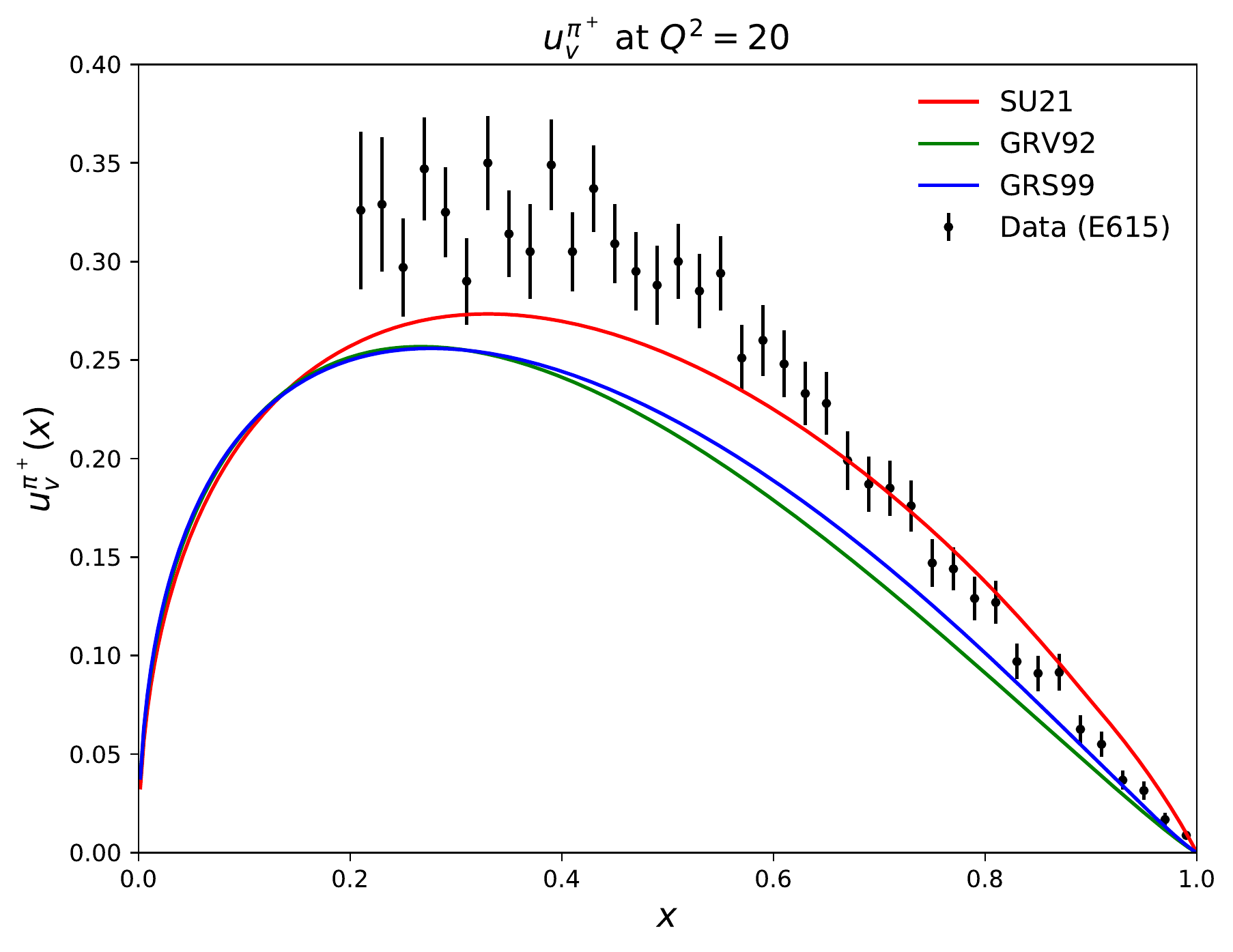}
  \includegraphics[width=0.49\linewidth]{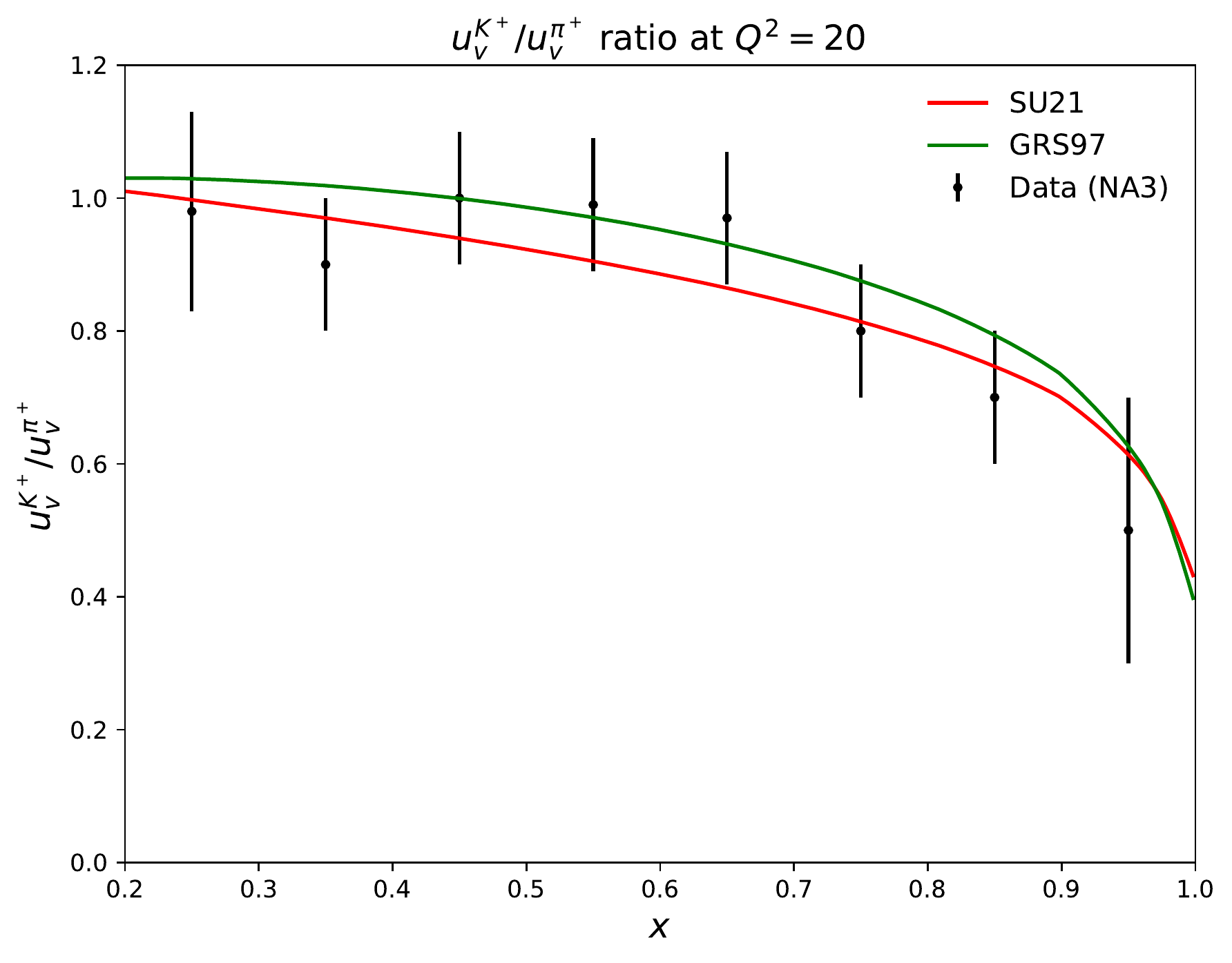}
  \caption{(a) Pion PDFs, comparing our simplified form to GRV92 and GRS99,
    and to data. 
    (b) $\ubar_v^{\K^-} / \ubar_v^{\pi^-}$ ratio, comparing our simplified
    parameterization on the form given in \eqref{eq:basicPDF} to the slightly
    more detailed description of GRS97, and comparing to data. Note that both
    cases use our simplified $\ubar_v^{\pi^-}$ shown in (a), and only differ 
    in $\ubar_v^{\K^-}$.}
  \label{fig:PDFdata}
\end{figure}

In our work, we make the ansatz that hadron PDFs can be
parameterized on the form given in \eqref{eq:basicPDF} at the initial
scale $Q_0$, but with $A = B = 0$ since there are no data or guiding
principles to fix them in the generic case. The $a$ and $b$ parameters
are allowed to vary with the particular parton and hadron in question,
while $N$ is fixed by \eqref{eq:pdfValenceSum} for valence quarks.
The deviations introduced by the  $A = B = 0$ assumption are illustrated
in \figref{fig:PDFdata}. In \figref{fig:PDFdata}a E615 data
\cite{Conway:1989fs} are compared with the $\pi$ PDFs as given
by GRV92 \cite{Gluck:1991ey}, by GRS99 \cite{Gluck:1999xe}, and
by our simplified description (labeled SU21) where $a$ has been adjusted
to give the same $\langle x \rangle$ as for GRS99. In \figref{fig:PDFdata}b
the $\ubar_v^{\K^-} / \ubar_v^{\pi^-}$ ratio is compared between data
\cite{Badier:1980jq}, GRS97 \cite{Gluck:1997ww} and our simplified model.
In both cases the model differences are comparable with the uncertainty
in data.

\begin{figure}
\begin{minipage}[c]{0.49\linewidth}
  \centering
  \includegraphics[width=\linewidth]{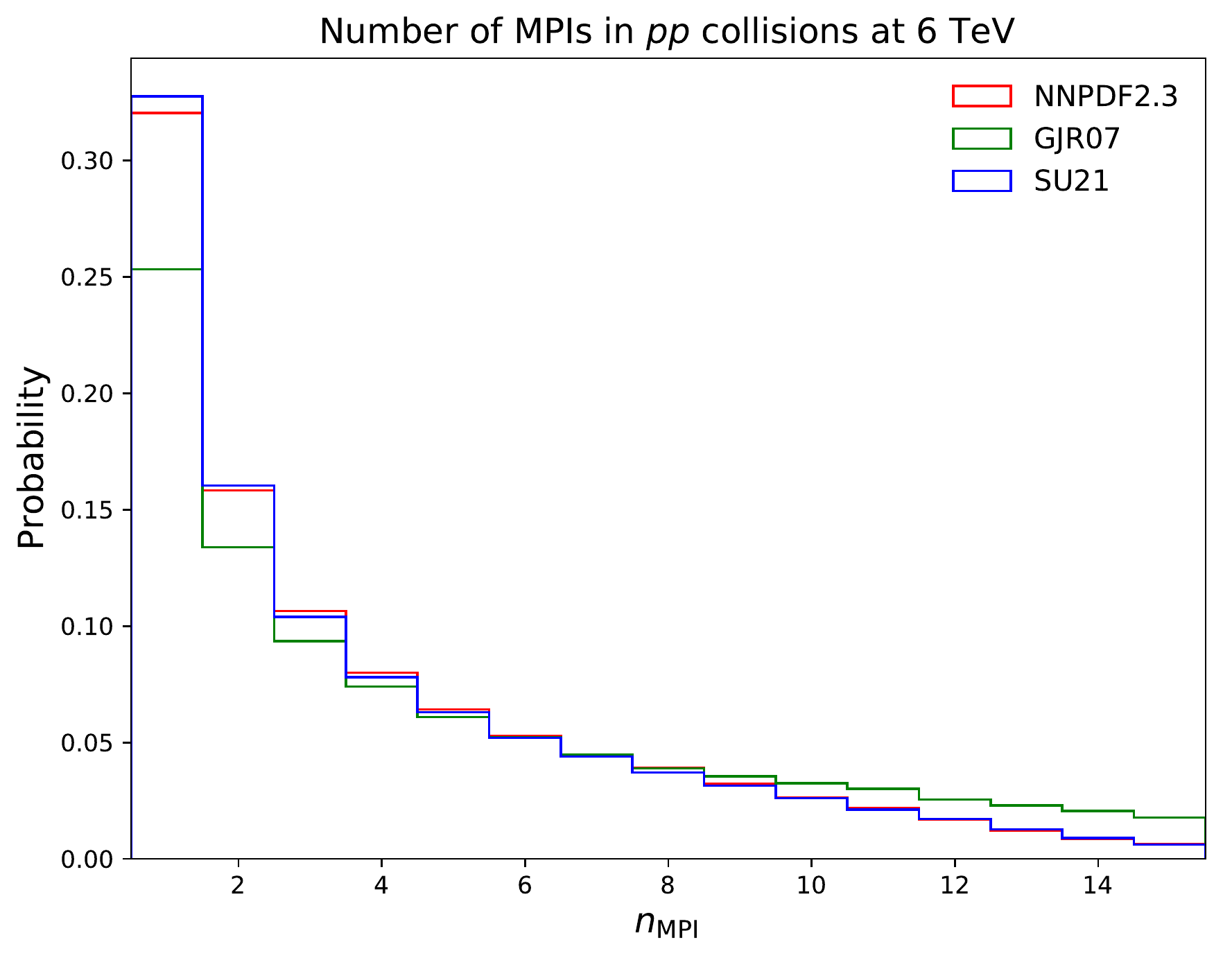}
  (a)
\end{minipage}
\begin{minipage}[c]{0.49\linewidth}
  \centering
  \includegraphics[width=\linewidth]{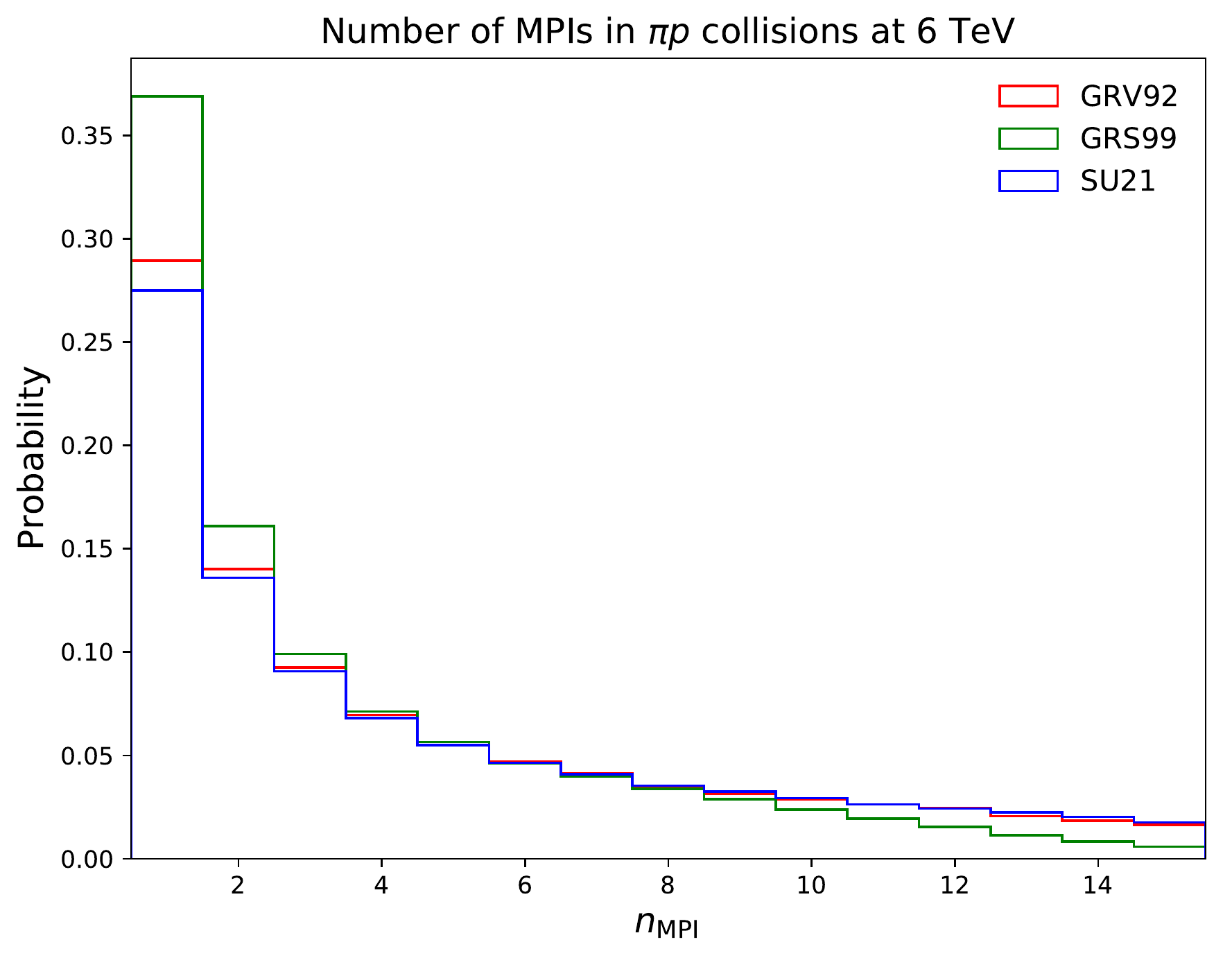}
  (b)
\end{minipage}
\begin{minipage}[c]{0.49\linewidth}
  \centering
  \includegraphics[width=\linewidth]{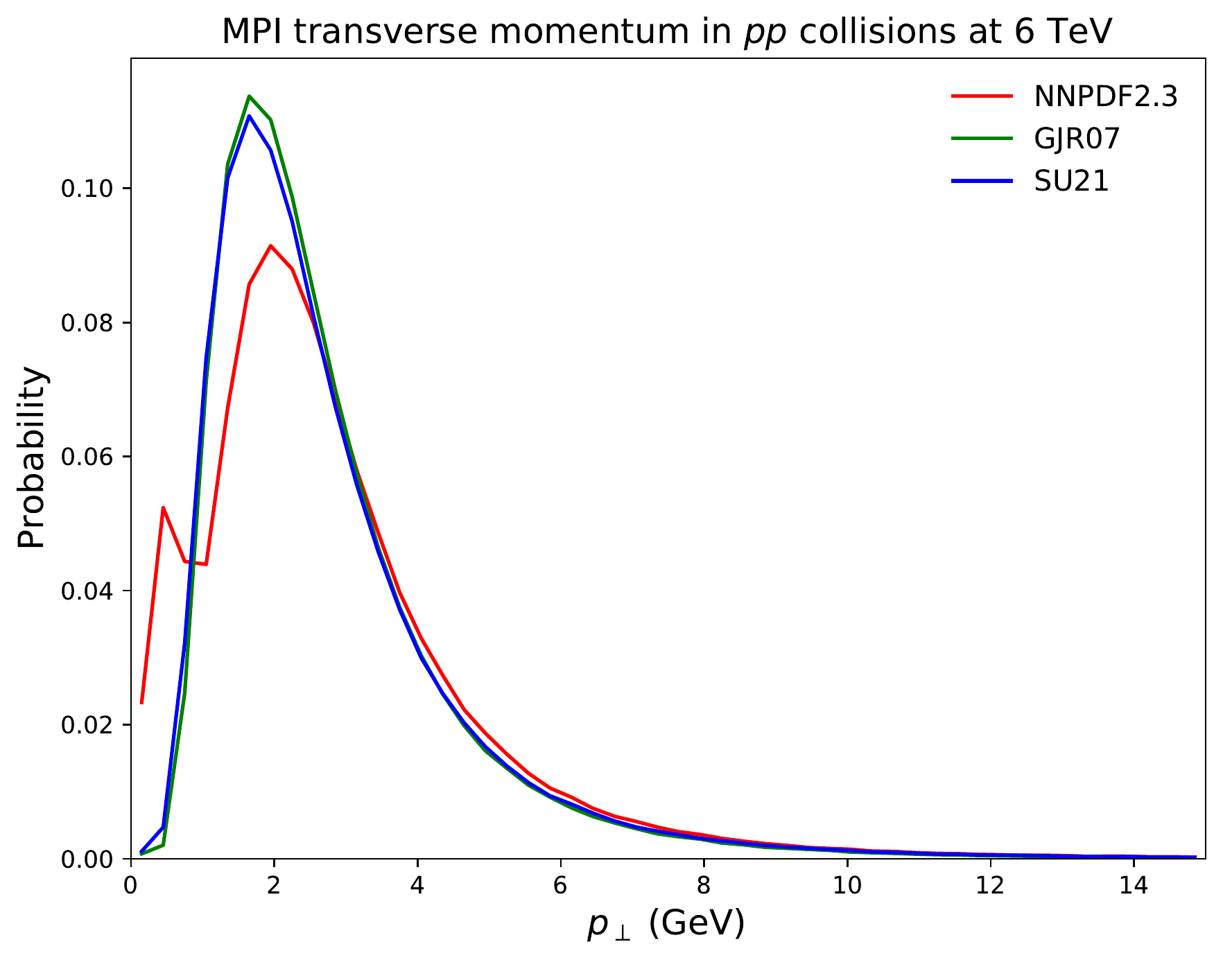}
  (c)
\end{minipage}
\begin{minipage}[c]{0.49\linewidth}
  \centering
  \includegraphics[width=\linewidth]{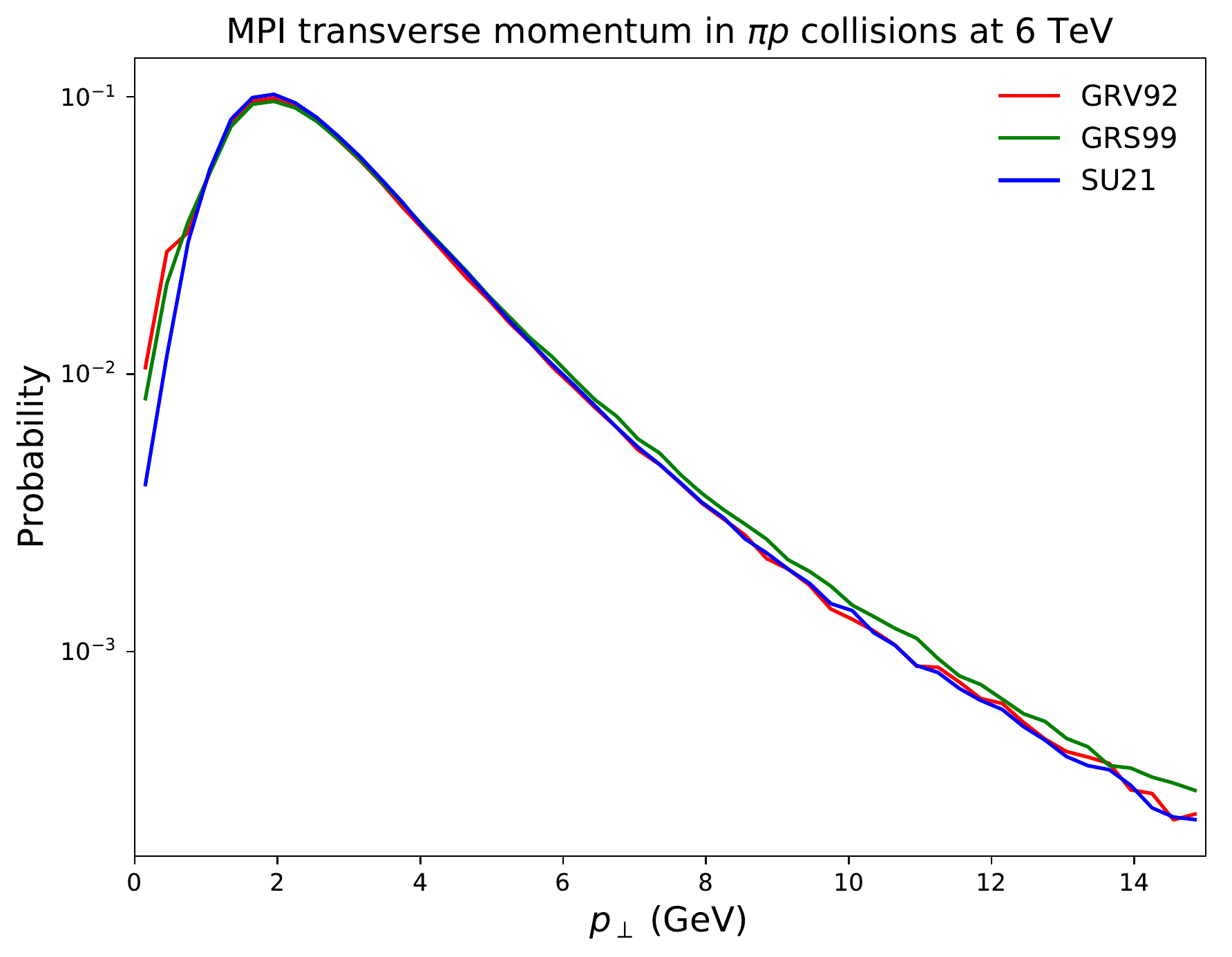}
  (d)
\end{minipage}
\caption{The (a,b) number and (c,d) transverse momentum spectrum of MPIs,
for (a,c) protons and (b,d) pions. Each PDF has a cutoff and is considered
constant below some $Q_0$, which leads to the bumps at low $p_\perp$,
especially noticeable for NNPDF2.3 distribution in (c), whose cutoff
is $Q = 0.5$~GeV.}
\label{fig:PDFppi}
\end{figure}

\begin{table} \centering
\begin{tabular}{|l|c c|}
  \hline
   & $\langle n_\mathrm{MPI} \rangle$ & $\langle p_{\perp,\mathrm{MPI}} \rangle$ \\
  \hline
  $\p$, NNPDF2.3 & 3.27 & 2.56 \\
  $\p$, GJR07   & 3.88 & 2.58 \\
  $\p$, SU21    & 3.24 & 2.54 \\
  \hline
  $\pi$, GRV92 & 3.70 & 2.67 \\
  $\pi$, GRS99 & 3.10 & 2.68 \\
  $\pi$, SU21  & 3.78 & 2.72 \\
  \hline
\end{tabular}
\caption{Average number and transverse momentum spectrum for MPIs with
different PDFs in $\p\p$ and $\pi\p$ collisions. The default $\p$ PDF
in \Pythia is NNPDF2.3 QCD+QED LO with $\alpha_S(M_\mathrm{Z}) = 0.130$
\cite{Ball:2013hta}. This default is used for the proton PDF in the $\pi\p$
collisions.}
\label{tab:PDFppi}
\end{table}

To further illustrate the changes introduced by setting $A = B = 0$,
\figref{fig:PDFppi} shows the number and transverse momentum of MPIs
for different (a) proton and (b) pion PDFs, with average values as in
\tabref{tab:PDFppi}. In both cases, our simplified SU21 ansatz leads
to a shift that is comparable to the difference between the two standard
PDFs. Thus we feel confident that our simplified ansatz is sufficient
also for other hadrons, where there are neither data nor detailed theory
calculations available. Nevertheless, for accuracy, we use the NNPDF2.3
QCD+QED LO distribution function for protons and GRS99 LO for pions in our
studies, and the SU21 ansatz only for hadrons beyond that.

Given that there is no solid theory for heavy hadron PDFs,
the specific choices of $a$ and $b$ necessarily are heuristic. Our guiding
principle is that all quarks should have roughly the same velocity,
as already mentioned, and thus heavier quarks must have a larger average
momentum fraction $\langle x \rangle$, and a smaller $b$, while gluons and
sea $\u$/$\d$ must be softer. The $\langle x \rangle$ choices do not exactly
agree with the assumed mass ratios in our AQM ansatz, \eqref{eq:nqAQM}, but
are somewhat less uneven than that. This is supported by the Kaon data
\cite{Badier:1980jq}, and also by some modelling \cite{Brodsky:1980pb}.

Except for some fine print to come later, our procedure to determine
PDFs at the $Q_0$ starting scale is as follows:
\begin{enumerate}
\item Let the valence quark distributions be given by $N x^a (1-x)^b$,
  \ie put $A = B = 0$.
  \item Choose sensible $b$ and $\langle x \rangle$ values for each
  valence quark, based on the principles above.
  \item Derive $a$ from $\langle x \rangle =
  \frac{\int_0^1 dx\,x\,f(x)}{\int_0^1 dx\,f(x)} = (a + 1) / (a + b + 2)$.
  \item Derive $N$ to satisfy \eqref{eq:pdfValenceSum}.
  \item For sea and gluon distributions, pick a $d$ and set
  $f(x) \propto x^d f^\pi(x)$ (here with $A$ and $B$ values as for
  the pion).
  \item Rescale the gluon and $\u/\d$ sea distributions by a common 
        factor to satisfy the momentum sum relation
    \begin{equation*}
      \int_0^1 dx~\sum\limits_q x f_q(x) = 1.
    \end{equation*}
  \item The $\s$, $\c$ and $\b$ contents are zero at the starting scale.
\end{enumerate}

\begin{figure}
\includegraphics[width=0.49\linewidth]{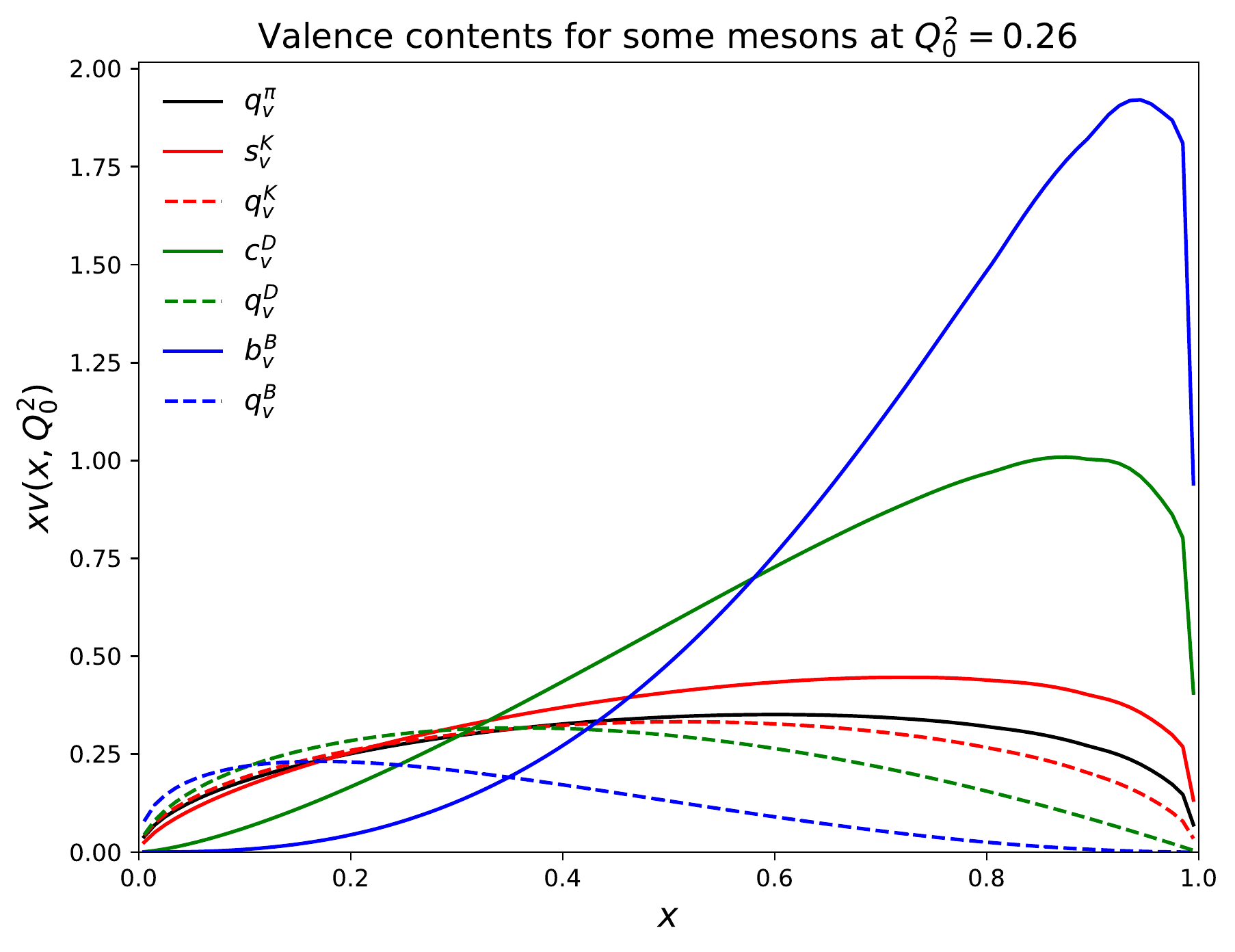}
\includegraphics[width=0.49\linewidth]{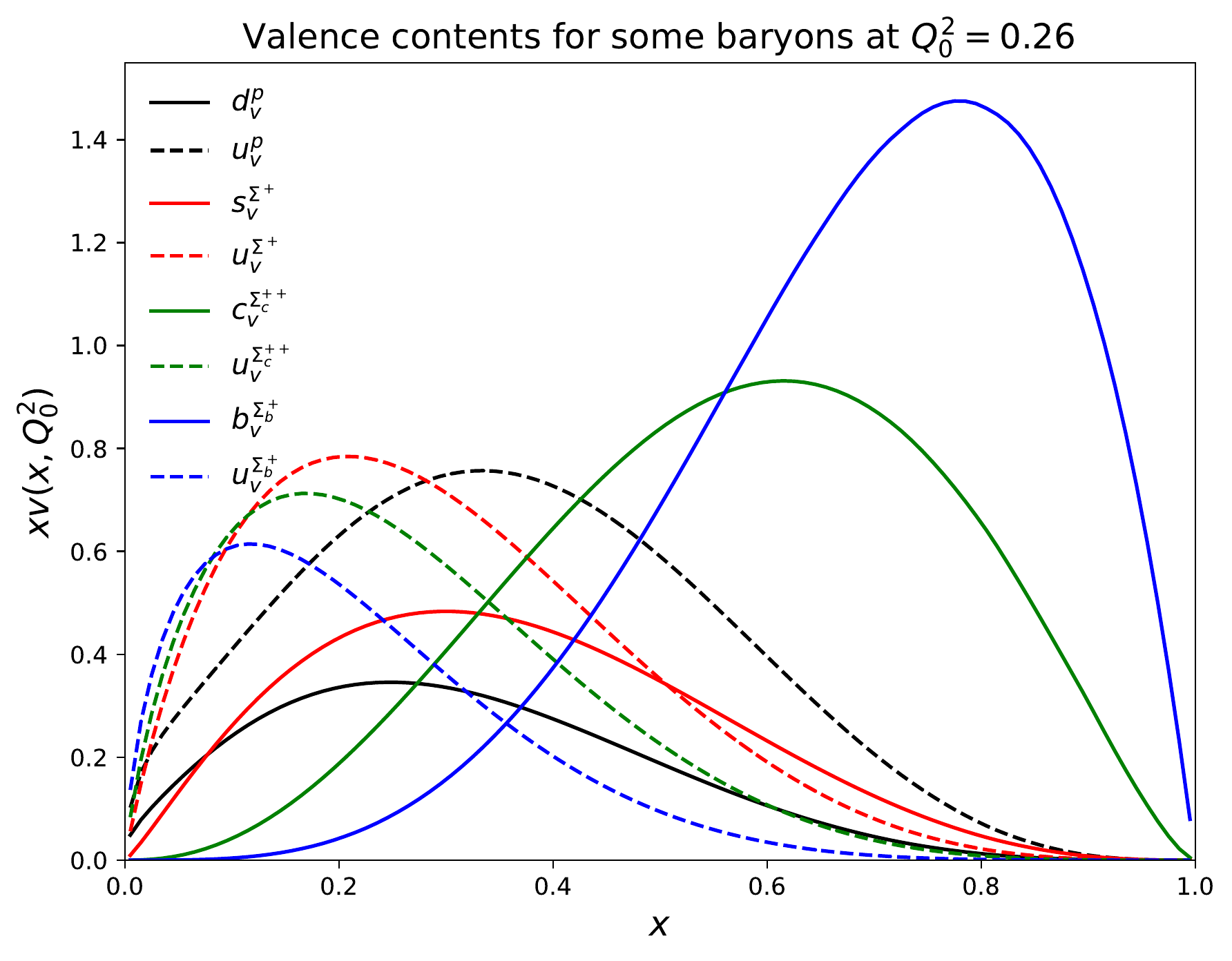}
\caption{Different valence PDFs at the initial scale $Q_0^2 = 0.26$~GeV$^2$,
for (a) $\pi$, $\K$, $\D$ and $\B$ mesons, showing the flavoured valence and
the $q$ ($=\d/\u$) valence contents; and for (b) $\u\u q$ baryons for
$q = \d$ (proton), $\s$ ($\Sigma^+$), $\c$ ($\Sigma_\c^{++}$) and $\b$ ($\Sigma_\b^+$).}
\label{fig:Q0-pdfs}
\end{figure}

\begin{table}[tp!] \centering
\begin{tabular}{|c | c c | c c | c c | c|}
\hline
  Particle & $b_1$ & $\langle x \rangle_1$ & $b_2$ & $\langle x \rangle_2$
  & $b_3$ & $\langle x \rangle_3$ & $d$ \\
    \hline
    $\pi$                  & 0.35 & 0.28  & 0.35 & 0.28  &  --  &  --  & 0.00 \\
    $\K$                   & 0.25 & 0.34  & 0.52 & 0.26  &  --  &  --  & 0.17 \\
    $\eta$                 & 0.32 & 0.30  & 0.32 & 0.30  &  --  &  --  & 0.17 \\
    $\phi$                 & 0.30 & 0.32  & 0.30 & 0.32  &  --  &  --  & 0.17 \\
    $\D$                   & 0.20 & 0.55  & 1.00 & 0.22  &  --  &  --  & 1.00 \\
    $\D_s$                 & 0.25 & 0.53  & 0.80 & 0.26  &  --  &  --  & 1.00 \\
    $\Jpsi$                & 0.30 & 0.43  & 0.30 & 0.43  &  --  &  --  & 2.00 \\
    $\B$                   & 0.15 & 0.70  & 2.00 & 0.12  &  --  &  --  & 2.00 \\
    $\B_\s$                & 0.20 & 0.68  & 1.60 & 0.16  &  --  &  --  & 2.00 \\
    $\B_\c$                & 0.25 & 0.64  & 1.00 & 0.24  &  --  &  --  & 3.00 \\
    $\Upsilon$             & 0.30 & 0.46  & 0.30 & 0.46  &  --  &  --  & 4.00 \\
    \hline
    $\Sigma/\Lambda$       & 2.8  & 0.24  &  3.5 & 0.17  &  3.5 & 0.17 & 0.17 \\
    $\Xi$                  & 3.0  & 0.235 &  3.0 & 0.235 &  3.8 & 0.15 & 0.17 \\
    $\Omega$               & 3.2  & 0.22  &  3.2 & 0.22  &  3.2 & 0.22 & 0.17 \\
    $\Sigma_\c/\Lambda_c$  & 1.5  & 0.49  &  4.0 & 0.14  &  4.0 & 0.14 & 1.00 \\
    $\Xi_\c$               & 1.6  & 0.475 &  3.9 & 0.16  &  4.5 & 0.14 & 1.00 \\
    $\Omega_\c$            & 1.7  & 0.46  &  3.8 & 0.16  &  3.8 & 0.16 & 1.00 \\
    $\Sigma_\b/\Lambda_\b$ & 1.0  & 0.64  &  5.0 & 0.10  &  5.0 & 0.10 & 2.00 \\
    $\Xi_\b$               & 1.1  & 0.625 &  4.8 & 0.12  &  5.0 & 0.10 & 2.00 \\
    $\Omega_\b$            & 1.2  & 0.61  &  4.8 & 0.12  &  4.8 & 0.12 & 2.00 \\
    \hline
  \end{tabular}
  \caption{Input parameters for the implemented hadron PDFs, as described
  in the text. Columns are ordered so that heavier quarks appear first.
  Excited hadrons are also implemented, using the same parameters as for
  a lighter hadron with the same flavour content.}
  \label{tab:pdfparams}
\end{table}

\noindent Our choices of $b$, $\langle x \rangle$ and $d$ are given in
\tabref{tab:pdfparams}. Excited particles use the same PDFs as their
unexcited counterparts. Some PDFs at the initial scale are shown in
\figref{fig:Q0-pdfs} for (a) mesons and (b) baryons, which clearly show
how heavier quarks are made harder. The baryons are normalized to two
$\u$ valence quarks, and still the $\c$/$\b$ peaks in 
$\Sigma_\c^{++}$/$\Sigma_\b^+$ stand out in the comparison.

Once the initial state has been set up, the DGLAP equations
\cite{Gribov:1972ri,Dokshitzer:1977sg,Altarelli:1977zs}
describe the evolution towards higher $Q^2$ scales. Any number
of implementations of these equations exist, both private and public,
such as \QCDNUM \cite{Botje:2010ay}, HERAFitter/xFitter
\cite{Alekhin:2014irh} and APFEL \cite{Bertone:2013vaa},
that in principle should be equivalent. We choose to use QCDNUM since
we find it well documented and well suited for our purposes.
Nevertheless there are some limitations that we had to circumvent.

One such is that the framework is not set up to handle $\c$ and $\b$
quarks below the respective thresholds $Q_\c^2$ and $Q_\b^2$. To handle their
presence, we map some flavours onto others during evolution. Consider \eg
a $\B^+ = \u\bbar$ meson, where we wish to evolve the bottom valence
$v_{\bbar}$ by $\bbar \to \bbar\g$ branchings starting from $Q_0^2$,
but allow $\g \to \b\bbar$ only above $Q_\b^2$. To handle this,
we can redefine the initial $\bbar$ valence as a contribution \eg to
the $\dbar$ content, \ie set $\tilde{f}_{\dbar}(x, Q_0^2) =
f_{\sea}(x, Q_0^2) + v_{\bbar}(x, Q_0^2)$. Since evolution is linear,
this relation also holds for $Q_0^2 \to Q^2 > Q_0^2$, while 
$f_\d(x, Q^2) = f_{\sea}(x, Q^2)$. For $Q^2 > Q_\b^2$ there will also
be a ``sea'' bottom content $\tilde{f}_\b(x, Q^2)$ from $g \to \b\bbar$ 
splittings. Then the correct $\dbar$ and $\bbar$ contents are 
reconstructed as
\begin{equation}
\begin{split}
  f_\bbar(x, Q^2) &= \tilde{f}_\b(x, Q^2) + (\tilde{f}_\dbar(x, Q^2) - f_\d(x, Q^2)), \\
  f_\dbar(x, Q^2) &= f_\d(x, Q^2).
\end{split}
\end{equation}
For doubly heavy flavoured mesons, like $\B_\c$, we place one valence
content in $\dbar$ and the other in $\u$, then use the same procedure. 
The same trick can be modified to work for flavour-diagonal mesons,
like $\phi$, $\Jpsi$ and $\Upsilon$, \eg by adding the valence content
to $\d$ and $\dbar$. Afterward the $\u = \ubar = \d = \dbar$ symmetry
of the unmodified sea can be used to shift the heavy flavour content back
where it belongs.

There is a further complication for $\eta$ and $\eta'$, which fluctuate
between $\u\ubar/\d\dbar/\s\sbar$ valence states. We handle this by
treating them as a  $\u\ubar$ state during evolution. After a specific quark
content is chosen during event generation, the valence part of the evolved
$\u\ubar$ is shifted to the corresponding distribution. For simplicity
both $\eta$ and $\eta'$ are assumed to have the same valence
$\langle x \rangle$ values, intermediate between $\pi$ and $\phi$,
whether in a $\s\sbar$ state or in a $\u\ubar/\d\dbar$ one.

Baryons are treated similarly to mesons, with the obvious exception that
they have three valence distributions. In this case the starting point
is the GJR07 proton. Despite the known asymmetry of the proton,
that the $\u$ valence is harder than the $\d$ one, we take $\u$ and
$\d$ distributions, where present, to be equal for all other baryons.
For heavy-flavoured baryons, the heavy valence is shifted into the $\d$
valence in analogy with the meson case. We have not implemented doubly-
or triply heavy-flavoured baryons, since these should be produced at a
negligible rate, and thus there are no further complications.

After having studied the PDFs resulting from this procedure, we make one additional
ad hoc adjustment for the $\Jpsi$, $\B_\c$ and $\Upsilon$ mesons, \ie
the ones that have exclusively $\c$ and $\b$ valence content. Using only the
procedure outlined so far, the average number of MPIs in interactions
involving these particles is much higher than for other hadrons.
This comes as no surprise in view of \eqref{eq:nMPI}; the absence of
light quarks makes for a small total and nondiffractive cross section,
while the normal evolution allows a non-negligible gluon and sea to
evolve right from the low $Q_0$ scale. But in real life one should expect
heavy quarks to have a reduced emission rate of gluons below their mass
scale. To compensate for this, increased $Q_0^2$ scales of 0.6~GeV$^2$,
0.75~GeV$^2$ and 1.75~GeV$^2$ are used for $\Jpsi$, $\B_\c$ and $\Upsilon$,
respectively. One could argue that similar shifts should be made for all
hadrons containing a $\c$ or $\b$ quark, but if there are also light
valence quarks then there should be some evolution already from small
scales. Any mismatch in the emission can then more easily be absorbed in
the overall uncertainty of the setup at the $Q_0$ starting scale.

\subsection{The forward region}

The fastest particles in the projectile region play a central role for
the shower evolution in the atmosphere, so the modelling of this region
is a topic of special interest. Traditionally \Pythia is more aimed
towards the modelling of the central region, and there are known issues
in the forward region \cite{LHCf:2012mtr,LHCf:2015nel,Kireyeu:2020wou}.
Briefly put, proton/neutron spectra are softer and pion spectra harder
than data. In \Sibyll, which normally uses Lund string fragmentation,
this has required a separate dedicated handling of leading-baryon
formation \cite{Riehn:2019jet}.

We are not here able to report a final resolution of these issues,
but a beginning has been made with two new options, both which modify the
fragmentation of a diquark in the beam remnant. The first is to disallow
popcorn handling \cite{Andersson:1984af}, \ie the mechanism
$\q_1\q_2 \to \q_1 \qbar_3 + \q_2\q_3$ whereby a meson can become the
leading particle of a jet. (If two valence quarks are kicked out from
the proton, which can happen in separate MPIs, the resulting junction
topology is unaffected by the popcorn handling.) The second is to set
the $a$ and $b$ parameters of the Lund symmetric fragmentation function
$f(z) = (1/z) \, z^a \, \exp(-b m_{\perp}^2 / z)$ separately from those
in normal hadronization. There is some support for such a deviation in a
few Lund studies \cite{Andersson:1980nj,Eden:1996xi,Eden:1996su}, where
it is argued that a drifting-apart of the two quarks of an original
dipole indirectly leads to a hardening of the baryon spectrum. 

\begin{figure}[t]
\includegraphics[width=0.49\linewidth]{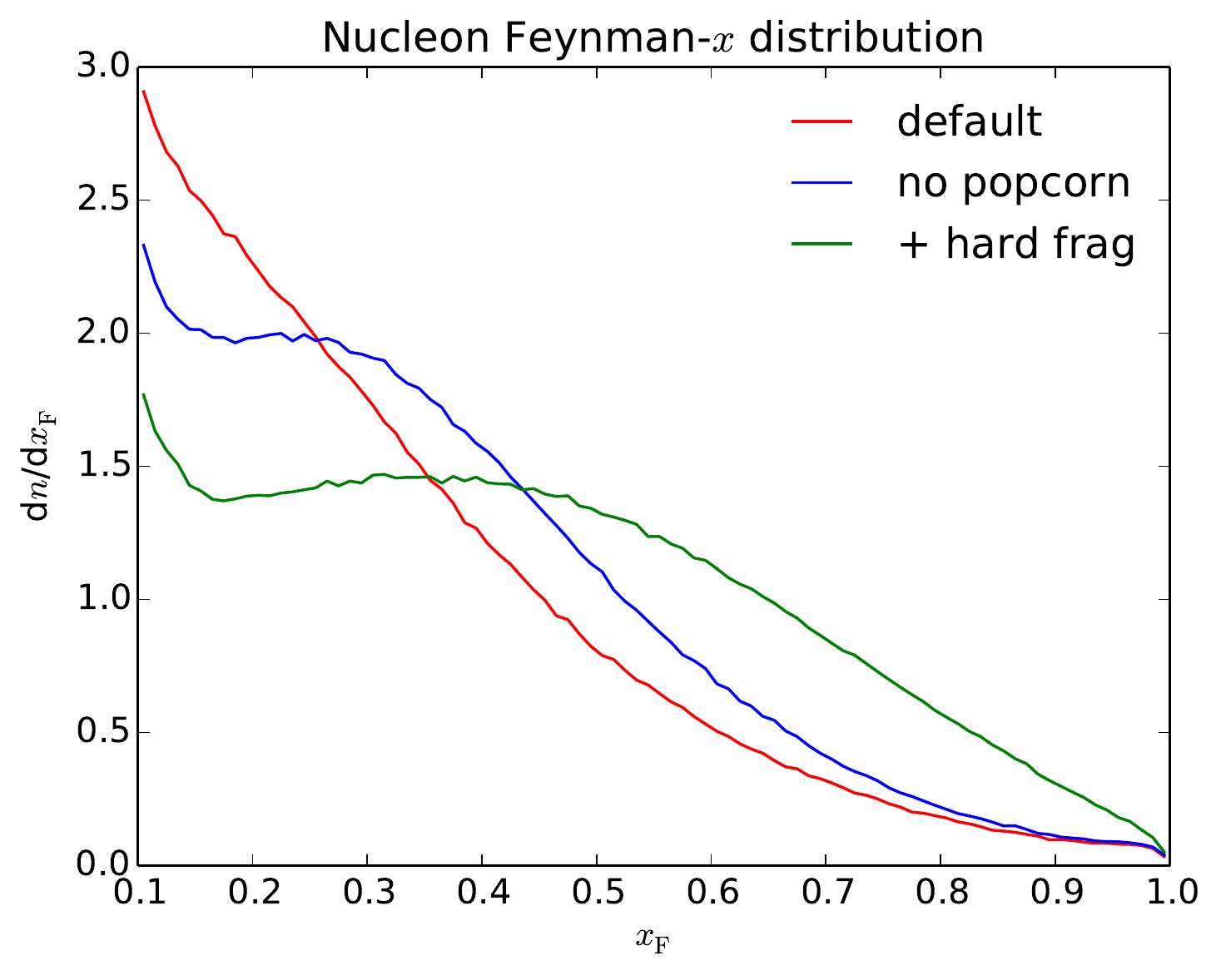}
\includegraphics[width=0.49\linewidth]{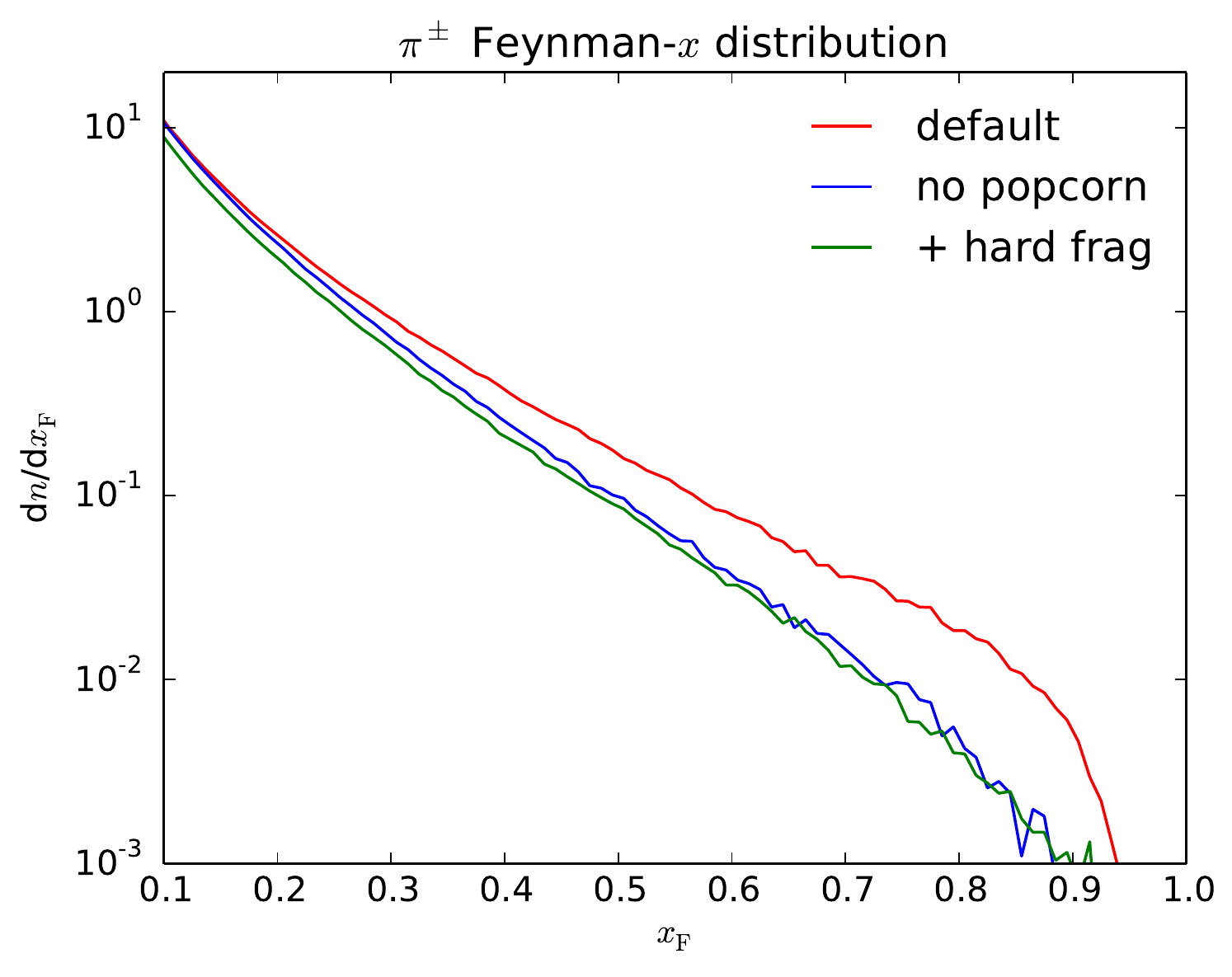}
\caption{Feynman-$x$ spectrum of (a) nucleons $\p/\n$ and (b) $\pi^{\pm}$
for 6~TeV $\p\p$ collisions with inelastic events, where the quasi-elastic
side of single diffraction is not considered.}
\label{fig:forwardxF}
\end{figure}

Some first results are shown in \figref{fig:forwardxF}. For the nucleon
production it can be noted that both steps are about equally important,
where $a = 0$ and $b = 2$~GeV$^{-2}$ in the second step. Obviously other
$a$ and $b$ values could have given a smaller or larger hardening of the
nucleon spectrum.
The composition is roughly 65\% $\p$ and 35\% $\n$, over the full phase space. 
The diffractive peak of protons near $x_{\mathrm{F}} = 1$ has here
been removed; in diffraction only the diffracted side of the event
is studied. Additional baryon--antibaryon pair production becomes
important in the central region, which is why only $x_{\mathrm{F}} > 0.1$
is shown. For pions the major effect comes from removing remnant-diquark
popcorn production, while the baryon fragmentation parameters here have a
lesser effect.  

Further modifications are likely to be necessary, and tuning studies
are in progress \cite{Kling:2021gos}.

\subsection{Technical details}
The new ``SU21'' PDFs will be included in an upcoming release of \Pythia
as LHAPDF-compatible \cite{Buckley:2014ana} files, using the
\texttt{lhagrid1} format, as a central grid. The grids go down to
$x = 10^{-9}$ and up to $Q = 10^4$~GeV.

It is already possible to have a variable energy for the collisions, 
if switched on at initialization. Then the MPI machinery is initialized
at a set of energies up to the maximal one, and later on it is possible 
to interpolate in tables to obtain relevant MPI values at the current energy.
This rather new feature is similar to what has existed a long time for
MPIs in diffractive systems, where the diffractive mass varies from event
to event even for fixed total energy. Note that it is only implemented
for the inclusive processes in the MPI framework, and not for rare
processes.

In a future release, it will also be possible to switch between different
beam particles on an event-by-event basis. It is assumed that $h\p$ and
$h\n$ cross sections are the same, and that $\p$ and $\n$ PDFs are
related by isospin symmetry. (The latter is not quite true when QED effects
are included, but it is close enough for our purposes.) Going one step further,
it is also possible to initialize MPIs for the average behaviour of
processes that have the same pomeron coefficient $X^{AB}$ in the total
cross section, where the PDFs are related by strong isospin, such
that the high-energy behaviour should converge. The main example is
$\pi^+\p$, $\pi^-\p$ and $\pi^0\p$. In detail, the MPI initialization is
then based on the average behaviour of $\sigma_{\mathrm{ND}}$ and
$\d \sigma/ \d p_{\perp}$ \eg in \eqref{eq:nMPI}, but the total cross
section for a collision to occur is still by individual particle combination.

There are other beam combinations that have different total cross sections
and PDFs that are not easily related to each other,
\cf \tabref{tab:XYcoef} and \tabref{tab:pdfparams}. In a study where
the user wishes to switch between such beam combinations during the run,
the PDFs and MPI data grids must be initialized for each
individual case. This may take tens of seconds
per species, and multiplied by twenty this may be annoyingly long for
simple test runs. The future release therefore introduces an option where
the MPI initialization data of an instance can be stored on file and reused
in a later run. This puts some responsibility on the user, since the new
run must then be under the same condition as the original one: same (or only
a subset of) allowed processes, same PDFs, same $p_{\perp 0}$, same (or lower)
maximum energy, and so on. These features are disabled by default and will
only be available if explicitly turned on by the user during initialization.

\section{Event properties and nuclear effects}

With the tools developed in \secref{sec:CSaPD} it is now possible
to generate a single hadron-nucleon collision for a wide selection
of hadrons and at an almost arbitrary energy. Some comparisons
between these hadron-beam options are first presented. But for a realistic
simulation of a full cascade, \eg in the atmosphere, we need
to consider nuclear effects. Here the \Angantyr model provides
some reference results for fixed topologies. Currently it is not
flexible enough for cascade simulation, however, so instead we
introduce a simplified approach. 

\subsection{Hadronic interaction properties}

One of the assumptions made above was that the changes in total
cross sections and in PDFs would match to some approximation,
such that event properties would be comparable over the range of
colliding hadrons. In this section we will briefly investigate how
this works out by studying non-diffractive hadron--proton
collisions at 6~TeV. There is no deep reason for this choice
of CM energy, except that any potential proton--oxygen
(or proton--nitrogen) run at the LHC is likely to be for a
nucleon--nucleon energy in that neighbourhood. The incoming
``projectile'' hadron will be moving in the $+z$ direction and the
``target'' proton in the $-z$ ditto.

\begin{figure}
  \begin{minipage}[c]{0.49\linewidth}
    \centering
    \includegraphics[width=\linewidth]{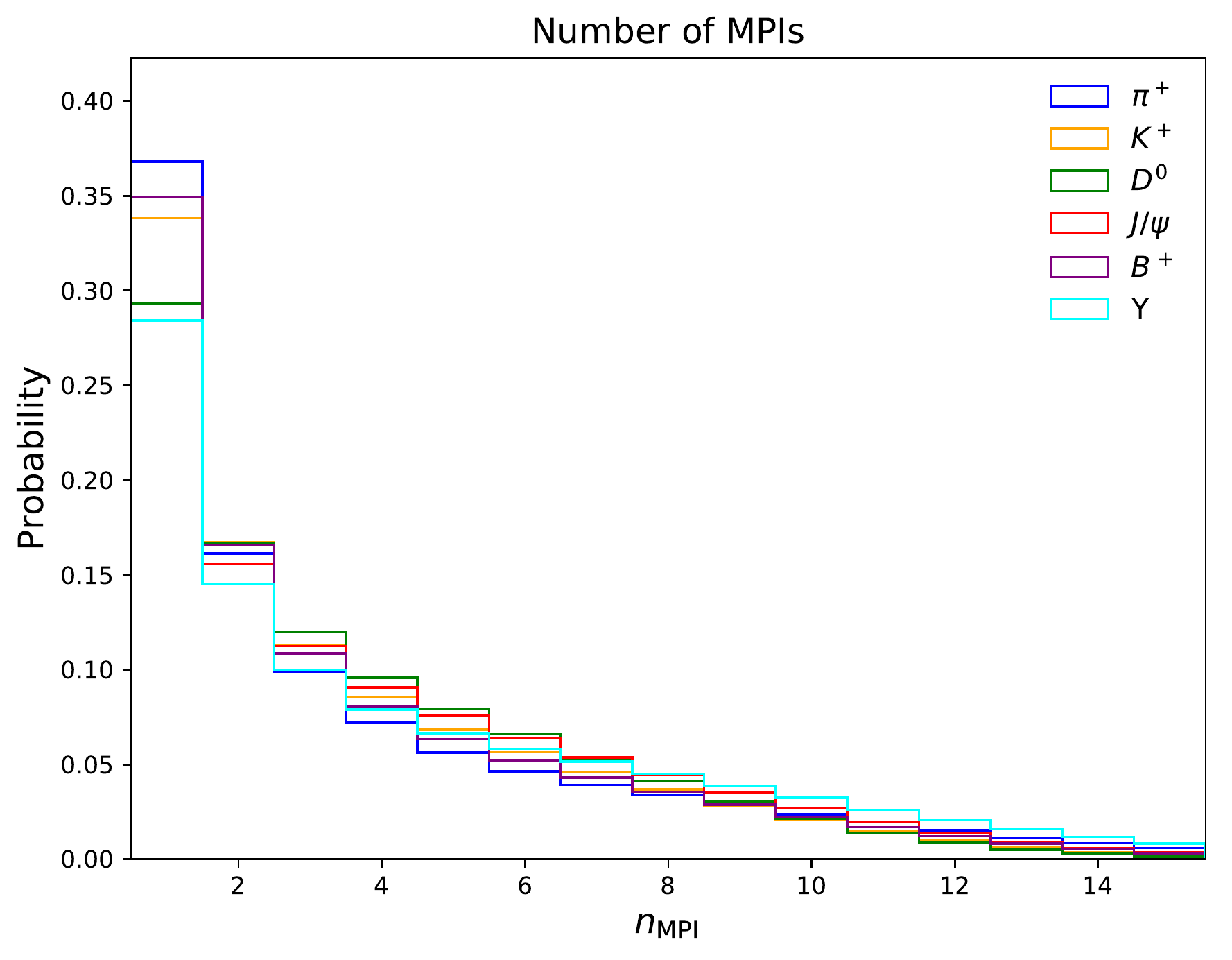}
    (a)
  \end{minipage}
  \begin{minipage}[c]{0.49\linewidth}
    \centering
    \includegraphics[width=\linewidth]{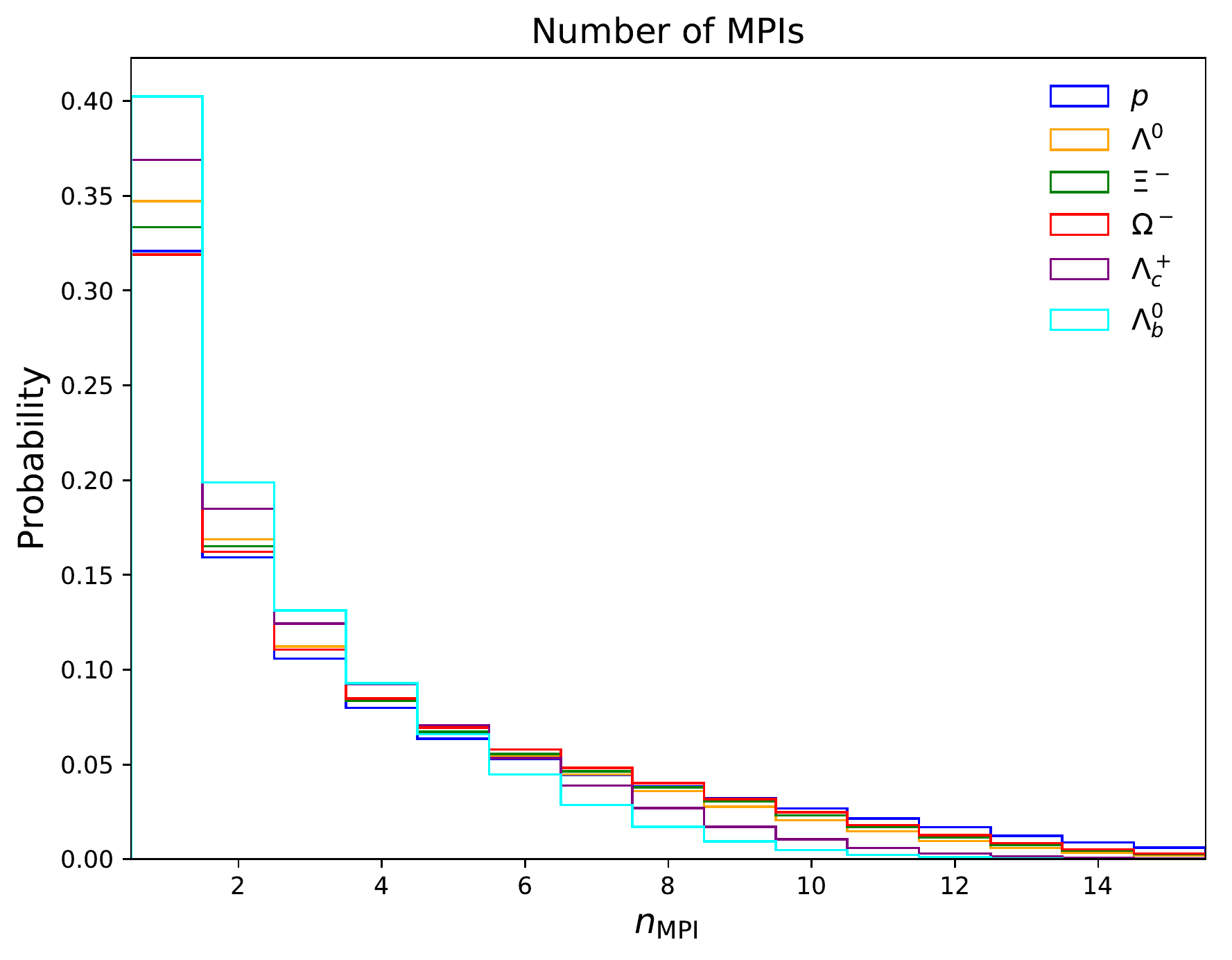}
    (b)
  \end{minipage}
  \begin{minipage}[c]{0.49\linewidth}
    \centering
    \includegraphics[width=\linewidth]{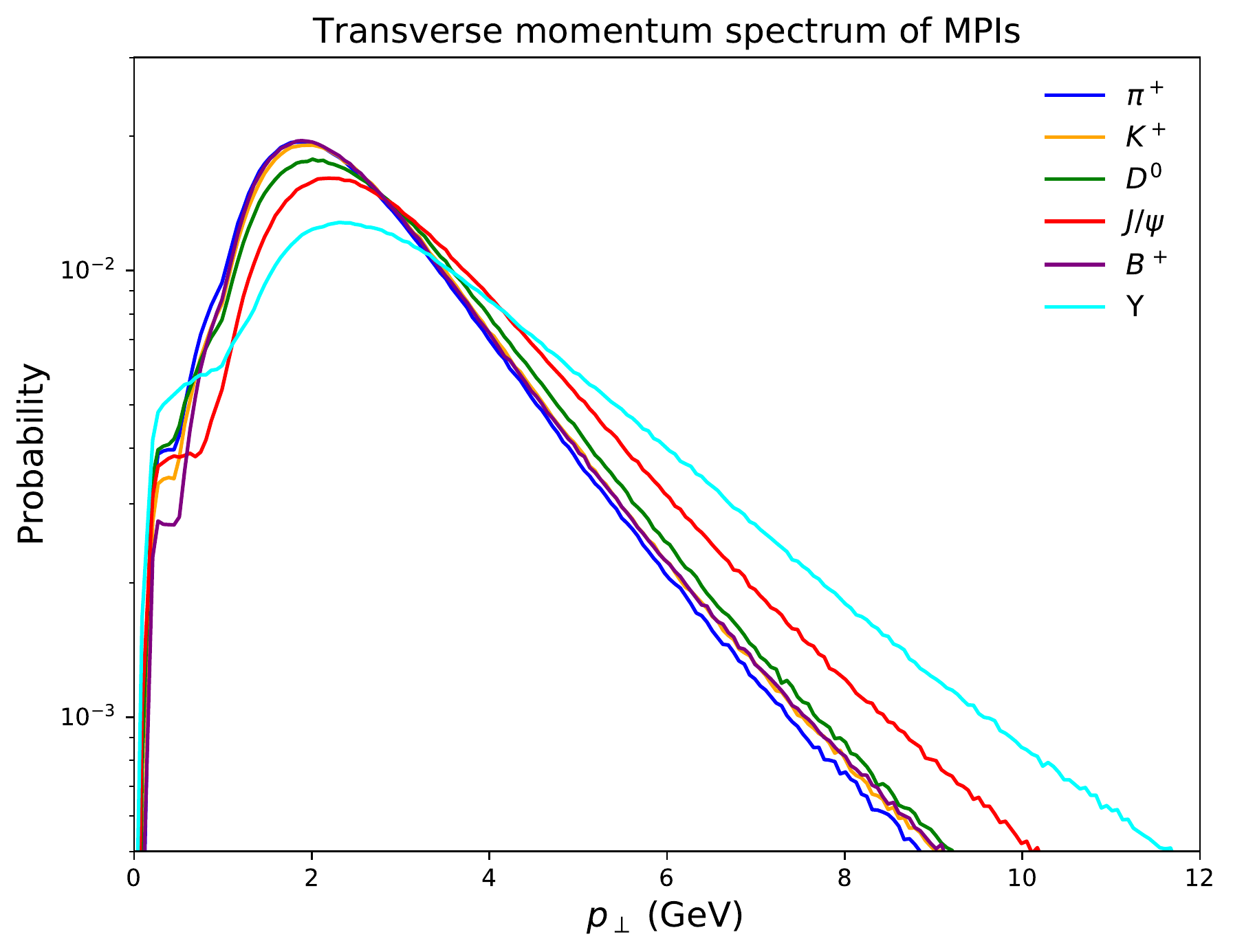}
    (c)
  \end{minipage}
  \begin{minipage}[c]{0.49\linewidth}
    \centering
    \includegraphics[width=\linewidth]{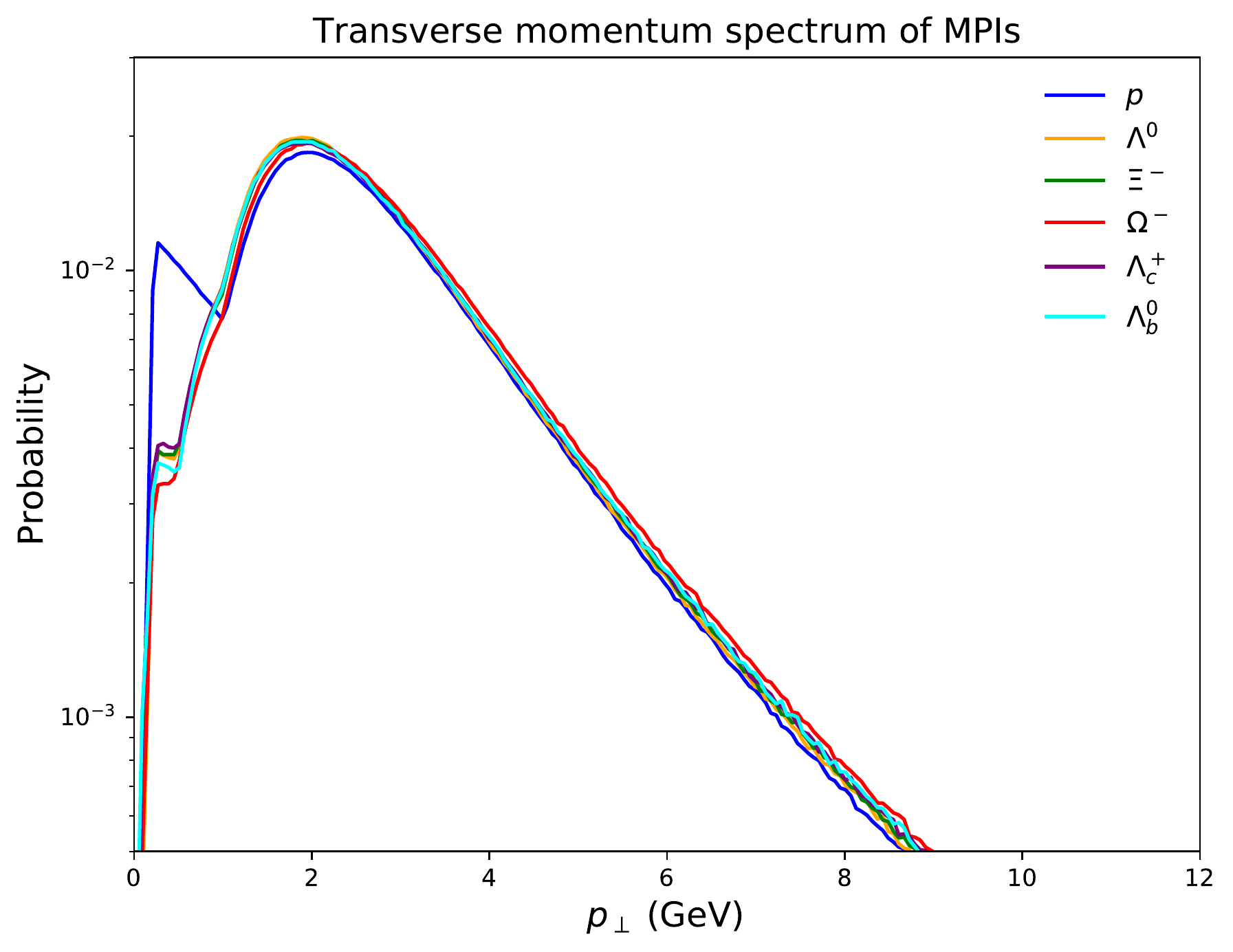}
    (d)
  \end{minipage}
  \caption{The (a,b) number and (c,d) transverse momentum spectrum of MPIs,
    for a selection of (a,c) meson--proton and (b,d) baryon--proton
    collisions at 6~TeV. Labels denote the respective hadron beam.}
  \label{fig:MPIs}
\end{figure}

\begin{figure}
\begin{minipage}[c]{0.49\linewidth}
  \centering
  \includegraphics[width=\linewidth]{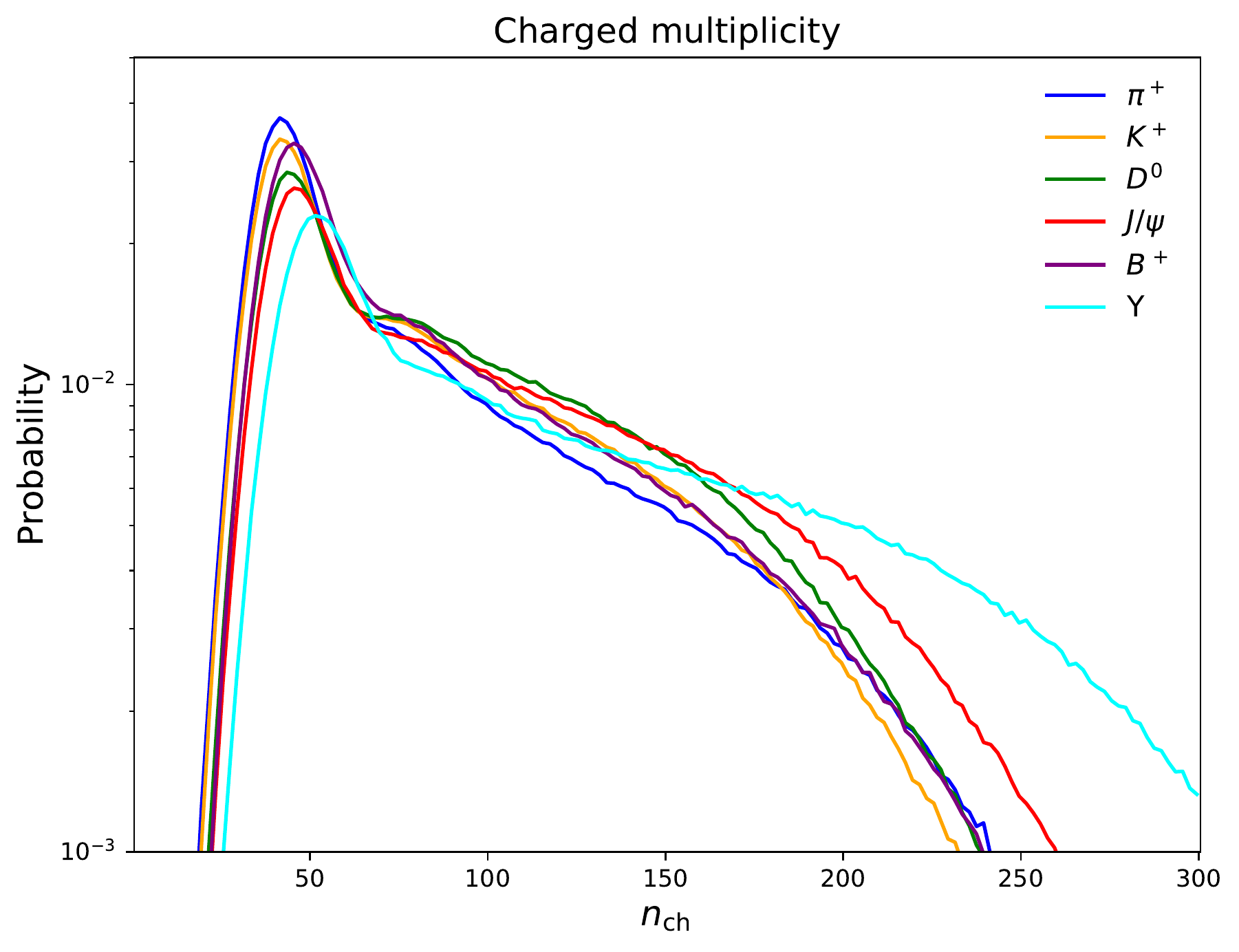}
  (a)
\end{minipage}
\begin{minipage}[c]{0.49\linewidth}
  \centering
  \includegraphics[width=\linewidth]{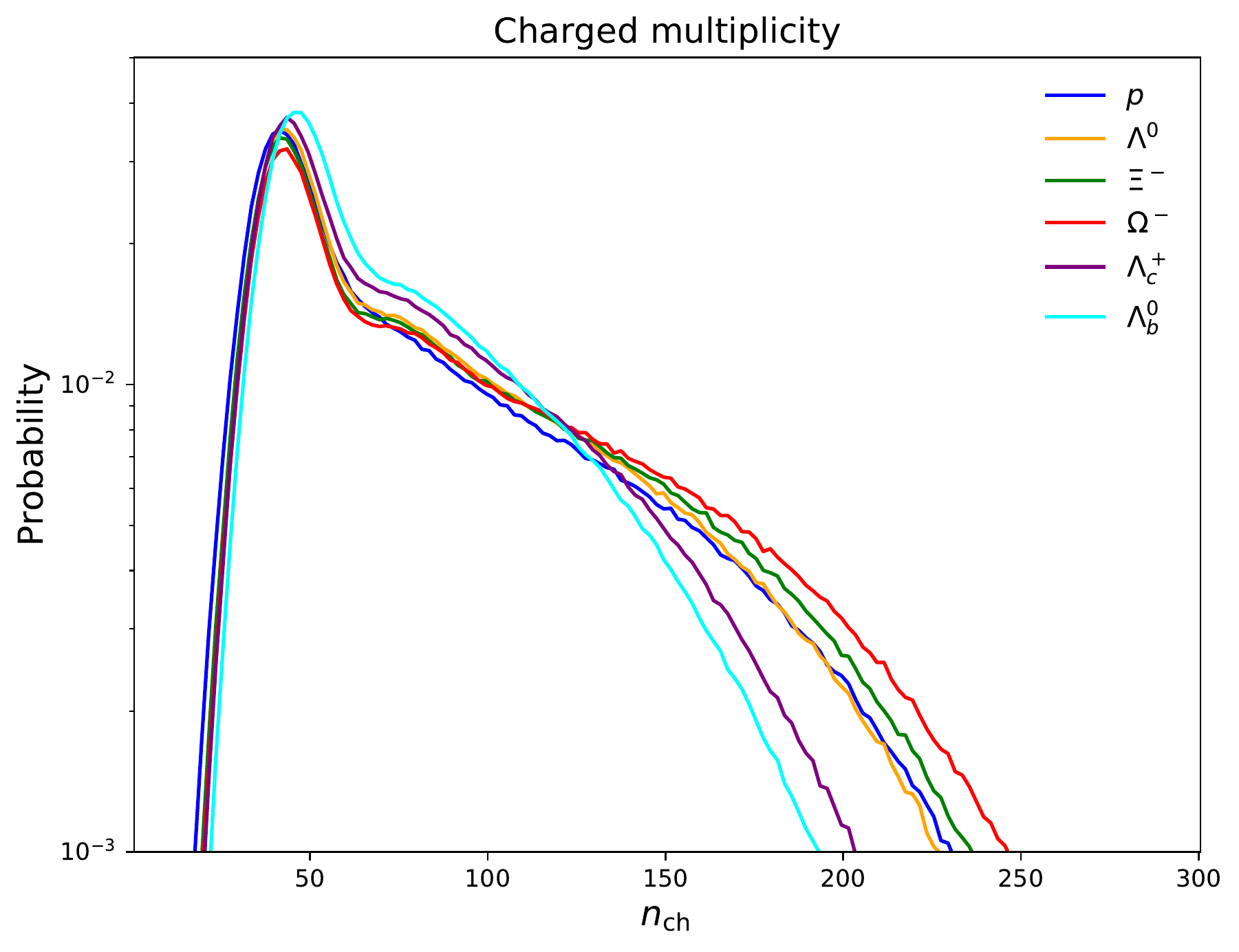}
  (b)
\end{minipage}
\\
\begin{minipage}[c]{0.49\linewidth}
  \centering
  \includegraphics[width=\linewidth]{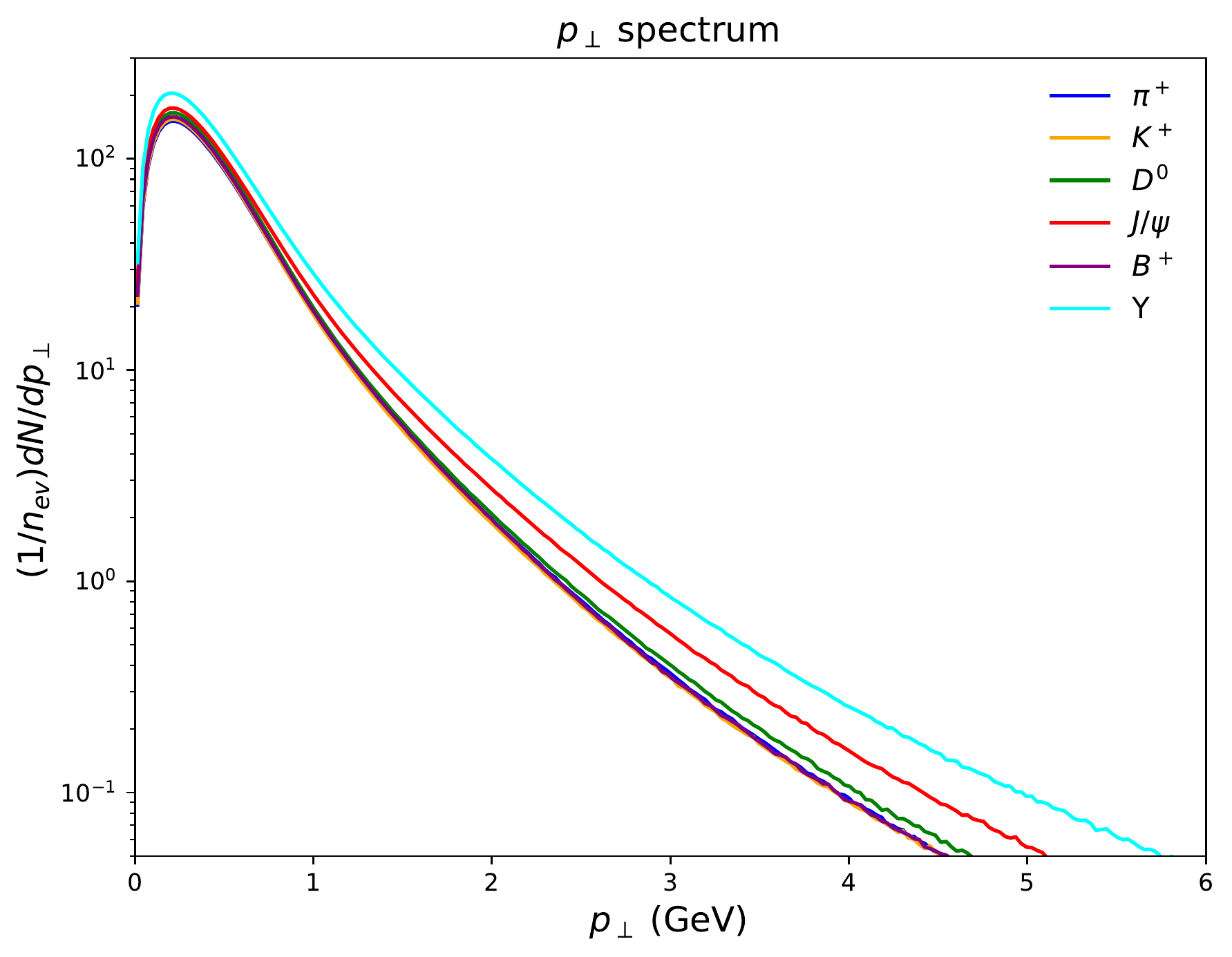}
  (c)
\end{minipage}
\begin{minipage}[c]{0.49\linewidth}
  \centering
  \includegraphics[width=\linewidth]{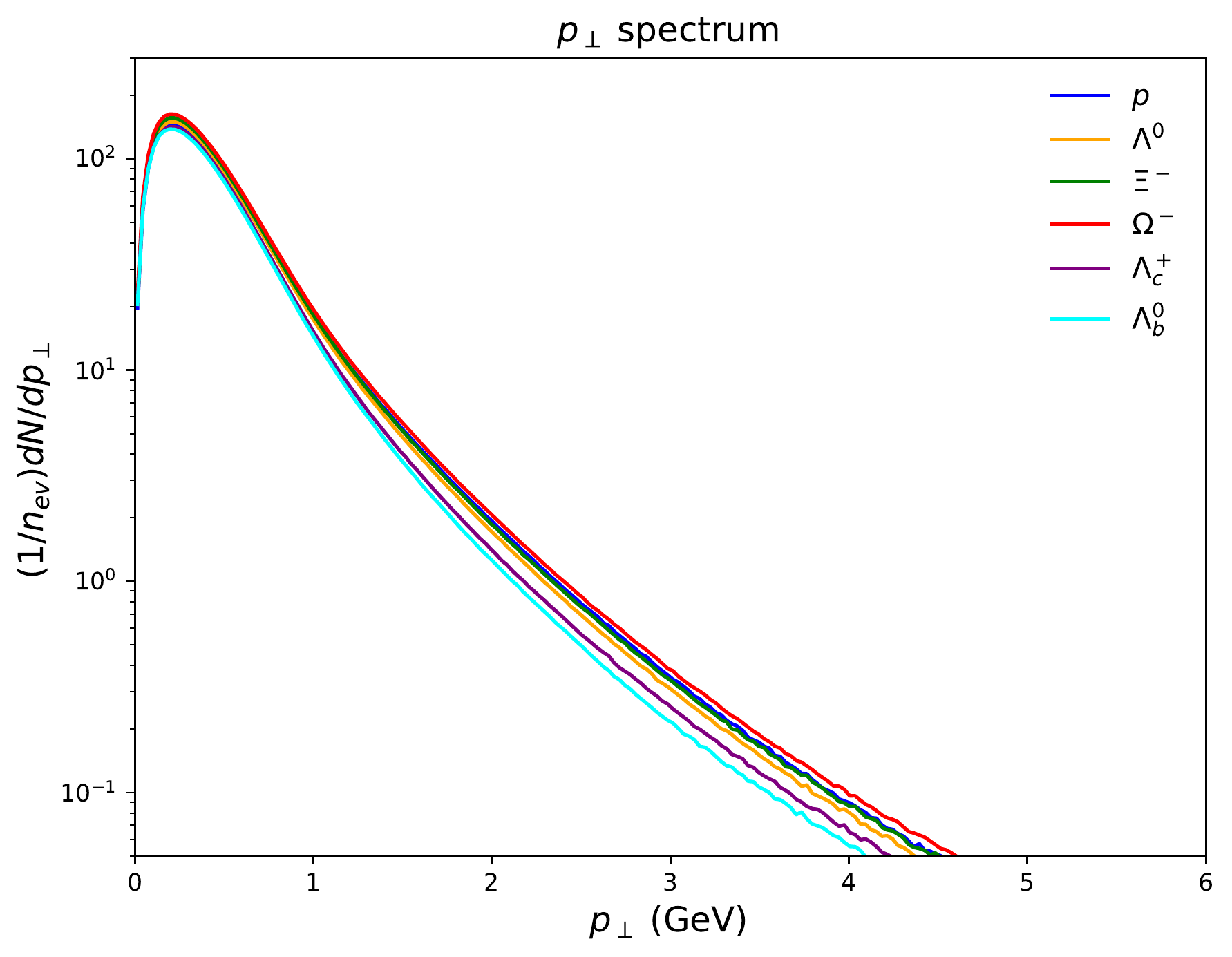}
  (d)
\end{minipage}
\\
\begin{minipage}[c]{0.49\linewidth}
  \centering
  \includegraphics[width=\linewidth]{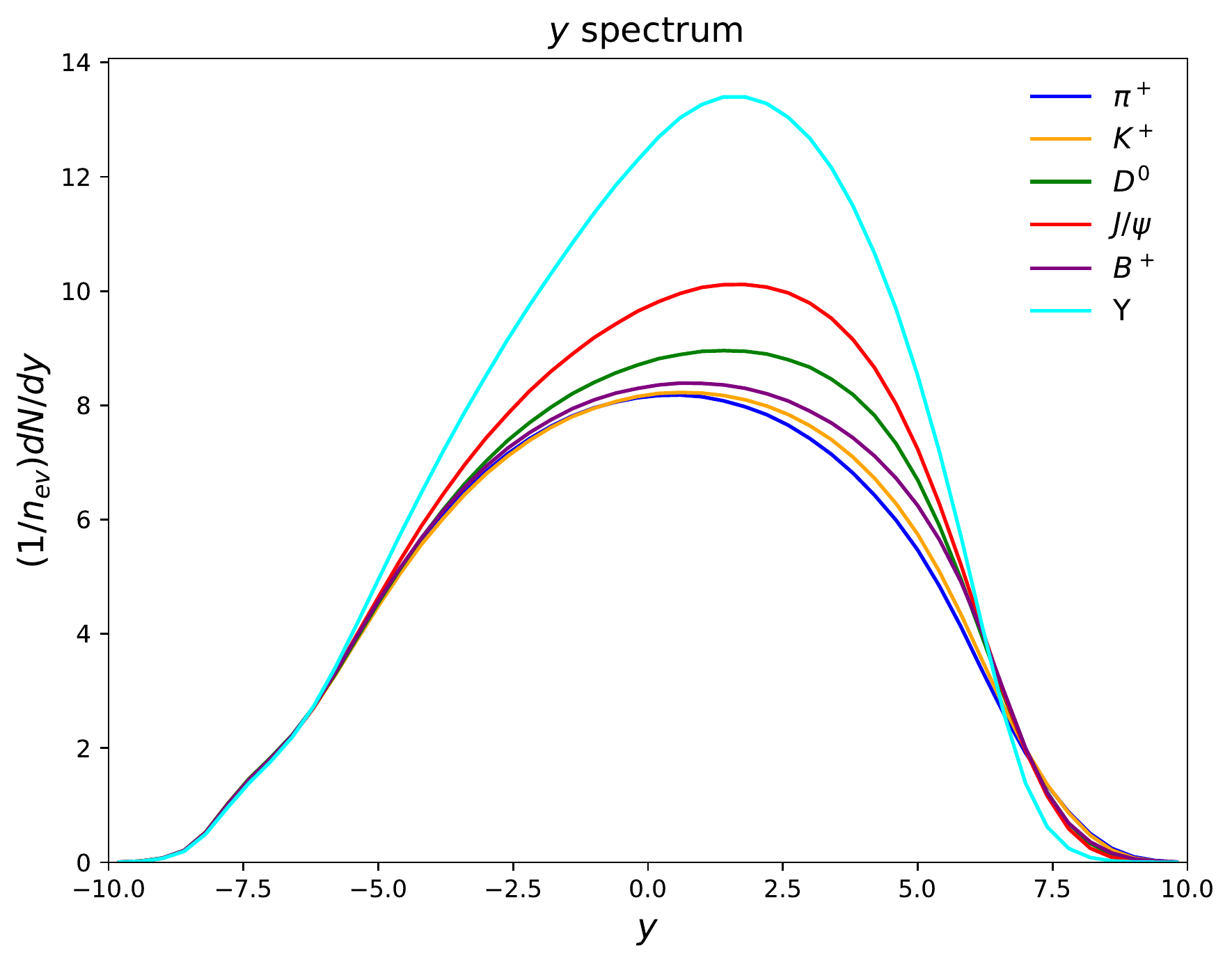}
  (e)
\end{minipage}
\begin{minipage}[c]{0.49\linewidth}
  \centering
  \includegraphics[width=\linewidth]{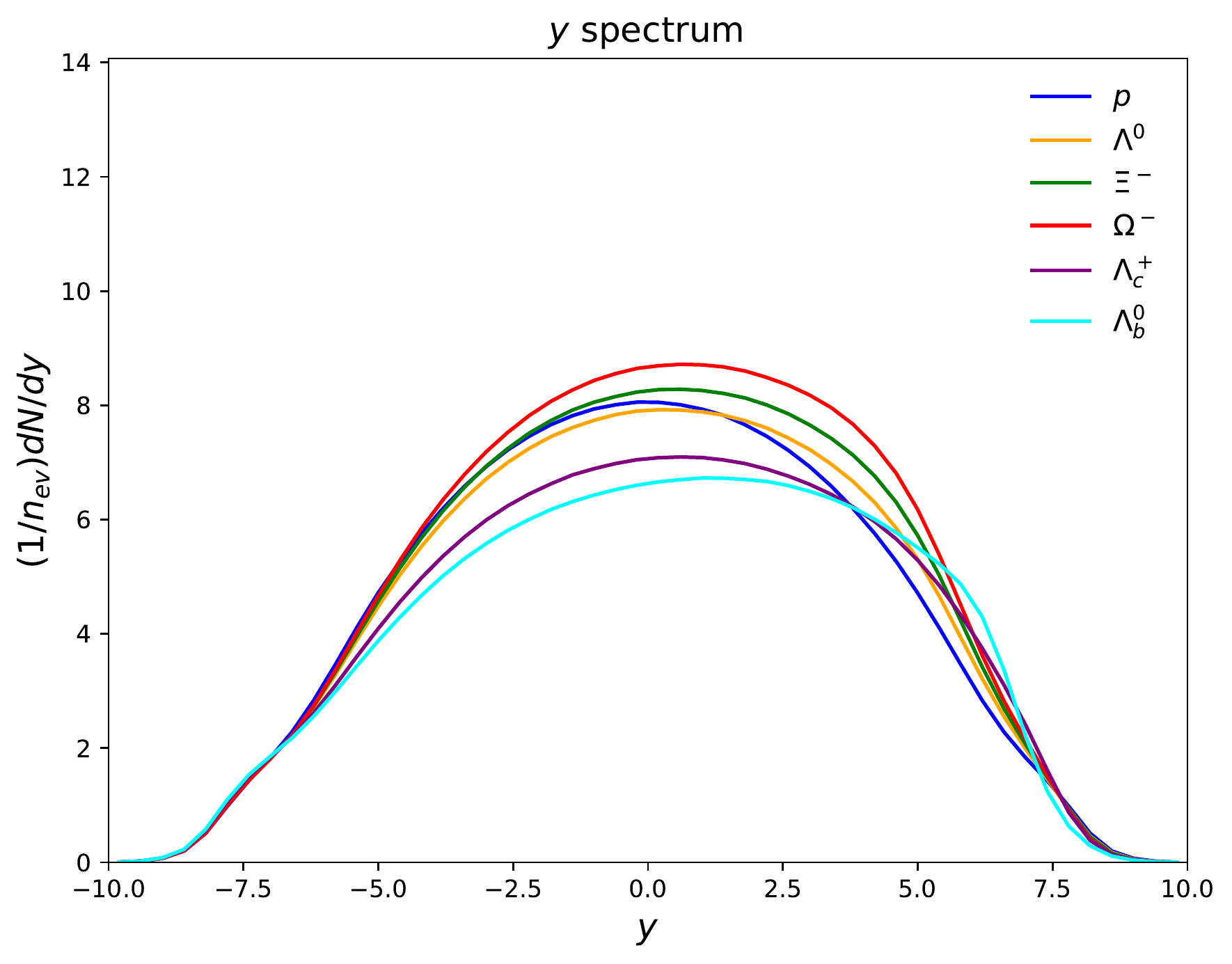}
  (f)
\end{minipage}
\caption{Charged-hadron (a,b) multiplicity distributions,
(c,d) $p_{\perp}$ spectra and (e,f) rapidity distributions for a
selection of (a,c,e) meson--proton and (b,d,f) baryon--proton
collisions. All results at a 6~TeV collision energy, and for
nondiffractive events only. Labels denote the respective hadron beam.}
\label{fig:simpleEventDistros}
\end{figure}

\begin{table}[tp!]
\centering
\begin{tabular}{|c|c c |c c|c c c|c c c|}
  \hline
  beam & $\sigma_\mathrm{total}$ & $\sigma_\mathrm{ND}$
  & $\left<n_\mathrm{MPI}\right>$ & $\left<p_{\perp,\mathrm{MPI}}\right>$
  & $\left<n_\mathrm{ch}\right>$ & $\left<p_\perp\right>$ & $\left<y\right>$ 
  & $\sigma_\mathrm{jet}$($\mu$b)
  & $\left<p_{\perp,\mathrm{jet}}\right>$ & $\left<y_{\mathrm{jet}}\right>$ \\
  \hline
  $\pi^+$        &  55.6 & 42.9 & 3.67 & 2.90 &  88.1 & 0.49 & 0.17 & 93.6 & 63.7 & 0.07 \\
  $\eta$         &  49.7 & 38.5 & 3.55 & 3.01 &  91.7 & 0.48 & 0.30 & 75.9 & 64.9 & 0.40 \\
  $\K^+$         &  48.2 & 37.6 & 3.57 & 2.97 &  89.3 & 0.48 & 0.23 & 77.6 & 64.8 & 0.35 \\
  $\phi$         &  40.8 & 32.2 & 3.75 & 3.02 &  93.6 & 0.48 & 0.26 & 75.2 & 64.6 & 0.45 \\
  $\D^0$         &  34.6 & 28.0 & 3.74 & 3.07 &  96.4 & 0.48 & 0.36 & 68.4 & 64.9 & 0.53 \\
  $\D_\s^+$      &  27.2 & 22.2 & 4.04 & 3.13 &  99.8 & 0.49 & 0.29 & 72.5 & 64.6 & 0.52 \\
  $\Jpsi$        &  13.6 & 11.3 & 4.04 & 3.40 & 103.9 & 0.50 & 0.41 & 63.1 & 64.8 & 0.70 \\
  $\B^+$         &  31.4 & 26.1 & 3.60 & 2.98 &  92.0 & 0.48 & 0.29 & 65.7 & 64.9 & 0.46 \\
  $\B_\s^0$      &  22.8 & 19.1 & 4.22 & 3.05 & 104.4 & 0.50 & 0.30 & 65.7 & 64.7 & 0.51 \\
  $\B_\c^+$      &   9.2 &  7.8 & 4.25 & 3.50 & 110.2 & 0.52 & 0.43 & 55.1 & 64.8 & 0.74 \\
  $\Upsilon$     &   4.8 &  4.1 & 4.37 & 3.73 & 120.1 & 0.52 & 0.47 & 53.7 & 65.5 & 0.86 \\
  \hline
  $\p$           &  88.5 & 68.3 & 3.96 & 2.77 &  85.9 & 0.49 & 0.00 & 92.9 & 63.2 & 0.00 \\
  $\Lambda^0$    &  76.7 & 59.9 & 3.52 & 2.88 &  86.6 & 0.47 & 0.15 & 89.8 & 64.0 & 0.14 \\
  $\Xi^-$        &  64.9 & 51.3 & 3.65 & 2.90 &  89.8 & 0.47 & 0.21 & 87.1 & 64.3 & 0.19 \\
  $\Omega^-$     &  53.1 & 42.6 & 3.80 & 2.98 &  94.4 & 0.48 & 0.26 & 82.1 & 64.3 & 0.17 \\
  $\Lambda_\c^+$ &  64.9 & 52.2 & 3.03 & 2.90 &  80.3 & 0.46 & 0.27 & 75.2 & 64.7 & 0.32 \\
  $\Xi_\c^0$     &  64.9 & 52.2 & 3.06 & 2.89 &  81.7 & 0.46 & 0.31 & 79.5 & 65.0 & 0.37 \\
  $\Omega_\c^0$  &  41.3 & 33.6 & 3.96 & 2.99 &  97.2 & 0.48 & 0.32 & 76.9 & 64.8 & 0.37 \\
  $\Lambda_\b^0$ &  61.1 & 50.6 & 2.69 & 2.91 &  78.0 & 0.45 & 0.35 & 71.6 & 64.9 & 0.43 \\
  $\Xi_\b^-$     &  61.1 & 50.6 & 2.72 & 2.91 &  79.5 & 0.45 & 0.40 & 73.9 & 64.7 & 0.44 \\
  $\Omega_\b^-$  &  37.5 & 31.7 & 3.65 & 2.99 &  96.0 & 0.47 & 0.40 & 72.0 & 64.4 & 0.40 \\
  \hline
\end{tabular}
\caption{Some basic numbers for various non-diffractive hadron--proton
collisions at 6~TeV: total and nondiffractive cross sections (in mb); 
average number and $p_\perp$ of MPIs in nondiffractive events; average charged
multiplicity,  charged $p_{\perp}$ and $y$; and jet cross section (in $\mu$b)
and $p_\perp$ and $y$ of jets with $p_{\perp,\textrm{min}} = 50$~GeV.}
\label{tab:simpleEventNumbers}
\end{table}

Distributions for the number and transverse momentum of MPIs are shown in
\figref{fig:MPIs} for a few different hadron types, while
\figref{fig:simpleEventDistros} shows charged hadronic multiplicity,
$p_\perp$ spectra, and rapidity spectra.
We note that, by and large, the various distributions follow suit
quite well, and notably the charged multiplicities are comparable.
A few key numbers are shown for a larger class
of collisions in \tabref{tab:simpleEventNumbers}, cementing the general
picture. This indicates that the joint handling of total
cross sections and PDFs are as consistent as can be expected.

Exceptional cases are $\Upsilon$, $\B_{\c}^+$ and $\Jpsi$ where, even
after adjusting the initial $Q_0^2$ scale for the PDF evolution to
reduce the number of MPIs, these particles give a higher activity than
others. The technical reason is that, even if the MPI cross section is
reduced to approximately match the smaller total cross, a non-negligible
fraction of the remaining MPIs now come from the heavy valence quarks at
large $x$ values. These thereby more easily can produce higher-$\p_{\perp}$
collisions (see $\left<p_{\perp,\mathrm{MPI}}\right>$ in
\tabref{tab:simpleEventNumbers}), which means more event activity in
general.

\begin{figure}
\begin{minipage}[c]{0.49\linewidth}
  \centering
  \includegraphics[width=\linewidth]{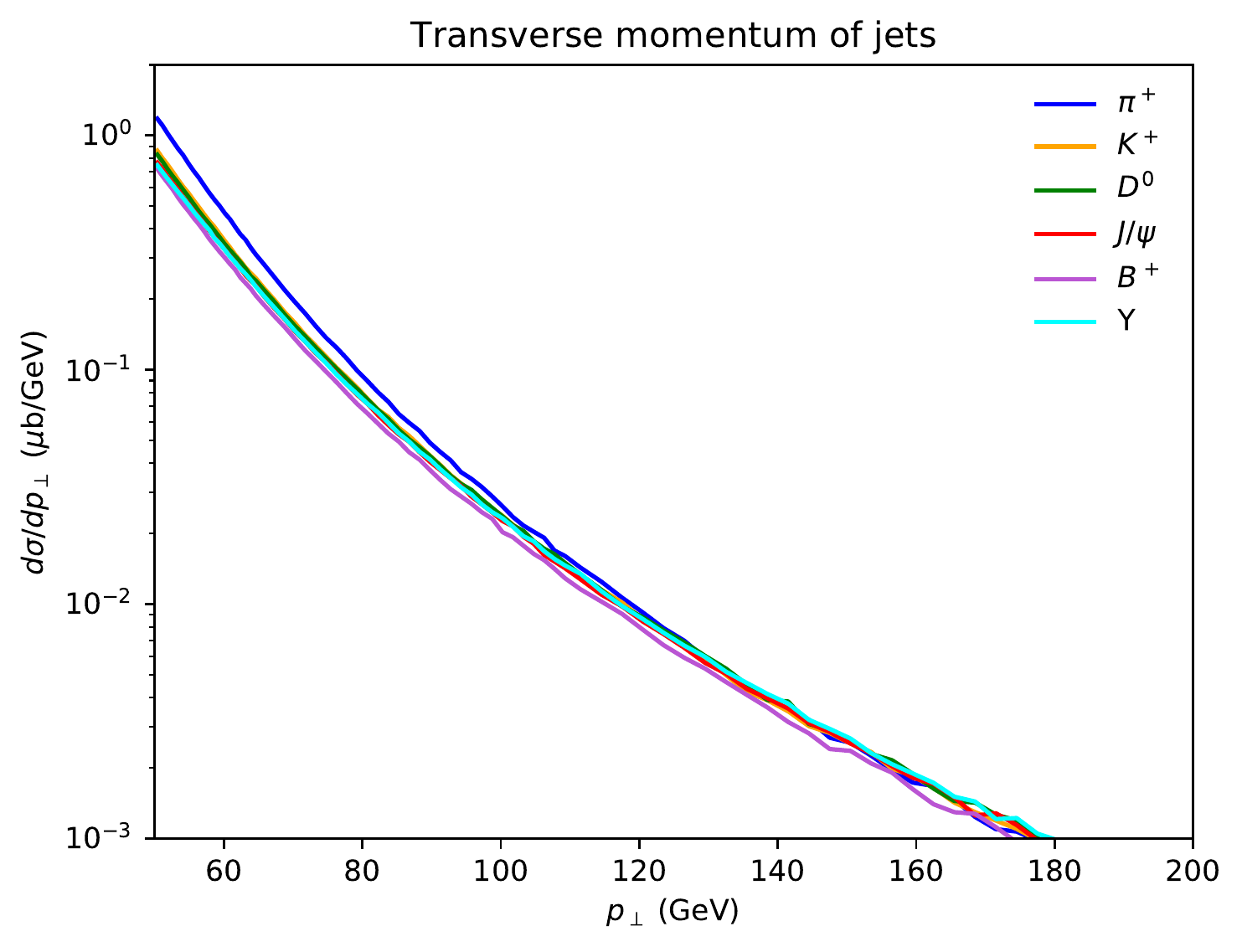}
  (a)
\end{minipage}
\begin{minipage}[c]{0.49\linewidth}
  \centering
  \includegraphics[width=\linewidth]{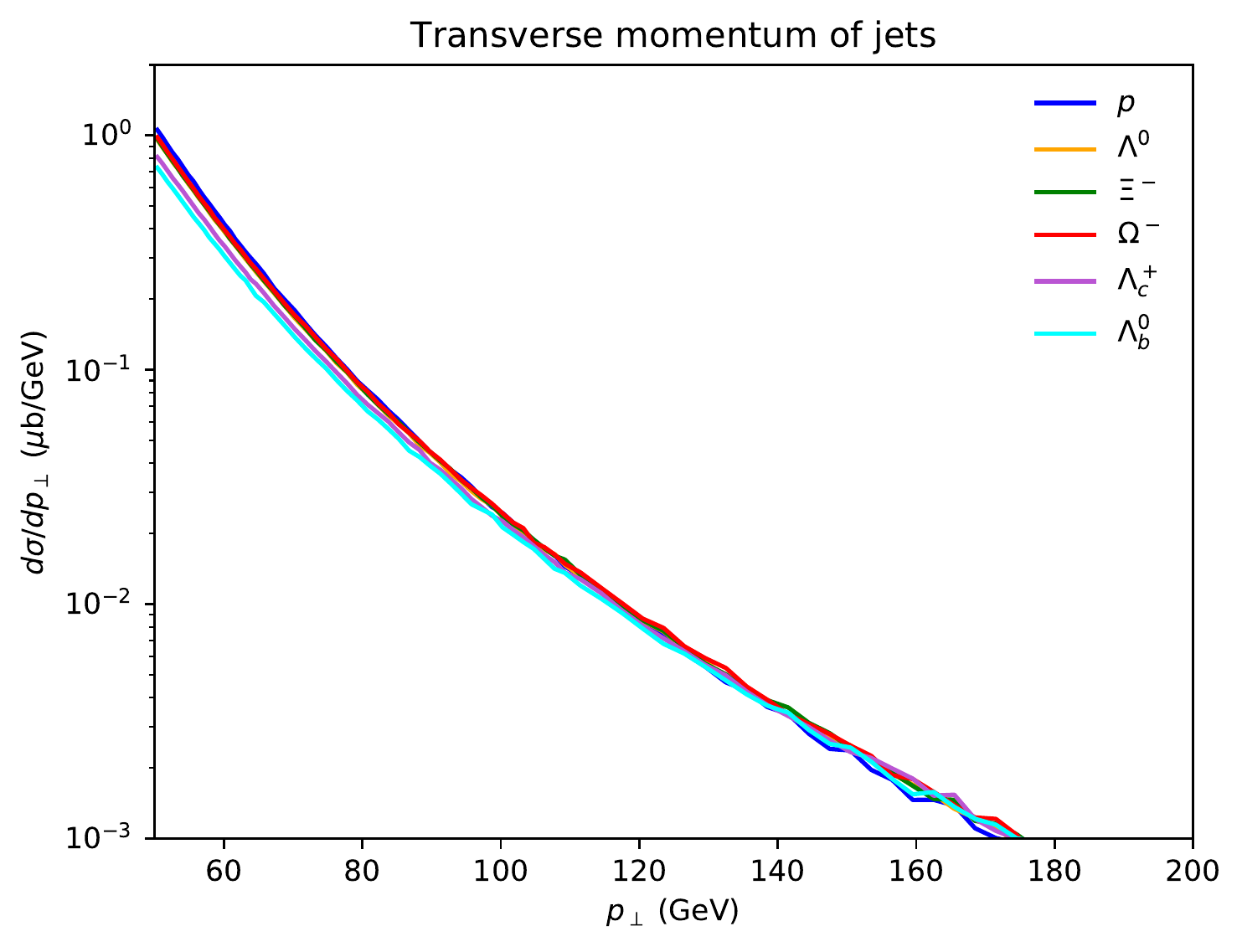}
  (b)
\end{minipage}
\\
\begin{minipage}[c]{0.49\linewidth}
  \centering
  \includegraphics[width=\linewidth]{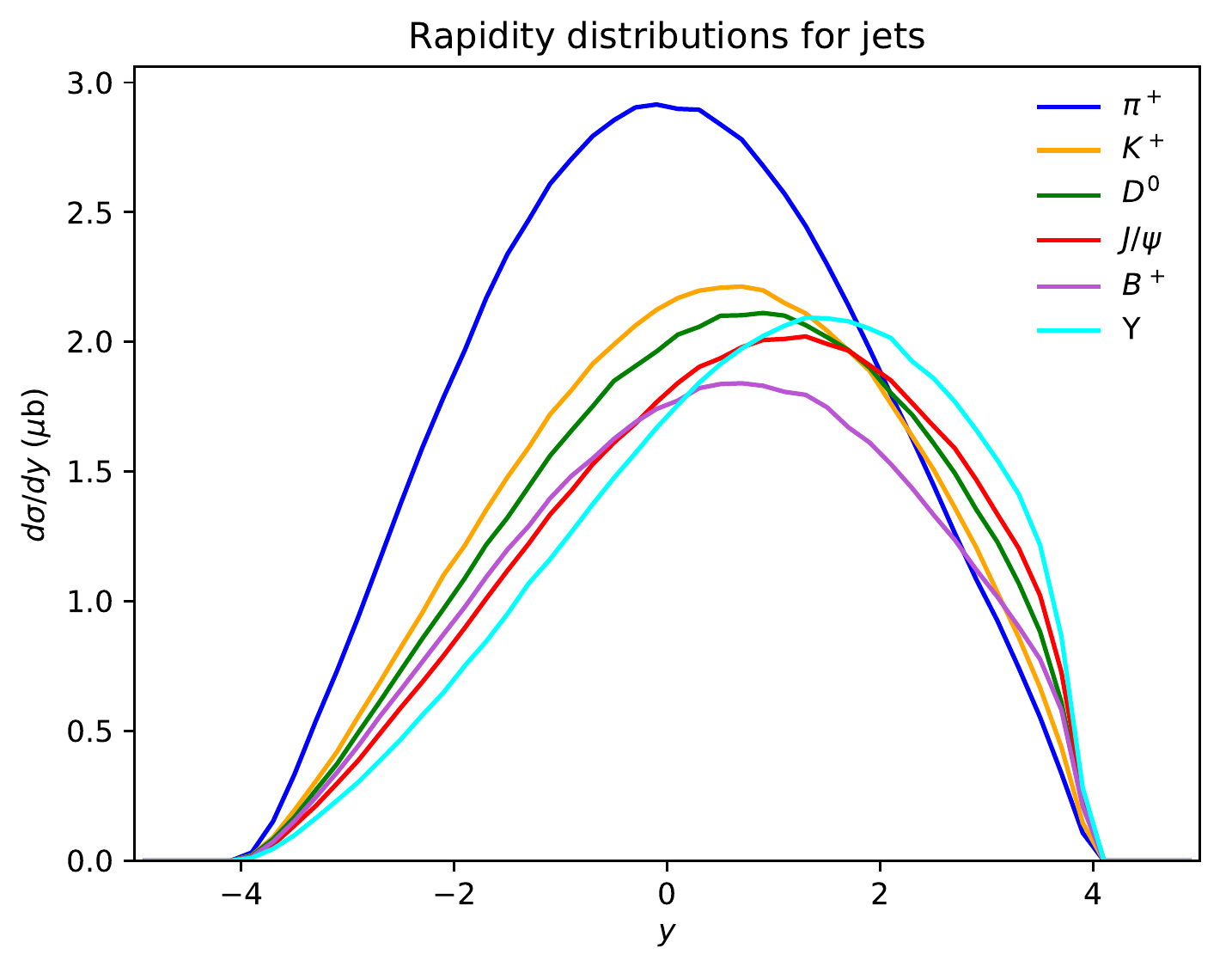}
  (c)
\end{minipage}
\begin{minipage}[c]{0.49\linewidth}
  \centering
  \includegraphics[width=\linewidth]{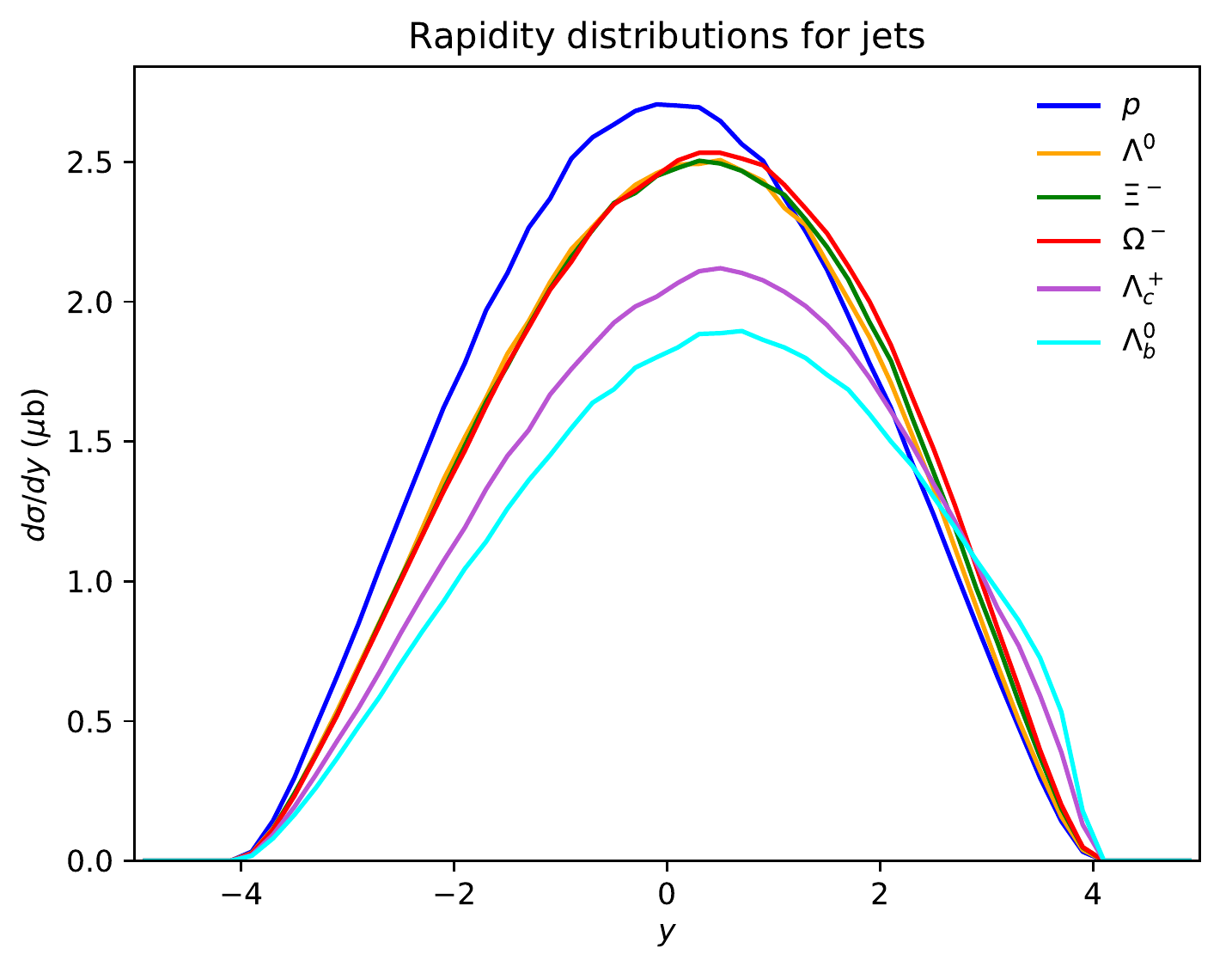}
  (d)
\end{minipage}
\caption{(a,b) Jet $p_\perp$ and (c,d) rapidity differential cross sections
(in units of $\mu$b), with $p_{\perp,\mathrm{min}} = 50$~GeV.
Average values are shown in \tabref{tab:simpleEventNumbers}.
Labels denote the respective hadron beam.}
\label{fig:simpleJetDistros}
\end{figure}

Studying the rapidity spectra closer, we find 
asymmetries around zero, depending on how different the
PDFs of the projectile are from those of the target proton.
That is, a projectile with harder PDFs should give a spectrum shifted
towards positive rapidities, and vice versa. Harder valence quarks
tend to be counteracted by softer gluons and sea quarks, however.
Possible effects also are partly masked by strings being stretched out
to the beam remnants, no matter the rapidity of the perturbative
subcollision. To better probe larger $x$ values, we also study jet
distributions, using the anti-$k_{\perp}$ algorithm
\cite{Cacciari:2008gp,Cacciari:2011ma} with $R = 0.7$ and
$p_{\perp\mathrm{jet}} > 50$~GeV. Some distributions are shown in 
\figref{fig:simpleJetDistros}, with average values again given in
\tabref{tab:simpleEventNumbers}. Indeed asymmetries now are quite visible.
We also note that the high-$p_{\perp}$ jet rate, normalized relative to
the total nondiffractive cross section, is enhanced in the cases with harder
PDFs, even if this is barely noticeable in the average $p_{\perp}$
of all hadrons.

\subsection{Nuclear collisions with Angantyr}

\Pythia comes with a built-in model for heavy-ion collisions,
\Angantyr \cite{Bierlich:2018xfw}, which describes much $p\A$
and $\A\A$ data quite well. It contains a model for the
selection of nuclear geometry and impact parameter of collisions.
In the Glauber formalism \cite{Glauber:1955qq} the nucleons are
assumed to travel along straight lines, and a binary nucleon--nucleon
subcollision can result anytime two such lines pass close to each other.
Any nucleon that undergoes at least one collision is called ``wounded''
\cite{Bialas:1976ed}. In our case the projectile is a single hadron,
so the number of subcollisions equals the number of wounded target
nucleons. 

In principle all of the
components of the total cross section can contribute for each subcollision,
but special consideration must be given to diffractive topologies.
Notably diffractive excitation on the target side gives rapidity
distributions tilted towards that side, a concept used already in
the older \textsc{Fritiof} model \cite{Andersson:1986gw} that partly
has served as an inspiration for \Angantyr. Alternatively, one
can view such topologies as a consequence of the \Pythia MPI machinery,
wherein not all colour strings from several target nucleons are
stretched all the way out to the projectile beam remnant, but some
tend to get ``short-circuited''. If such colour connections occur
flat in rapidity, then this is equivalent to a $\d M^2/M^2$ diffractive
mass spectrum. To first approximation, an \Angantyr hadron--nucleus collision
can be viewed as one ``normal'' subcollision plus a variable number
of diffractive events on the target side.

In its basic form, \Angantyr event generation is quite fast,
about as fast as ordinary \Pythia hadron--hadron collisions
per hadron produced. That is, the overhead from nuclear geometry
considerations and energy--momentum sharing between partly
overlapping nucleon-nucleon collisions is negligible.
The program becomes much slower if the more sophisticated features
are switched on, such as ropes \cite{Bierlich:2014xba,Bierlich:2017sxk},
shove \cite{Bierlich:2017vhg,Bierlich:2020naj}, and hadronic
rescattering \cite{Bierlich:2021poz}. These contribute aspects
that only become apparent in a more detailed scrutiny of events,
beyond what is needed for our purposes. With minor
modifications to the \Angantyr code itself, \ie on top of the
\Pythia-generic ones we have already introduced in this article,
it is also possible to allow any hadron to collide with a
nucleus.

There are two severe limitations, however. Firstly, a 
time-consuming recalculation of hadronic geometry parameters
is required anytime the collision energy or incoming hadron
species is to be changed in \Angantyr. Potentially this could be
fixed in the future, \eg by interpolation in a grid of initializations
at different energies, but it appears to be less simple than
what we have introduced for the MPI framework. And secondly,  
\Angantyr is only intended to be valid for nucleon--nucleon collision
energies above roughly 100~GeV.

A less severe limitation is that the handling of nuclear remnants is very
primitive. All non-wounded nucleons are lumped into a new nucleus,  
without any possibility for it to break up into smaller fragments.
For a fixed target the new nucleus is essentially at rest, however,
which means that it does not contribute to the continued evolution
of the hadronic cascade. Therefore even the simple approach is good
enough. In a cascade initiated by a primary nucleus with a fixed energy,
it would have been conceivable to handle at least the primary collision using
the full \Angantyr, if it was not for this last limitation.

\begin{figure}[t!]
\begin{minipage}[c]{0.49\linewidth}
\centering
\includegraphics[width=\linewidth]{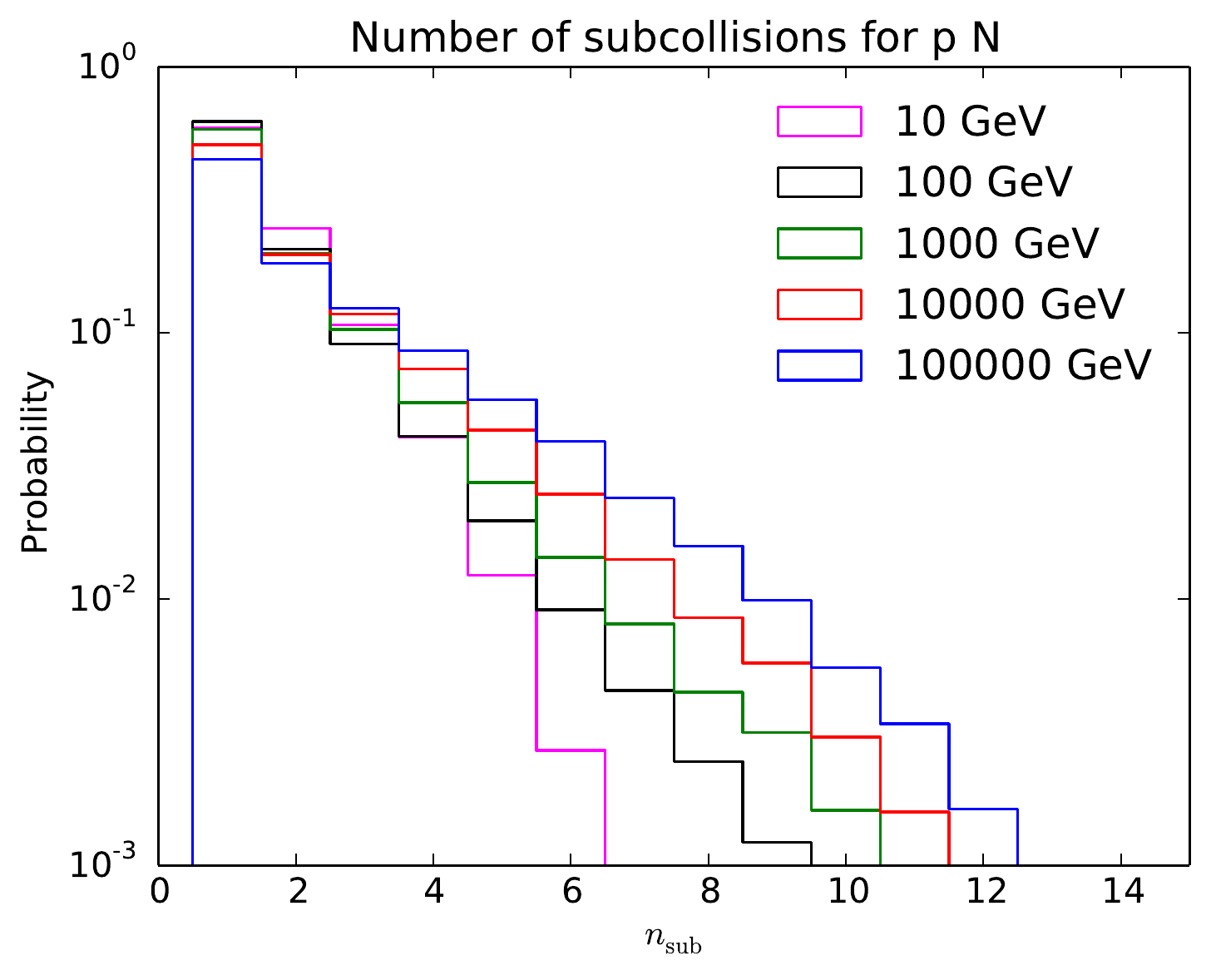}\\
(a)
\end{minipage}
\begin{minipage}[c]{0.49\linewidth}
\centering
\includegraphics[width=\linewidth]{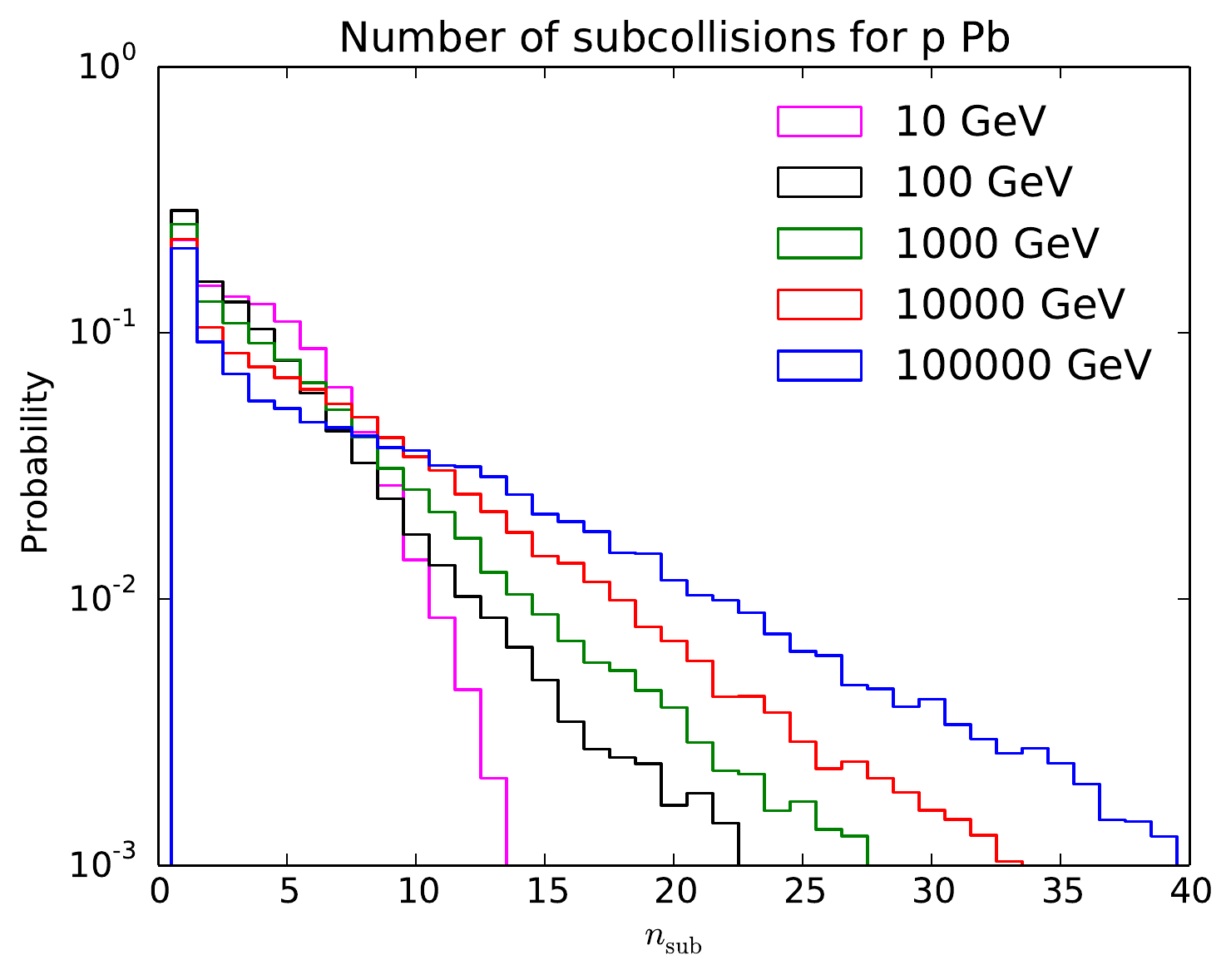}\\
(b)
\end{minipage}\\
\begin{minipage}[c]{0.49\linewidth}
\centering
\includegraphics[width=\linewidth]{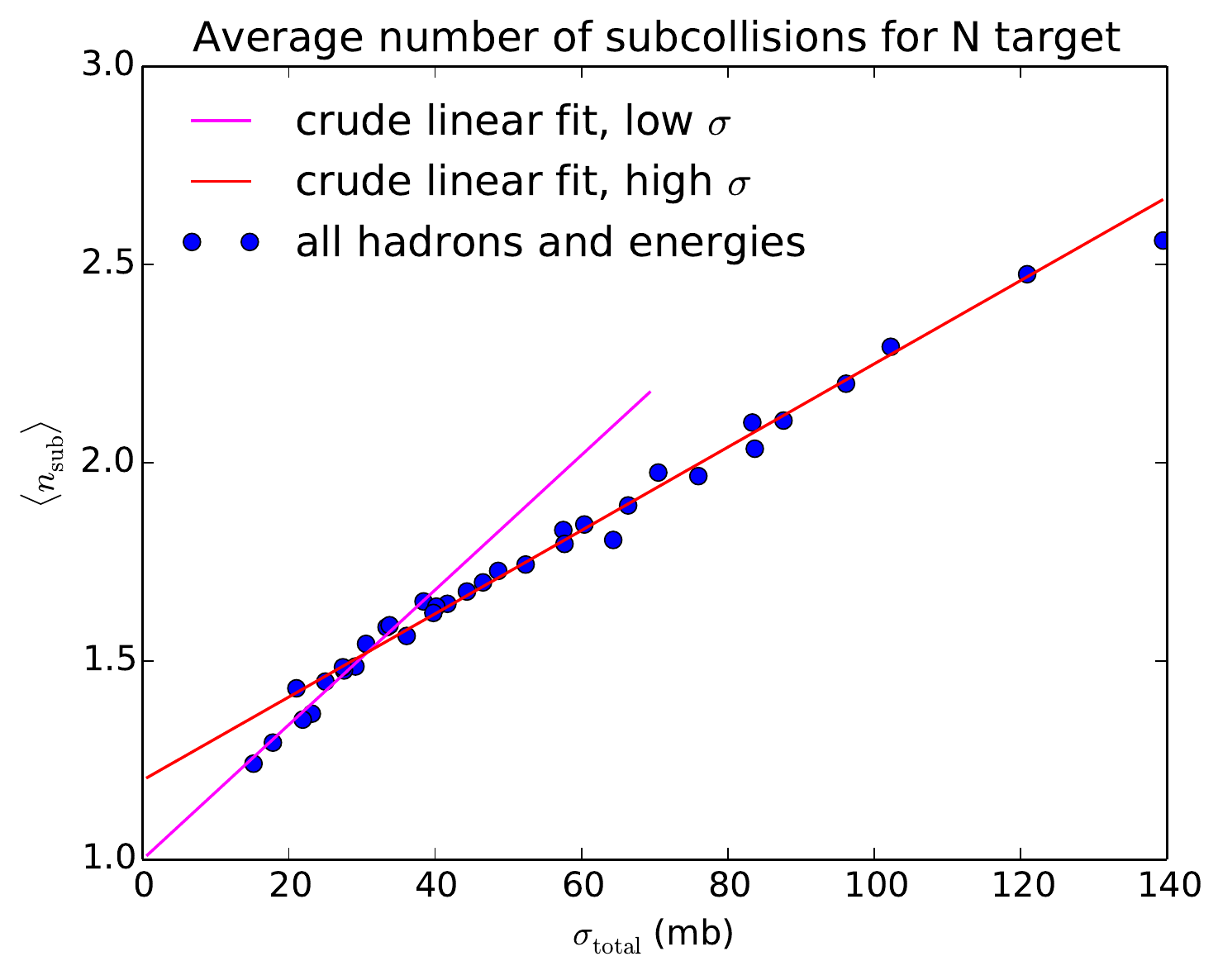}\\
(c) 
\end{minipage}
\begin{minipage}[c]{0.49\linewidth}
\centering
\includegraphics[width=\linewidth]{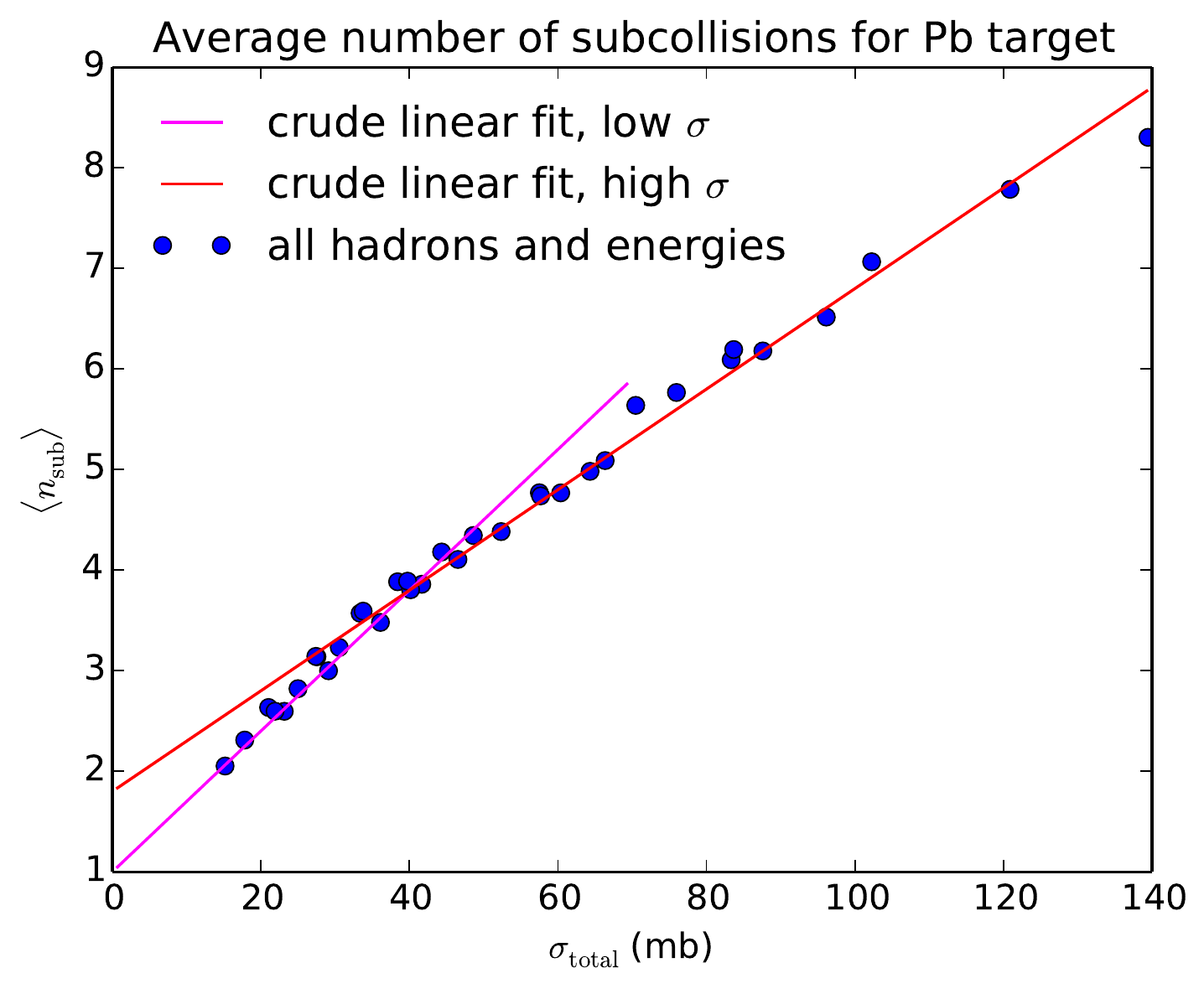}\\
(d)
\end{minipage}\\
\caption{Number of subcollisions in (a) proton--nitrogen and (b) proton--lead
collisions at five different subcollision energies. The average subcollision
number in (c) hadron--nitrogen and (d) hadron--lead as a function of the
total cross section, with some fits, see text for details.}
\label{fig:angantyrProperties}
\end{figure}

Even given the limitations, \Angantyr offers a useful reference
when next we come up with a simplified framework. Firstly, the
number of subcollisions in $\p\N$ collision roughly follows a 
geometric series, \figref{fig:angantyrProperties}a. This is largely a
consequence of geometry, where peripheral collisions are common and
usually only give one subcollision, while central ones are rare but
give more activity. To reach the highest multiplicities one also
relies on rare chance alignments of target nucleons along the projectile
trajectory. The deviations from an approximate geometric series are
larger if one instead considers $\p\Pb$ collisions,
\figref{fig:angantyrProperties}b, but not unreasonably so.  

An approximate geometric behaviour is also observed for other $h\N$ and
$h\Pb$ collisions. The average number of subcollisions depends on the
hadron species and the collision energy, but mainly via the total  
cross section, as can be seen from \figref{fig:angantyrProperties}c,d.
Here the results are shown
for seven different incoming hadrons ($\p,\pi^+,\K^+, \phi^0, \Lambda^0,
\Xi^0, \Omega^-$) at five different energies (10, 100, 1000, 10000,
100000 GeV). In the limit of $\sigma_{\mathrm{total}} \to 0$ one would 
never expect more than one subcollision. Given this constraint, a
reasonable overall description is obtained as
\begin{align}
\langle n_{\mathrm{sub}}^{h\N} \rangle &= \left\{
\begin{tabular}{ll}
1 + 0.017 $\sigma_{\mathrm{tot}}$ & for $\sigma_{\mathrm{tot}} < 31$\\
1.2 + 0.0105 $\sigma_{\mathrm{tot}}$ & else 
\end{tabular} \right.
\label{eq:avgnsubhN}\\
\langle n_{\mathrm{sub}}^{h\Pb} \rangle &= \left\{
\begin{tabular}{ll}
1 + 0.07 $\sigma_{\mathrm{tot}}$ & for $\sigma_{\mathrm{tot}} < 40$\\
1.8 + 0.05 $\sigma_{\mathrm{tot}}$ & else 
\end{tabular} \right.
\label{eq:avgnsubhPb}
\end{align}
with $\sigma_{\mathrm{tot}}$ in mb, as plotted in the figures.

\begin{figure}[t!]
\begin{minipage}[c]{0.49\linewidth}
\centering
\includegraphics[width=\linewidth]{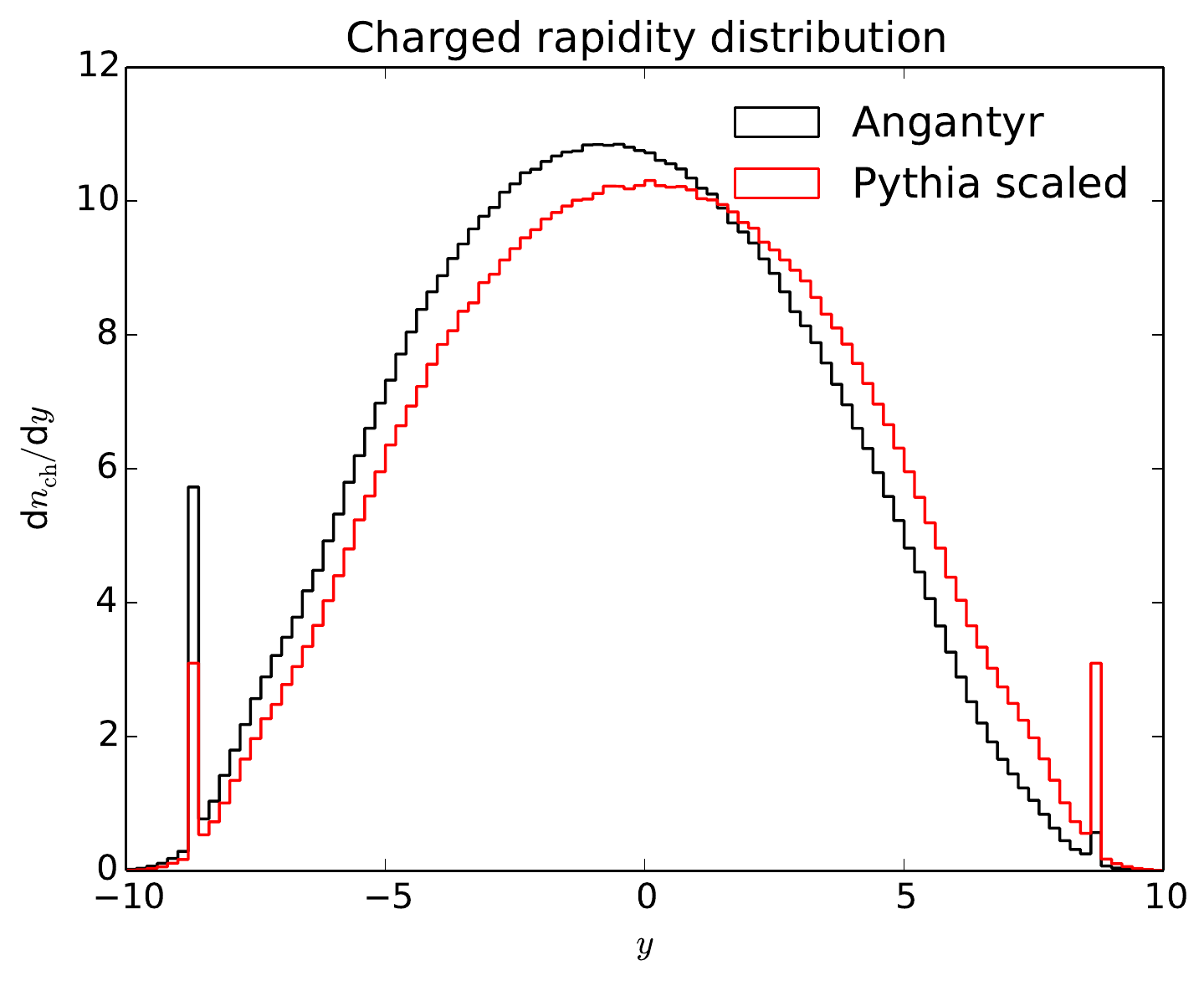}\\
(a)
\end{minipage}
\begin{minipage}[c]{0.49\linewidth}
\centering
\includegraphics[width=\linewidth]{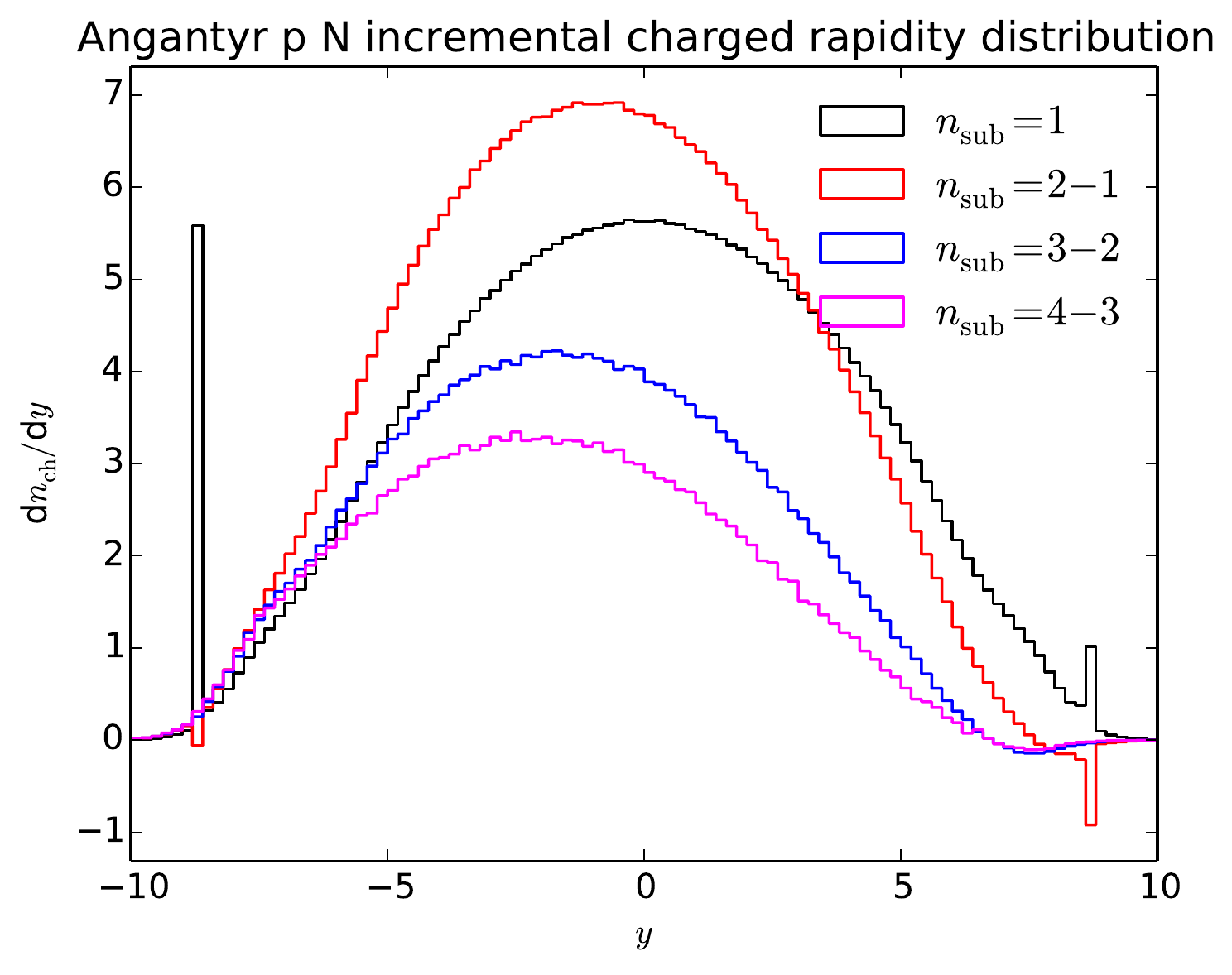}\\
(b)
\end{minipage}\\
\begin{minipage}[c]{0.49\linewidth}
\centering
\includegraphics[width=\linewidth]{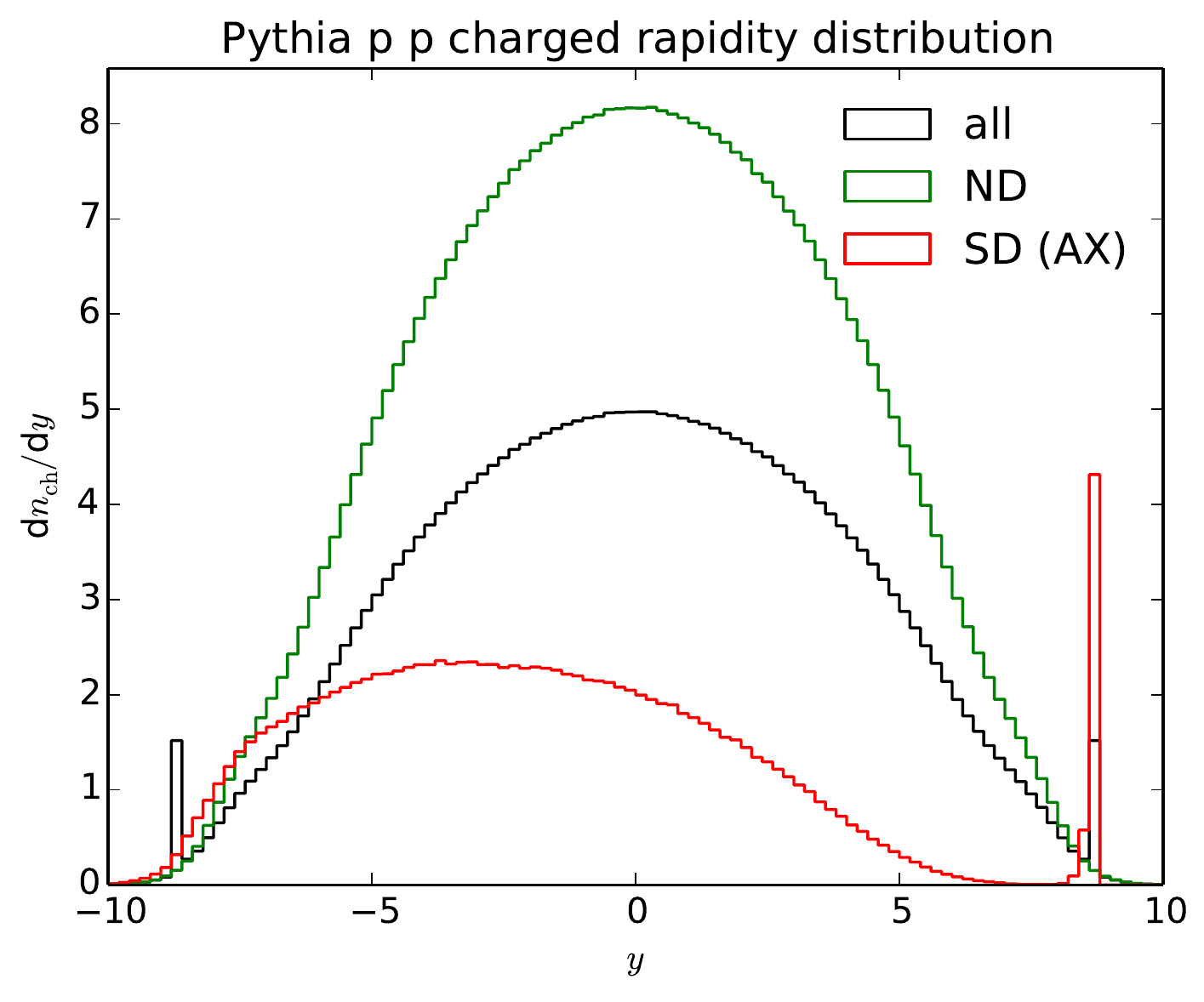}\\
(c) 
\end{minipage}
\begin{minipage}[c]{0.49\linewidth}
\centering
\includegraphics[width=\linewidth]{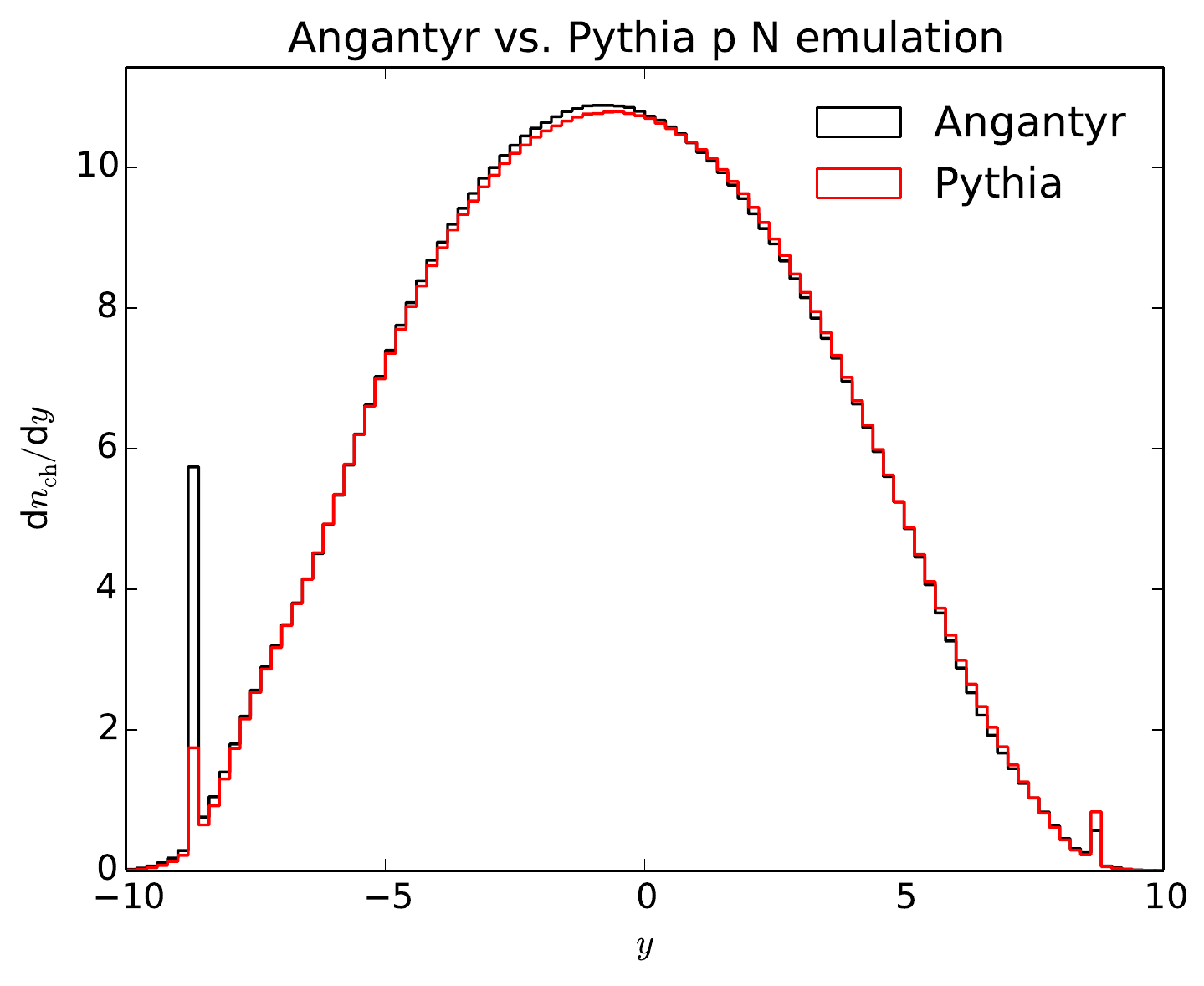}\\
(d)
\end{minipage}\\
\caption{Charged rapidity distributions $\d n_{\mathrm{ch}}/\d y$
for a nucleon--nucleon collision energy of 6 TeV.
(a) Inclusive \Angantyr $\p\N$ events relative to \Pythia events
scaled up with the average number of subcollisions in \Angantyr.
(b) \Angantyr for different number of subcollisions, relative 
to results for one less subcollision.
(c) \Pythia inclusive, the nondiffractive (ND) component only,
and the target-side single diffractive one (SD (AX)) in $\p\p$
collisions.
(d) The \Pythia emulation of $\p\N$.}
\label{fig:angantyrPythiaDiff}
\end{figure}

Secondly, \Angantyr may also be used as a reference for expected
final-state properties, such as the charged rapidity distribution,
$\d n_{\mathrm{ch}}/\d y$. As a starting point,
\figref{fig:angantyrPythiaDiff}a compares the \Angantyr $\p\N$
distribution with the \Pythia $\p\p$ one at a 6~TeV collision energy.
The \Pythia curve has beeen scaled up by a factor of 2.05, which
corresponds to the average number of subcollisions in $\p\N$.
While the total charged multiplicities are comparable, there are three
differences of note. (1) The \Angantyr distribution is shifted into
the target region, while the \Pythia one by construction is
symmetric around $y = 0$. (2) \Pythia has a large peak at around
$y \approx 8.7$ from elastic scattering, and some single diffraction,
that is much smaller in \Angantyr. (3) At $y \approx -8.7$ instead
\Angantyr has a narrow peak from the not wounded nucleons that
together create a new nucleus. 

The description of nuclear effects on hadronization is nontrivial.
At \Angantyr initialization the relative composition of different
subprocesses is changed. This means \eg that the elastic rate in
$\p\N$ with one single subcollision is reduced relative to \Pythia
$\p\p$, \figref{fig:angantyrPythiaDiff}b,c. Each further subcollision
in \Angantyr involves the addition of a diffractive-like system on
the target side, but the step from 1 to 2 also includes \eg a drop
in the elastic fraction, \figref{fig:angantyrPythiaDiff}b, and the
correlations arising from nucleons being assumed to have fluctuating
sizes event-to-event. The diffractive 
systems have a higher activity and are more symmetric than corresponding
\Pythia ones, \figref{fig:angantyrPythiaDiff}c.
As already mentioned the \Angantyr mechanism is not quite equivalent 
with that of ordinary diffraction, which is reflected in the choice
of PDF for the ``pomeron'', as used for the MPI activity inside the
diffractive system. In \Angantyr a rescaled proton PDF is used, while
by default the H1 2006 Fit B LO \cite{H1:2006zyl} pomeron PDF is
used in \Pythia. If MPIs (and parton showers) are switched off, the
diffractive systems become quite similar, and have a marked triangular
shape, as expected for a $\d M^2 / M^2$ diffractive mass  distribution.
That is, without MPIs the asymmetry of \Angantyr events is dramatically
larger. It is possible to use the rescaled-proton approach also in the
\Pythia simulation to obtain better agreement, but this is not quite good
enough, so next we will present a slightly different solution.

\subsection{Simplified nuclear collisions}

\begin{figure}
\centering
\includegraphics[width=0.5\textwidth]{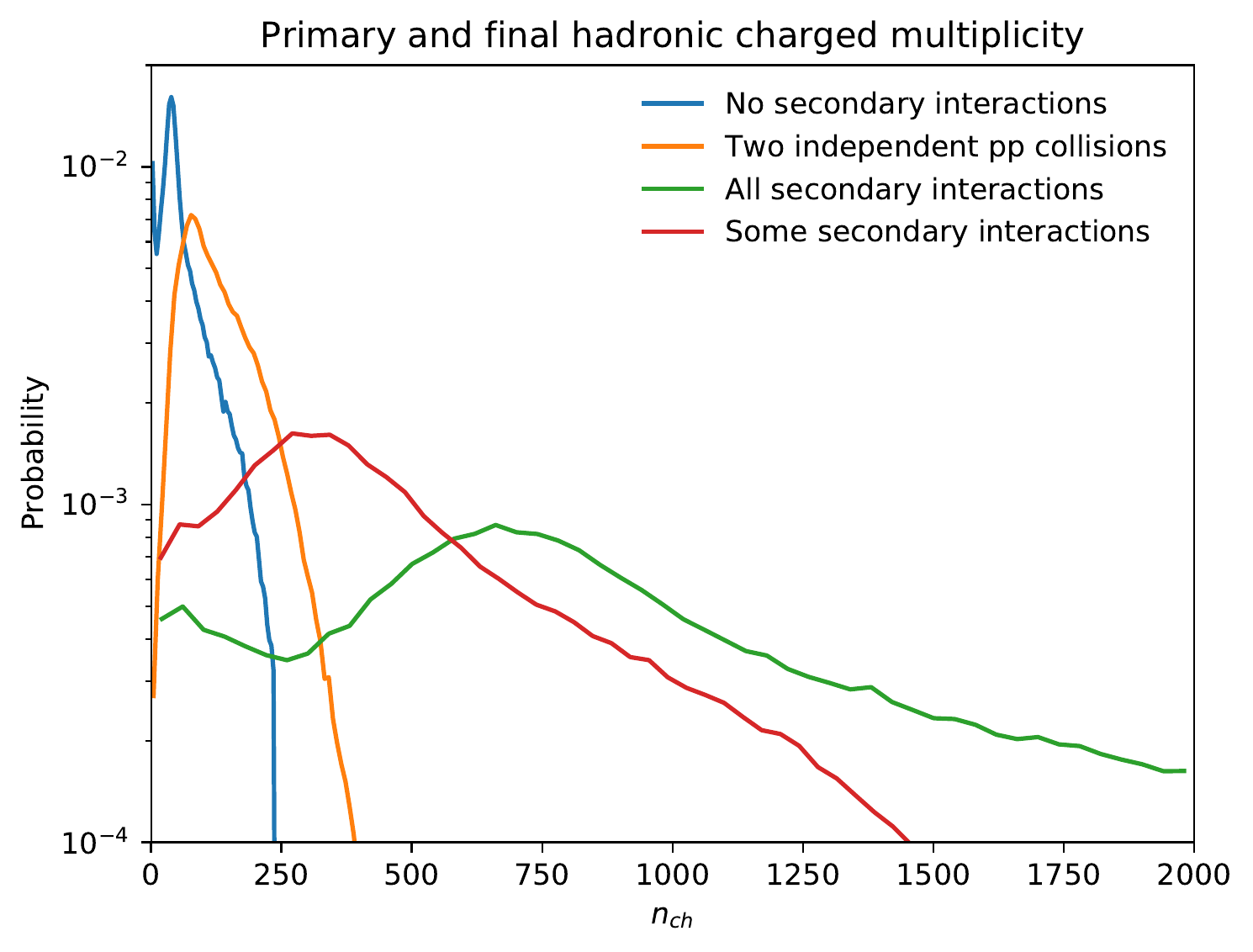}
\caption{Two superimposed $\p\p$ collisions compared with a single $\p\p$
collision, as well as two collisions where the outgoing hadrons interact
with a $\p$ medium. Either all the outgoing hadrons interact, or only some
of them as outlined in the text.}
\label{fig:simpleHI}
\end{figure}

$\p\N$ events with two wounded nucleons in the target produce a rapidity
distribution not so different from two separate $\p\p$ collisions,
\figref{fig:angantyrPythiaDiff}a. It would thus seem that a modelling
with a naive atmosphere consisting of separate protons and neutrons
would be a decent approximation. What one gets wrong in such a picture
is the multiplication factor. That is, in the naive atmosphere, each of
the hadrons produced in a first interaction can go on and interact in
their turn. But in the correct one, with the incoming beams
Lorentz-contracted pancakes, there is no time for any new hadrons to
form in the passage of the proton through the nitrogen nucleus.
To understand the effect of multiplication, we have studied the case
where each of the products of a primary 6~TeV $\p\p$ event,
represented by a 19,200 TeV beam on a fixed target,  can interact with
one further $\p$ in the target, \figref{fig:simpleHI}. For this
toy study the cascade stops after the second step. One should
note that the primary $\p\p$ collision at the maximal energy corresponds
to the largest cross section in the cascade, while other hadrons at lower
energies have smaller cross sections. To compensate for this we also
show an option where the probability for a secondary interaction of each
hadron from the primary collision is given by its cross section,
normalized to the primary $\p\p$ one. Also note that outgoing particles
with low momenta may interact with protons in very soft elastic collisions.
This introduces new slow-moving protons in the event record that presumably
would not be detectable. We partly avoid this effect by only allowing hadrons
to interact if the kinetic energy of the collision is larger than
$E_{\mathrm{kin,min}}$, by default 0.2~GeV. This is just at the border where
an inelastic collision with one additional outgoing pion is possible.  
Even with such corrections it is clear that a lumping of individual nucleon
collision into fewer but bigger nuclear ones makes a difference, reducing
the average charged multiplicity by about a factor of five in this case.

Having noted the effect of excessive multiplication, we have introduced
a simple model that allows us to keep it in check, and approximately
emulate the main features of the \Angantyr model for a hadron $h$
impinging on a $\N$ target at rest.
\begin{enumerate}
\item Calculate the invariant mass for the $h\p$, which is the same as
the one for $h\n$ to first approximation.
\item Evaluate the $h\p$ total cross section $\sigma^{h\p}_{\mathrm{tot}}$
as already described; again assumed equal to the $h\n$ one.
\item Evaluate $\langle n_{\mathrm{sub}}^{h\N} \rangle$ from
\eqref{eq:avgnsubhN}.
\item Define $r = 1 - 1 / \langle n_{\mathrm{sub}}^{h\N} \rangle$, such that
the geometric distribution
$P_n = r^{n-1} / \langle n_{\mathrm{sub}}^{h\N} \rangle$,
$n\geq 1$ has $\sum P_n = 1$ and
$\langle n \rangle = \langle n_{\mathrm{sub}}^{h\N} \rangle$.
\item Decide with equal probability that a proton or a neutron in the
target is wounded, and correspondingly generate an inclusive $h\p$ or
$h\n$ event. Do not do any decays, however.
\item Continue the generation with probability $r$. If not go to point 10.
Also go there if there are no more nucleons that can be wounded, or
if another user-set upper limit has been reached.
\item Find the newly produced hadron that has the largest longitudinal
momentum along the direction of the mother hadron $h$. Redefine $h$ to
be this newfound hadron.
\item Pick a new target proton or neutron among the remaining ones,
and generate a corresponding $h\p$ or $h\n$ event. Below 10~GeV, all
low energy processes are allowed. Above 10~GeV, only allow a mix
between nondiffractive and target single diffractive topologies, the
former with probability $P_{\mathrm{ND}} = 0.7$. Again omit
decays at this stage.
\item Loop back to point 6.
\item At the end allow all unstable hadrons to decay.
\end{enumerate}

The procedure is to be viewed as a technical trick, not as a physical
description. Obviously there is no time for hadronization during the
passage of the original $h$ through the $\N$. Rather the idea is that the
new $h$ in each step represents the original one, only with some loss
of momentum. This is not too dissimilar from how a $\p$ projectile in 
\Angantyr has to give up some of its momentum for each subcollision
it is involved in, even if that particular loss is calculated before
hadronization. The new procedure also leads to the central rapidity of
each further subcollision being shifted towards the target region. 

The specific value of $P_\mathrm{ND}$ in step 8 has no deep physical meaning,
but is ``tuned'' such that this mix of the relevant curves in 
\figref{fig:angantyrPythiaDiff}c, combined with the rapidity shift
procedure just explained, gives the same average behaviour as the
consecutive steps in \figref{fig:angantyrPythiaDiff}b. Notably a high
$P_{\mathrm{ND}}$ is needed to reproduce the high activity and low
$\langle y \rangle$ shift in \Angantyr ``diffractive'' systems.

A simplified test is shown in \figref{fig:angantyrPythiaDiff}d,
where the new hadron $h$ and the target nucleon is always assumed
to be a proton, such that only one collision kind is needed for the 
simulation at this stage. We have checked that an almost equally good
description is obtained also for a lead target, and for a range of
collision energies. A warning is in place, however, that the picture
is a bit more impressive than warranted.
Specifically, the charged multiplicity distributions show non-negligible
discrepancies, where the \Pythia machinery gives somewhat more
low-multiplicity events, reflected in the forward elastic peak region,
which then is compensated elsewhere to give the same average.
It should be good enough to get some reasonable understanding of
nuclear effects on cascade evolution, however. Later on, for such studies,
we will use the full framework.

The Angantyr model is not intended to be applied at very low energies,
so there we have no explicit guidance. The same approach with successive
subcollisions is still used, with the hardest hadron allowed to go on
to the next interaction, meaning that the CM frame is gradually shifted
in the target direction. But there are two modifications. Firstly, below
10 GeV in the CM frame, all allowed (low-energy) processes are mixed in
their normal fractions. And secondly, no further subcollisions are
considered once the kinetic energy of the hardest hadron falls below
$E_{\mathrm{kin,min}}$.

As a final comment, if each initial $h\p$ or $h\n$ collision results in
an average of $\langle n_{\mathrm{sub}}^{h\N} \rangle$ subcollisions, then
the effective $\sigma_{\mathrm{tot}}^{h\p}$ cross section must be reduced
by the same factor. That is, relative to a gas of free $\p/\n$, the
incoming hadron will travel longer in a same-nucleon-density $\N$ gas
before interacting, but produce more subcollisions each time it interacts.
More generally, for a nucleus $\A$ with atomic number $A$, the ansatz is
that $\sigma_{\mathrm{tot}}^{h\A} \, \langle n_{\mathrm{sub}}^{h\A} \rangle =
A \, \sigma_{\mathrm{tot}}^{h\p} $. Nontrivial nuclear effects, such as
the fluctuating nuclear sizes or shadowing, could modify this. For the
28 cases studied in \figref{fig:angantyrProperties} the ratio
$\sigma_{\mathrm{tot}}^{h\A} \, \langle n_{\mathrm{sub}}^{h\A} \rangle /
(A \, \sigma_{\mathrm{tot}}^{h\p})$ lands in the range $1 - 1.2$ for N and
$1.2 - 1.4$ for Pb, with no obvious pattern. For now such a possible
correction factor is left aside.

\section{Modelling hadronic cascades}

In the previous sections we have developed and tested the tools needed
to described hadron--nucleon interactions, and in an approximate manner
hadron--nucleus ones, over a wide range of energies. In this section we
will present some simple studies making use of the resulting framework.
To this end we introduce a toy model of the atmosphere and study the
evolution of a cascade. At the end we also study a cascade in a slab
of lead, to go from a dilute to a dense medium, and from a light to a
heavy nucleus. These examples are intended to point to possibilities
rather than to give any definitive answers.

\subsection{Medium density}

The simplest possible medium density distribution is a uniform
density $\rho$. While not accurate for the atmosphere, it may be
applicable \eg for simulating solid particle detectors. In such a
medium, the mean free path of a particle is
\begin{equation}
l_0 = \frac{1}{\sigma \rho},
\end{equation}
where $\sigma$ is the cross section for an interaction between the beam
particle and a medium particle. The distance traveled before an
interaction then follows an exponential distribution with mean value $l_0$.
In \Pythia, lengths are given in units of mm
and cross sections in units of mb. For particle densities, we use the
common standard g/cm$^3$. The conversion reads
\begin{equation}
l_0 =  1.78266 \cdot 10^4 \, \mathrm{mm} \; \frac{1 \, \mathrm{mb}}{\sigma}
\; \frac{1 \, \mathrm{g/cm}^3}{\rho} \; \frac{m c^2}{1 \, \mathrm{GeV}} ~,
\end{equation}
where $m$ is the target nucleus mass.

If decays of long-lived particles can be neglected, the evolution of
the cascade is only a function of the g/cm$^2$ interaction depth traversed, 
and we will use this as standard horizontal axis along which to present
several results. When decays are to be included, however, it becomes
important to model density variations. Let the atmospheric density
$\rho(z)$ depend on height $z$ above the surface. The incoming particle
enters the medium at $z = z_0$ with a zenith angle $\theta$, and the
earth curvature is neglected. Then the naive probability for an interaction
at height $z$ is given by 
\begin{equation}
\frac{\d\mathcal{P}_{\mathrm{naive}}(z)}{\d z} =
  \frac{\sigma \rho(z)}{\cos\theta} \equiv f(z).
\end{equation}
But we are only interested in the first time this particle interacts, and
this gives the conventional ``radioactive decay'' exponential damping, that
the particle must not have interacted between $z_0$ and $z$:
\begin{equation}
\frac{\d\mathcal{P}(z)}{\d z} = f(z) \;
\exp \left( - \int_z^{z_0} f(z') \, \d z' \right) ~.      
\end{equation}
If $f(z)$ has an invertible primitive function $F(z)$ then the Monte Carlo
solution for the selection of $z$ is
\begin{equation}
z = F^{-1}(F(z_0) - \log\mathcal{R}) ~,
\label{eq:samplez}
\end{equation}
where $\mathcal{R}$ is a random number uniformly distributed between 0 and 1.
If not, then the veto algorithm \cite{Sjostrand:2006za} can be used.
Once the first interaction has been picked at a height $z_1$, then the
same algorithm can be used for each of the particles produced in it, 
with $z_0$ replaced by $z_1$ as starting point, and each particle having
a separate $\sigma$ and $\theta$. This is iterated as long as needed.
For unstable particles also a decay vertex is selected, and the decay wins
if it happens before the interaction. A particle reaches the ground
if the selected $z < 0$.

In our simple study, we model the atmospheric density at altitude $h$ starting
from the ideal gas law, $\rho(h) = p M / RT$,
where $T$ and $p$ are temperature and pressure at $h$, $M$ is the molar mass
of dry air, and $R$ is the ideal gas constant. Assuming a linear drop-off
for temperature (as is the case for the troposphere \cite{NWS:atmosphere}),
we have $T = T_0 - Lh$, where $L$ is the temperature lapse rate.
From the hydrostatic equation, $dp/dh = -g\rho$, the pressure is
\begin{equation}
p = p_0 \left(1 - Lh/T_0 \right)^{gM/RL}.
\end{equation}
Then
\begin{equation}
\rho(h) = \frac{p_0 M}{R T_0} \left(1 - \frac{Lh}{T_0}\right)^{\frac{gM}{RL} - 1}
  \approx \rho_0 e^{-h/H},
\label{eq:atmosdrop}
\end{equation}
where the approximation holds for $Lh \ll T_0$, and
\begin{equation}
\frac{1}{H} = \frac{gM}{RT_0} - \frac{L}{T_0}.
\end{equation}
Using International Standard Atmosphere values for the atmospheric 
parameters (ISO 2533:1975 \cite{ISO2533:1975})
gives $\rho_0 = 1.225$~g/m$^3$ and $H = 10.4$~km.
This approximation is good up until around $L = 18$~km (near the equator),
but in our simplified framework, we assume the entire atmosphere follows 
this shape. In this case, it is possible to sample $z$ according
to \eqref{eq:samplez}, specifically,
\begin{equation}
  z = -H \log\left( e^{-z_0/H} - \frac{\cos\theta}{H \sigma \rho_0}
    \log\mathcal{R} \right) ~.
\end{equation}
In our model, we make the additional assumption that the particle will
never interact above $z_0 = 100$~km. This is a good approximation since,
applying the exponential approximation in \eqref{eq:atmosdrop} to infinity,
the probability of such an
interaction is of order $10^{-5}$. Dedicated programs use a more
detailed description, \eg \CORSIKA has an atmosphere with five
different layers.

\subsection{Some atmospheric studies}

In this section we study the cascade initiated by a proton hitting
the model atmosphere above. In the future it would be useful to study
also \eg an incoming iron nucleus, but here it is still not clear
how to handle the nuclear remnants. The simulation includes
hadronic cascades and decays, plus muon decays. But there is no 
simulation of electromagnetic cascades, nor electromagnetic energy loss
of charged particles, nor bending in the earth magnetic field,
nor a multitude of other effects. Photoproduction of hadronic states
could be added, either for primary cosmic-ray photons or for secondary
ones, but would require some more work, and is not considered important
enough for now. The atmosphere is assumed to consist solely of nitrogen,
which is a reasonable first approximation, given that \eg oxygen has
almost the same atomic number.

Four atmospheric scenarios will be compared:
\begin{enumerate}
\item A constant-density atmosphere, like on earth surface,   
consisting of free protons and neutrons. It starts 10.4~km up,
so has the same total interaction depth as the normal atmosphere
at vanishing zenith angle. 
\item Also a constant-density atmosphere of same height, but using
the already described emulation of collisions with nitrogen. 
\item An exponentially attenuated ``nitrogen'' atmosphere, with vanishing
zenith angle. Upper cutoff at 100~km.
\item An exponentially attenuated ``nitrogen'' atmosphere, with
45$^{\circ}$ zenith angle. This means a factor $\sqrt{2}$ larger
interaction depth than the other three scenarios. Again a 100~km
upper cutoff.
\end{enumerate}
The first option has already been rejected as providing too much (early)
multiplication, but is kept here as a reference. What is expected
to differ between the other atmospheres is the competition between decays
and secondary interactions. An early interaction high up in an attenuated
atmosphere gives the produced hadrons more time to decay before they can
interact. 
For simplicity we assume an incoming proton energy of $10^8$ GeV.
This corresponds to a $\p\p$ collision CM energy of 13.7~TeV, \ie just
at around the maximal LHC energy. Only hadrons with a kinetic energy
above $E_{\mathrm{kin,min}} = 0.2$~GeV are allowed to interact.
Below that scale they can still decay, but those that do not are
assumed to have dissipated and will not be counted in our
studies below.

\begin{figure}[t!]
\begin{minipage}[c]{\linewidth}
\centering
\includegraphics[width=0.5\linewidth]{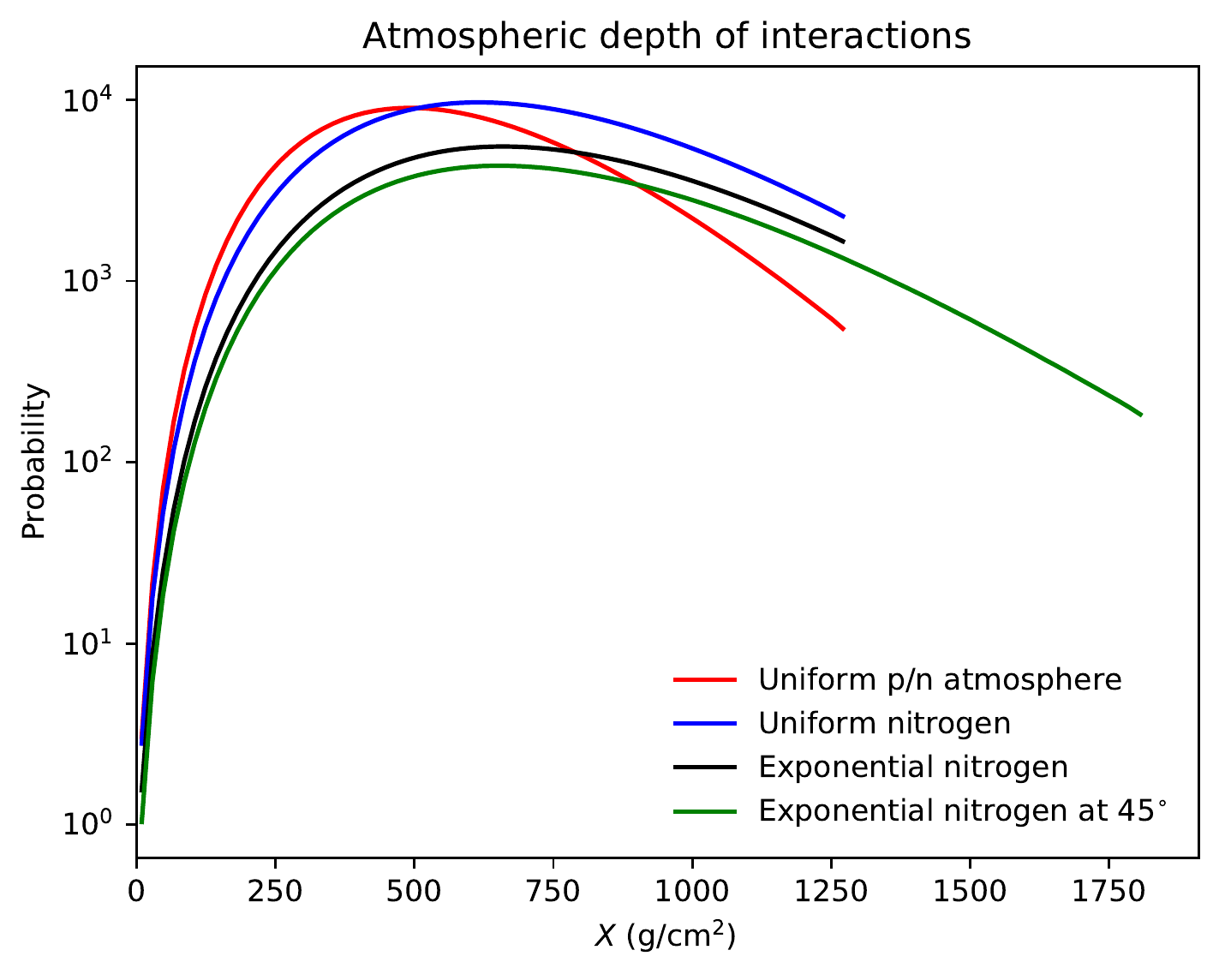}\\
(a)
\end{minipage}\\
\begin{minipage}[c]{0.49\linewidth}
\centering
\includegraphics[width=\linewidth]{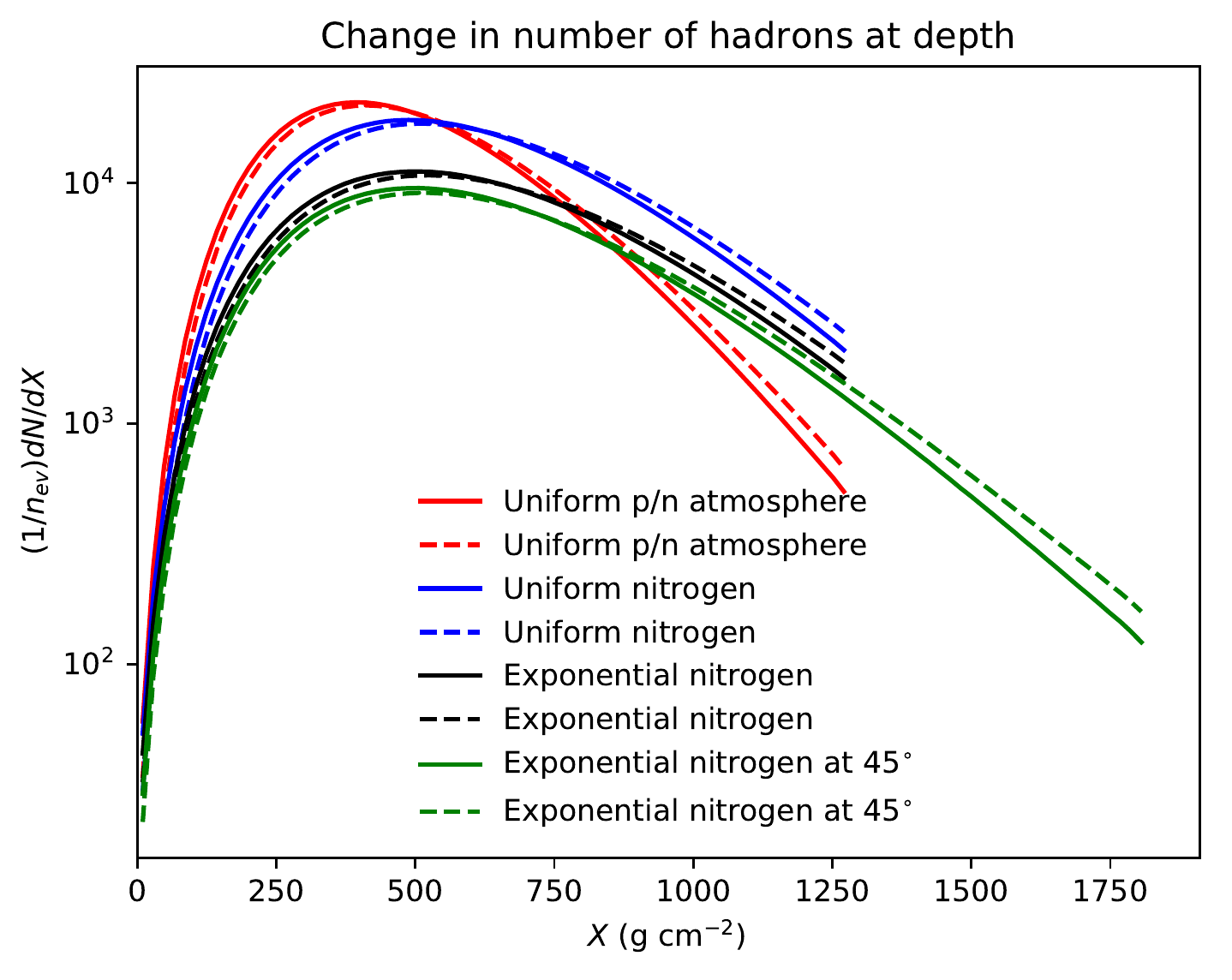}\\
(c) 
\end{minipage}
\begin{minipage}[c]{0.49\linewidth}
\centering
\includegraphics[width=\linewidth]{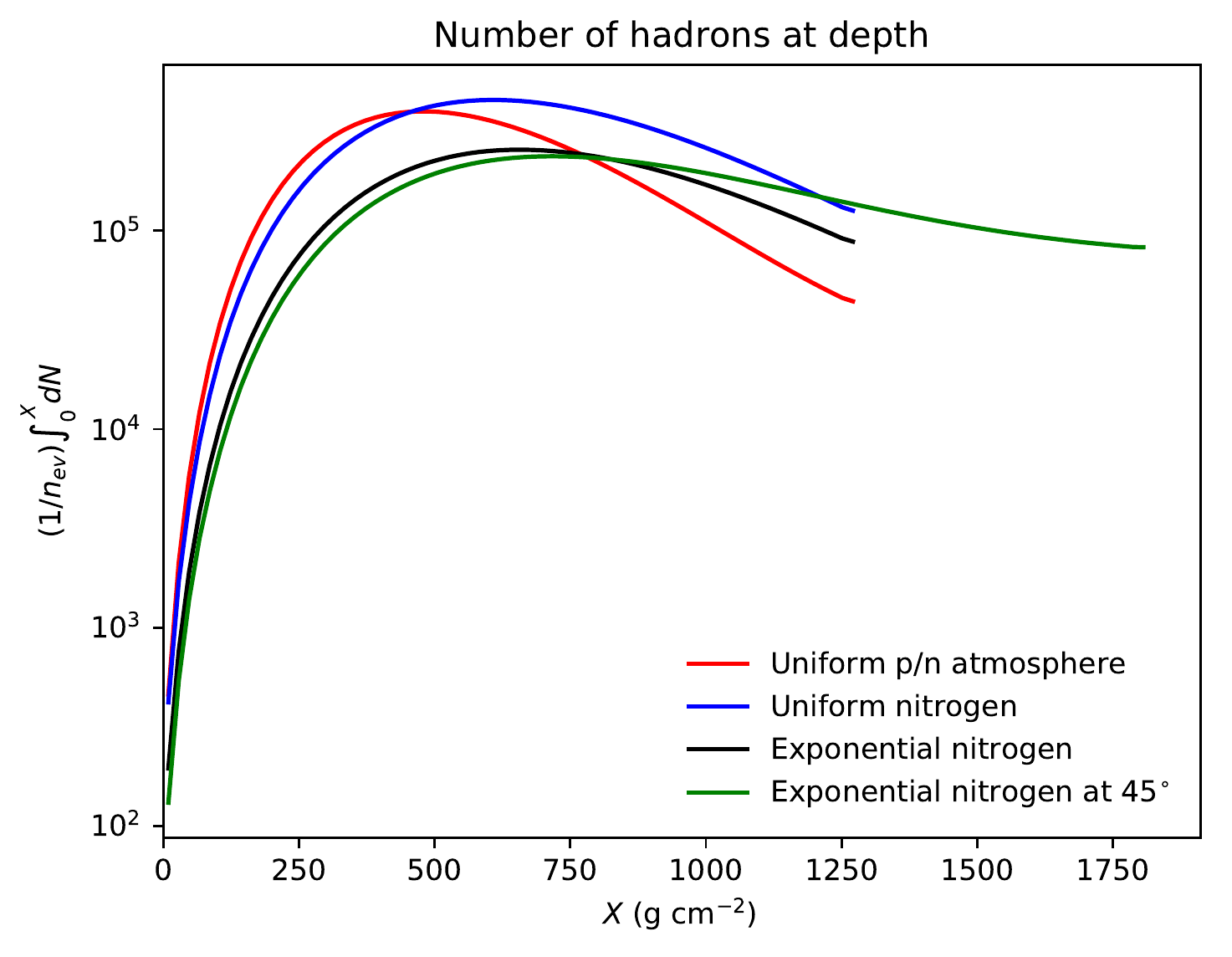}\\
(d)
\end{minipage}\\
\caption{Evolution of a cascade initiated by a $10^8$~GeV proton
travelling through four different simple models of the atmosphere,
as described in the text, as a function of atmospheric depth in
g/cm$^2$. The models stop when reaching the surface. Shown is the
number of (a) interactions, (b) hadrons produced (full) and decayed
(dashed), and (c) hadrons remaining. Hadrons that fall below the
$E_{\mathrm{kin,min}}$ threshold are removed from the numbers in
(c), but have not been counted as decays in (b).}
\label{fig:atmospherehad}
\end{figure}

\begin{figure}[t!]
\begin{minipage}[c]{0.49\linewidth}
\centering
\includegraphics[width=\linewidth]{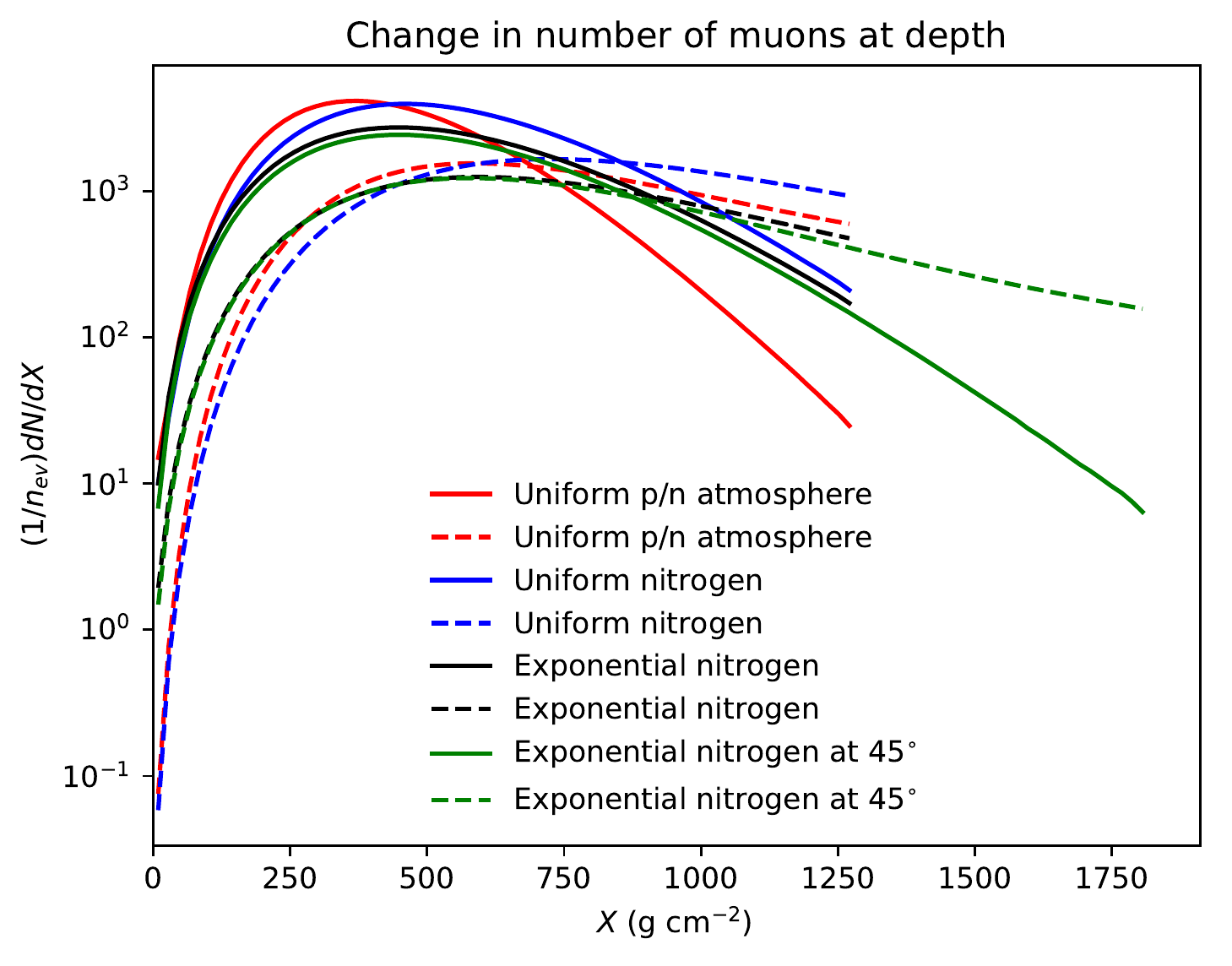}\\
(a)
\end{minipage}
\begin{minipage}[c]{0.49\linewidth}
\centering
\includegraphics[width=\linewidth]{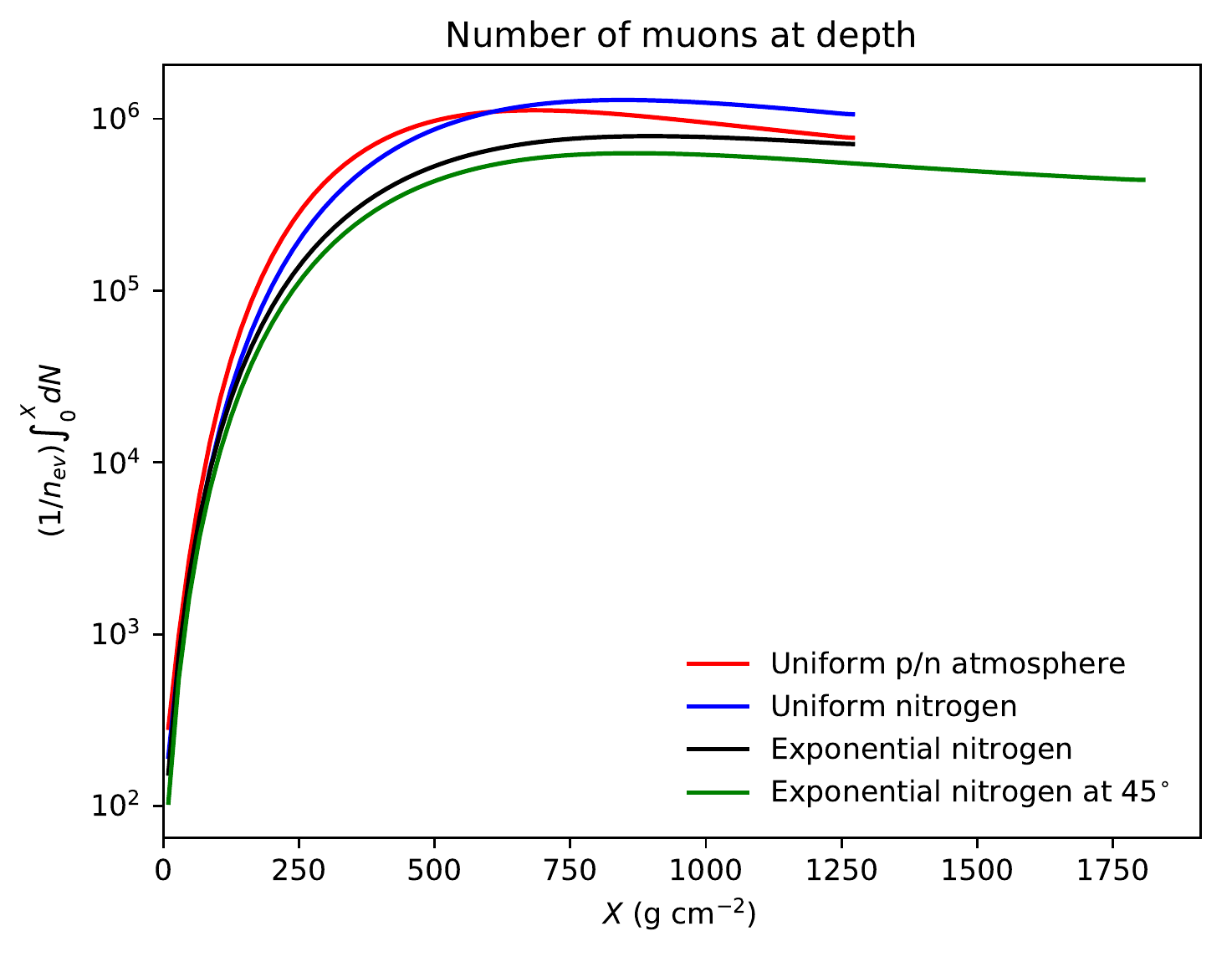}\\
(b)
\end{minipage}\\
\begin{minipage}[c]{0.49\linewidth}
\centering
\includegraphics[width=\linewidth]{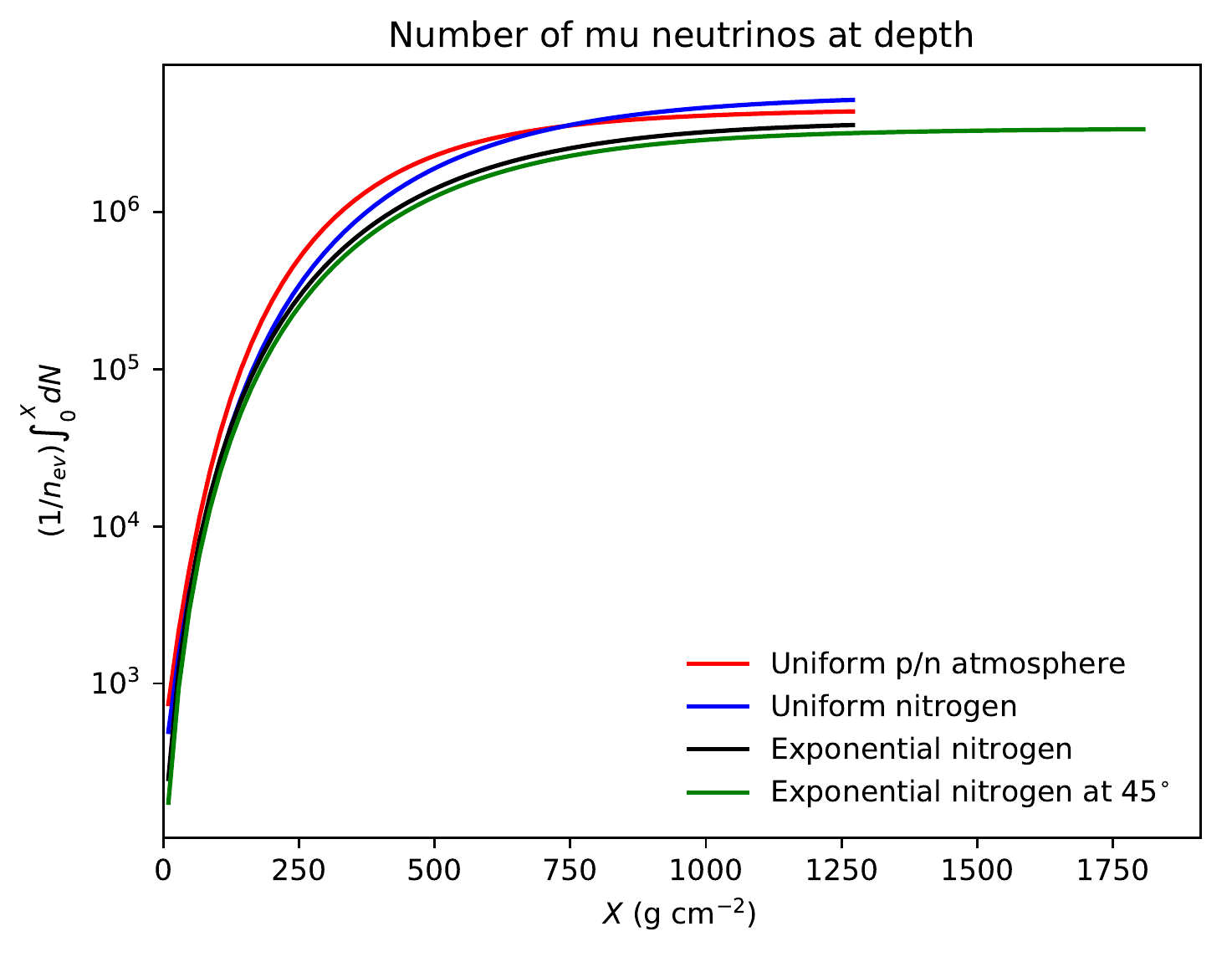}\\
(c)
\end{minipage}
\begin{minipage}[c]{0.49\linewidth}
\centering
\includegraphics[width=\linewidth]{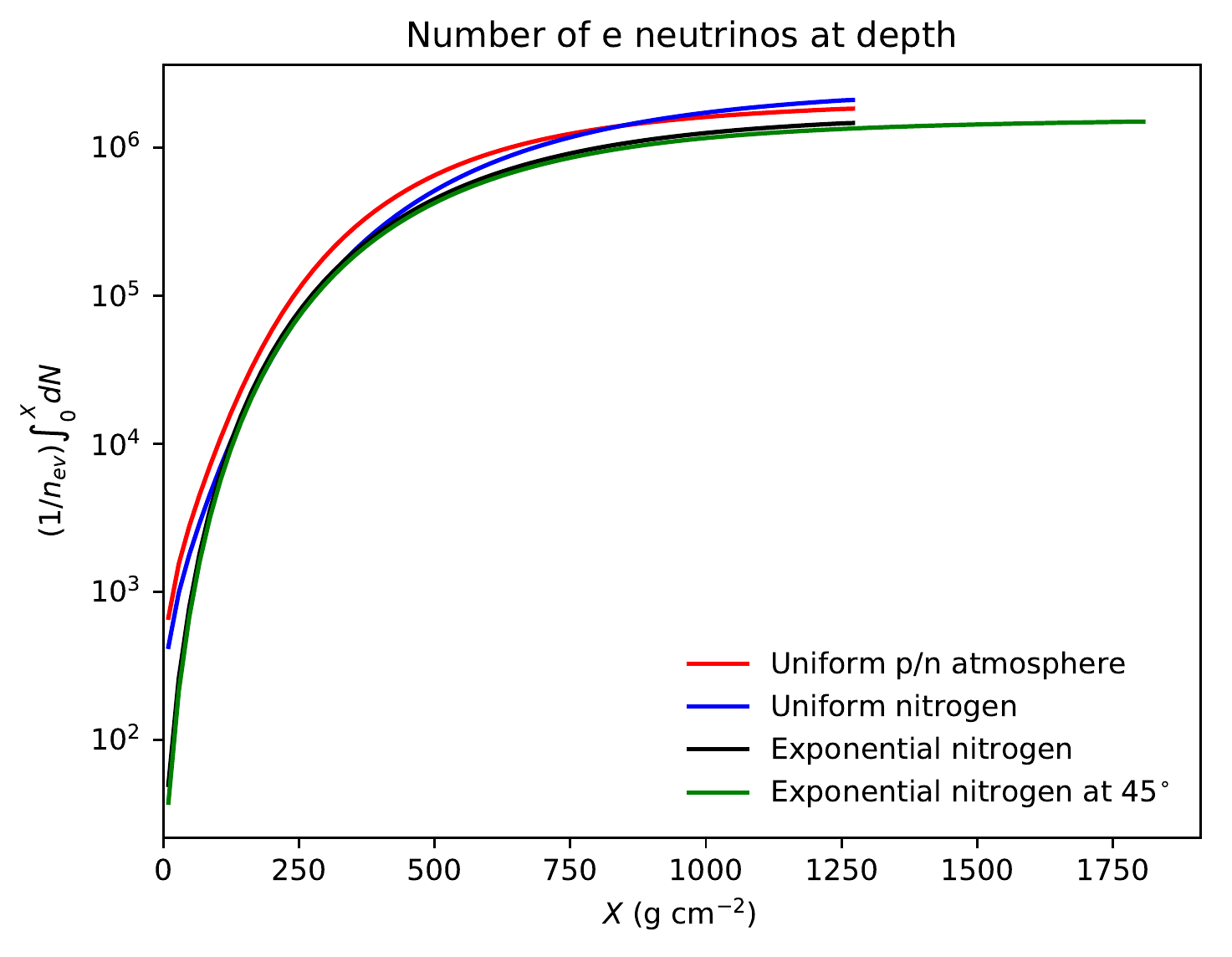}\\
(d)
\end{minipage}\\
\caption{Evolution of muons and neutrinos for the same cascades as in
\figref{fig:atmospherehad}. Shown is the number of (a) muons produced
(full) and decayed (dashed), (b) muons remaining, (c) muon neutrinos
remaining and (d) electron neutrinos remaining.}
\label{fig:atmospherelep}
\end{figure}

\begin{figure}[t!]
\begin{minipage}[c]{0.49\linewidth}
\centering
\includegraphics[width=\linewidth]{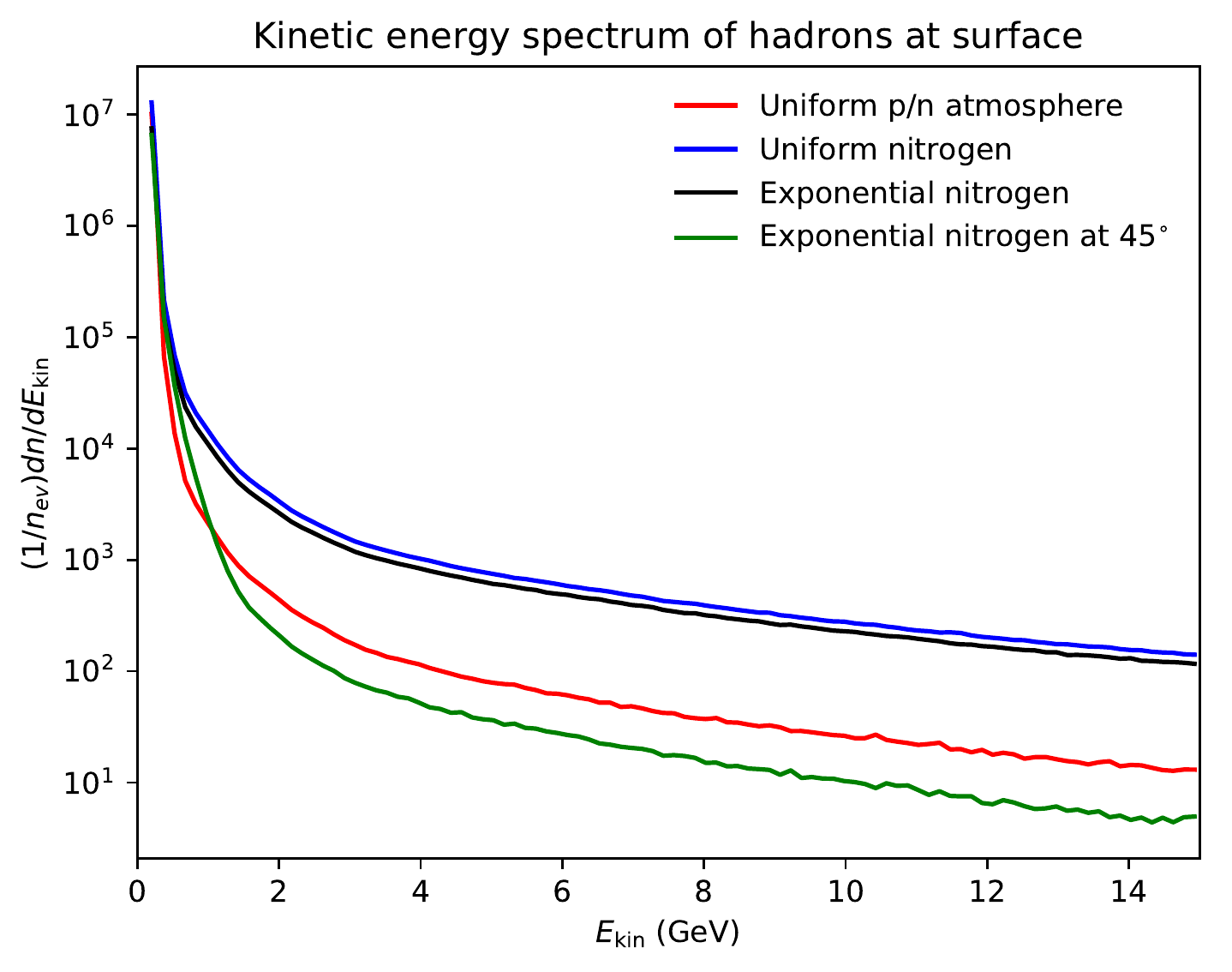}\\
(a)
\end{minipage}
\begin{minipage}[c]{0.49\linewidth}
\centering
\includegraphics[width=\linewidth]{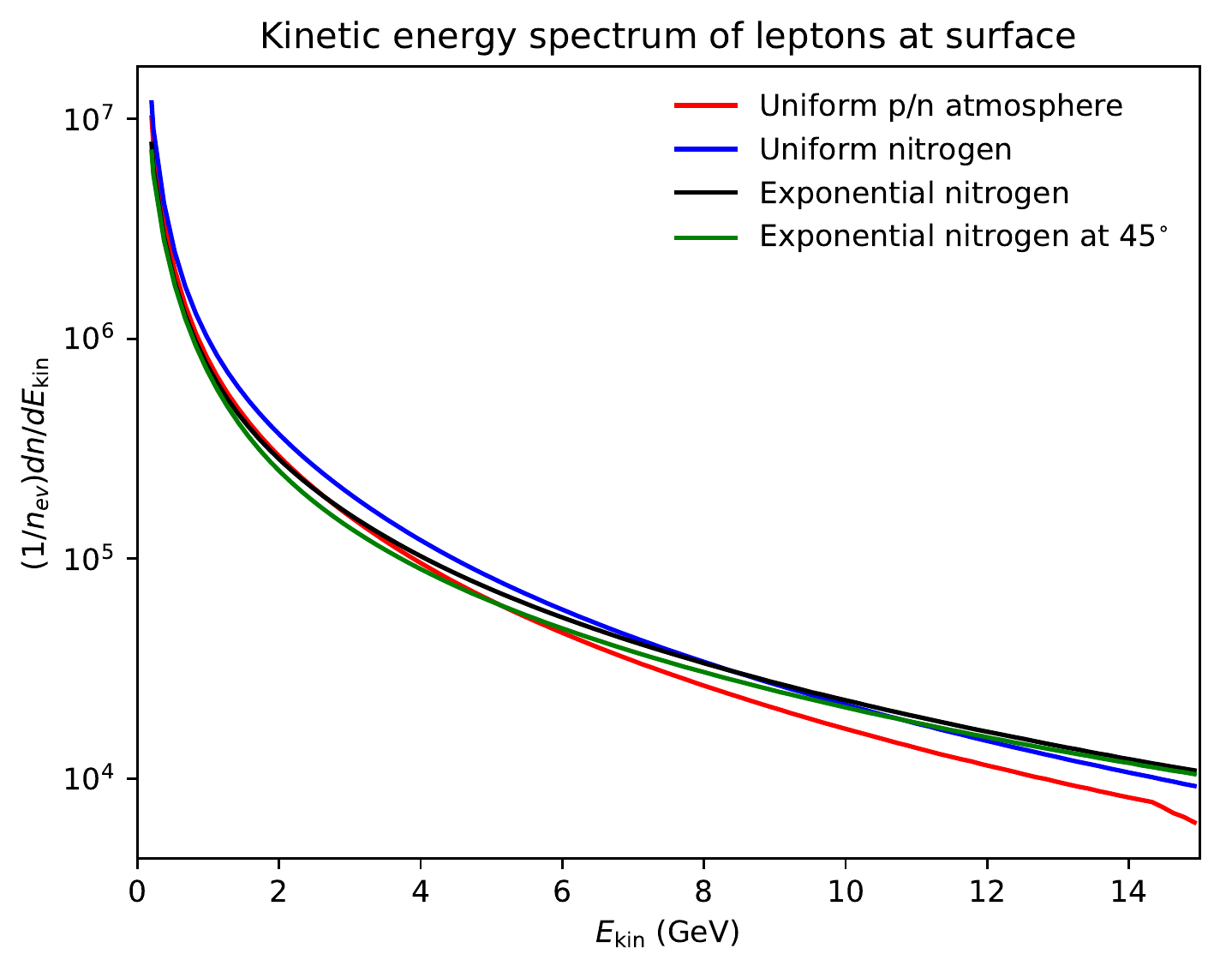}\\
(b)
\end{minipage}\\
\begin{minipage}[c]{0.49\linewidth}
\centering
\includegraphics[width=\linewidth]{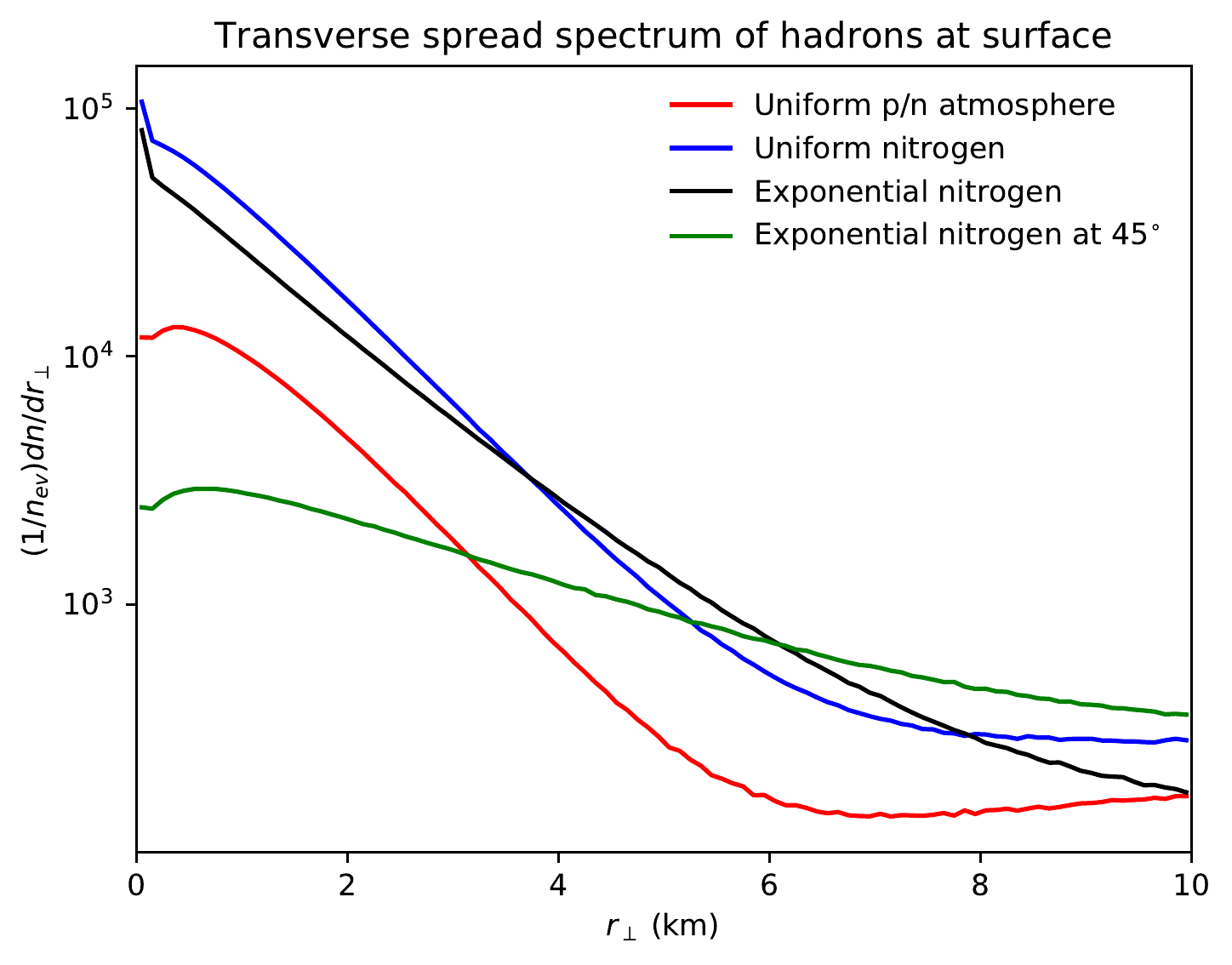}\\
(c)
\end{minipage}
\begin{minipage}[c]{0.49\linewidth}
\centering
\includegraphics[width=\linewidth]{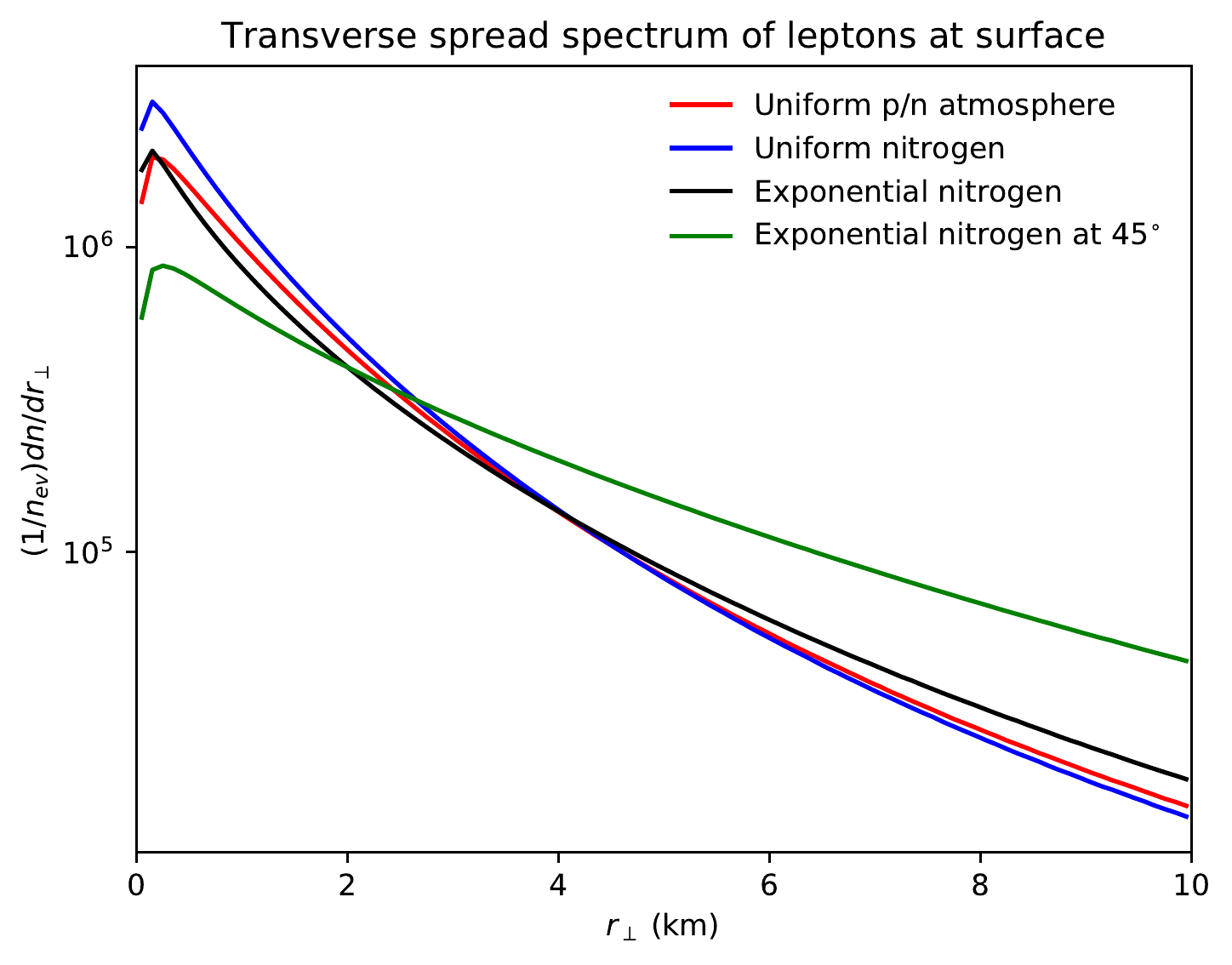}\\
(d)
\end{minipage}\\
\caption{Kinetic energy spectra (in a,b) and transverse spread (in c,d)
of the particles that reach the surface for the same cascades as in
\figref{fig:atmospherehad}. Shown are the spectra of (a,c) hadrons
and (b,d) muons and neutrinos.}
\label{fig:surface}
\end{figure}

The evolution of the cascade is shown in \figref{fig:atmospherehad}
and \figref{fig:atmospherelep}, and the energy spectra and transverse
spread of particles reaching the surface in \figref{fig:surface}.

Considering \figref{fig:atmospherehad}a, the hadron--nucleon interaction
rate as a function of atmospheric depth, we note that interactions begin
earlier in the $\p/\n$ atmosphere than in the $\N$ one, but then also
peters out earlier, when most hadrons have low energies. Moving on to
the exponential atmosphere, more hadrons (notably pions) decay before
they can interact, which reduces the interaction rate. Even more so
in the case of a 45$^{\circ}$ zenith angle, where more of the
early evolution takes place in a thin atmosphere. The production rate of
hadrons, \figref{fig:atmospherehad}b, correlates rather well with the
interaction rate, although the number of hadrons produced per collision
would gradually decrease as each hadron gets to have a lower energy.
The early hadron production is higher for the free $\p/\n$ atmosphere,
consistent with expectations but not quite as dramatic as
\figref{fig:simpleHI} might have led one to expect. Most of the hadrons
decay reasonably rapidly, leaving mainly protons and neutrons to carry on.
\figref{fig:atmospherehad}c shows how the number of such undecayed hadrons
increases, following the pattern of the previous plots. Specifically, the
exponential atmospheres give a reduced number of final hadrons.

The long lifetime of muons, $c\tau = 659$~m, means that muon decays lag
behind production, \figref{fig:atmospherelep}a. The number of muons
reaches a plateau, where production and decay roughly balance,
\figref{fig:atmospherelep}b. The total number of muons follows the same
pattern between the four atmospheric scenarios as noted for hadrons.
The production of muon and electron neutrinos, \figref{fig:atmospherelep}c,d,
is dominated by pion and muon decays, but also receives contributions
from other weak decays. Neutrino oscillations are not considered here.

The bulk of hadrons that apparently reach the ground have very low
kinetic energies, even given the cut
$E_{\mathrm{kin}} = E - m > E_{\mathrm{kin,min}}$, \figref{fig:surface}a.
In reality most of these would be stopped or bent away by the earth
magnetic field, so the figure should be viewed as a study of the
consequences of hadronic cascades on their own. The uniform and
exponential nitrogen atmospheres have comparable rates of higher-energy
hadrons. These hadrons are dominated by $\p$ and $\n$, which are not
affected by decays. That the higher-energy hadron rate is reduced for
a non-vanishing zenith angle is to be expected. Also the $\p/\n$
atmosphere gives a lower rate, presumably as a consequence of the
faster split of the original energy into several lower-energy collision
chains. The kinetic energy spectra of muons and neutrinos,
\figref{fig:surface}b, again are peaked at lower energies, though not
quite as dramatically. The four atmospheric models also come closer
to each other for leptons, though the $\p/\n$ one remains an outlier.

The cascades disperse particles in quite different directions, implying
large footprints on the earth surface. In \figref{fig:surface}c,d we
show the distributions of hadrons or muons/neutrinos as a function of the
distance $r_{\perp}$ away from the point where the original proton would
have hit if it had not interacted. Recall that the relevant area element
is $\d^2 r_{\perp} = 2\pi \, r_{\perp} \, \d r_{\perp}$, while
\figref{fig:surface}c,d plots $\d n / \d r_{\perp}$, so the number of
particles per area is strongly peaked around $r_{\perp} = 0$. The area
argument is also the reason why two of the curves can turn upwards at large
$r_{\perp}$. Not unexpectedly a non-vanishing zenith angle increases
the spread, both by having interactions further away and by the elongation
of a fictitious shower cone hitting the surface at a tilt. Conversely, the
uniform nitrogen gives less spread, by virtue of cascades starting closer
to the surface. It should be mentioned that kinetic-energy-weighted
distributions  (not shown) are appreciably more peaked close to
$r_{\perp} = 0$, as could be expected. Occasionally an event can have a
large energy spike close to but a bit displaced from the origin. 
We have not studied this phenomenon closer, but assume it relates to
an early branching where a high-energy particle is produced with a 
non-negligible transverse kick relative to the event axis.

\subsection{A lead study}

\begin{figure}[t!]
\begin{minipage}[c]{0.49\linewidth}
\centering
\includegraphics[width=\linewidth]{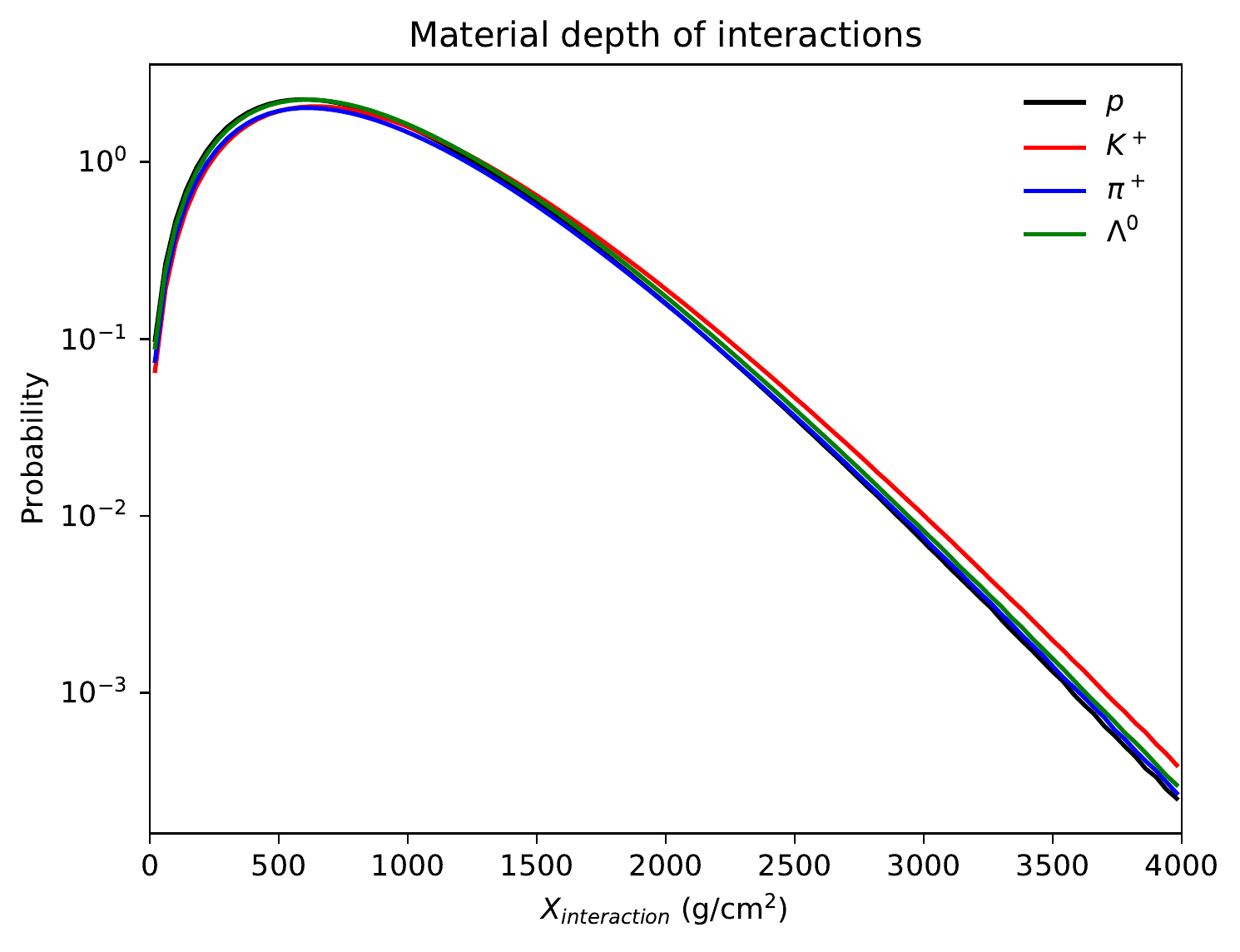}\\
(a)
\end{minipage}
\begin{minipage}[c]{0.49\linewidth}
\centering
\includegraphics[width=\linewidth]{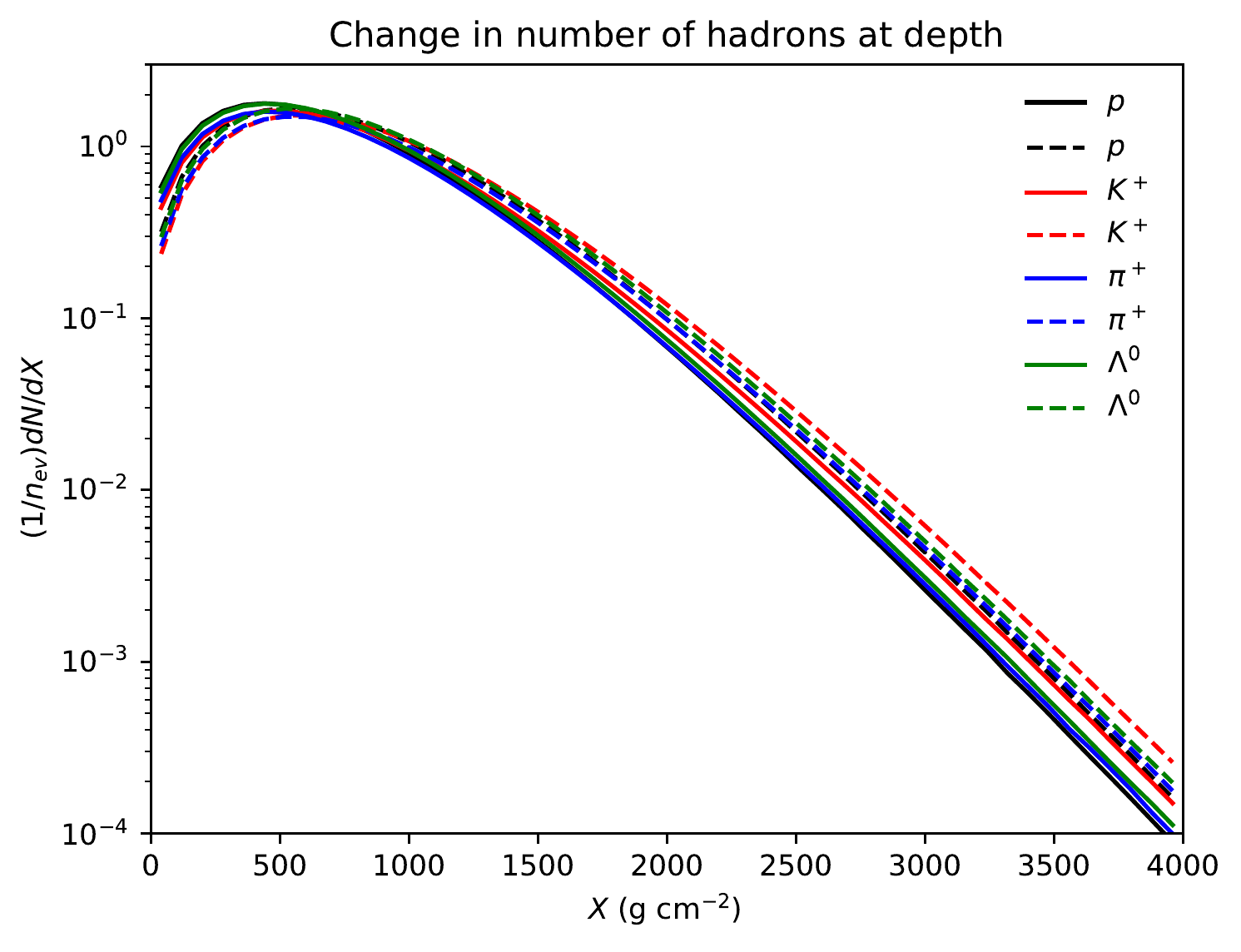}\\
(b)
\end{minipage}\\
\begin{minipage}[c]{0.49\linewidth}
\centering
\includegraphics[width=\linewidth]{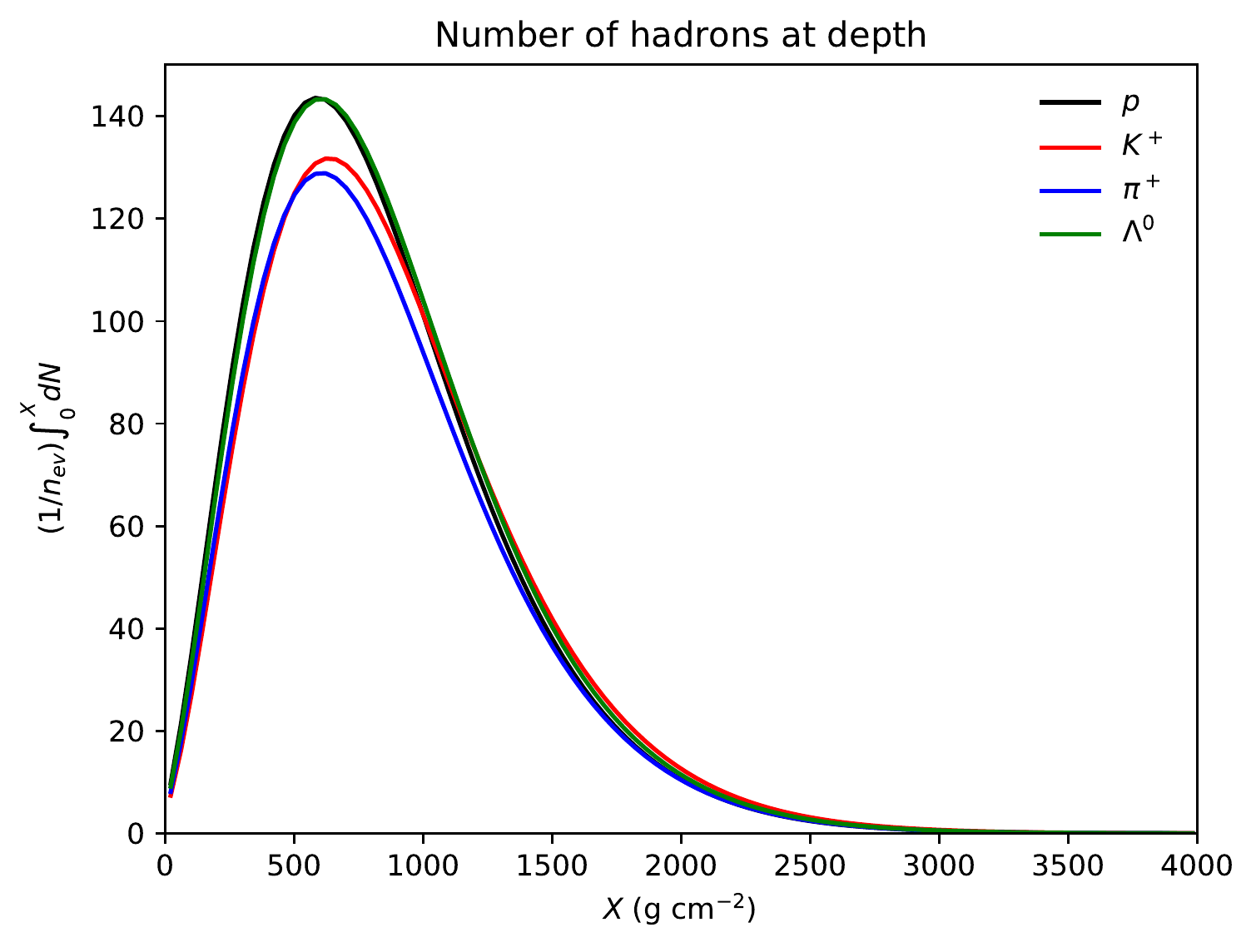}\\
(c) 
\end{minipage}
\begin{minipage}[c]{0.49\linewidth}
\centering
\includegraphics[width=\linewidth]{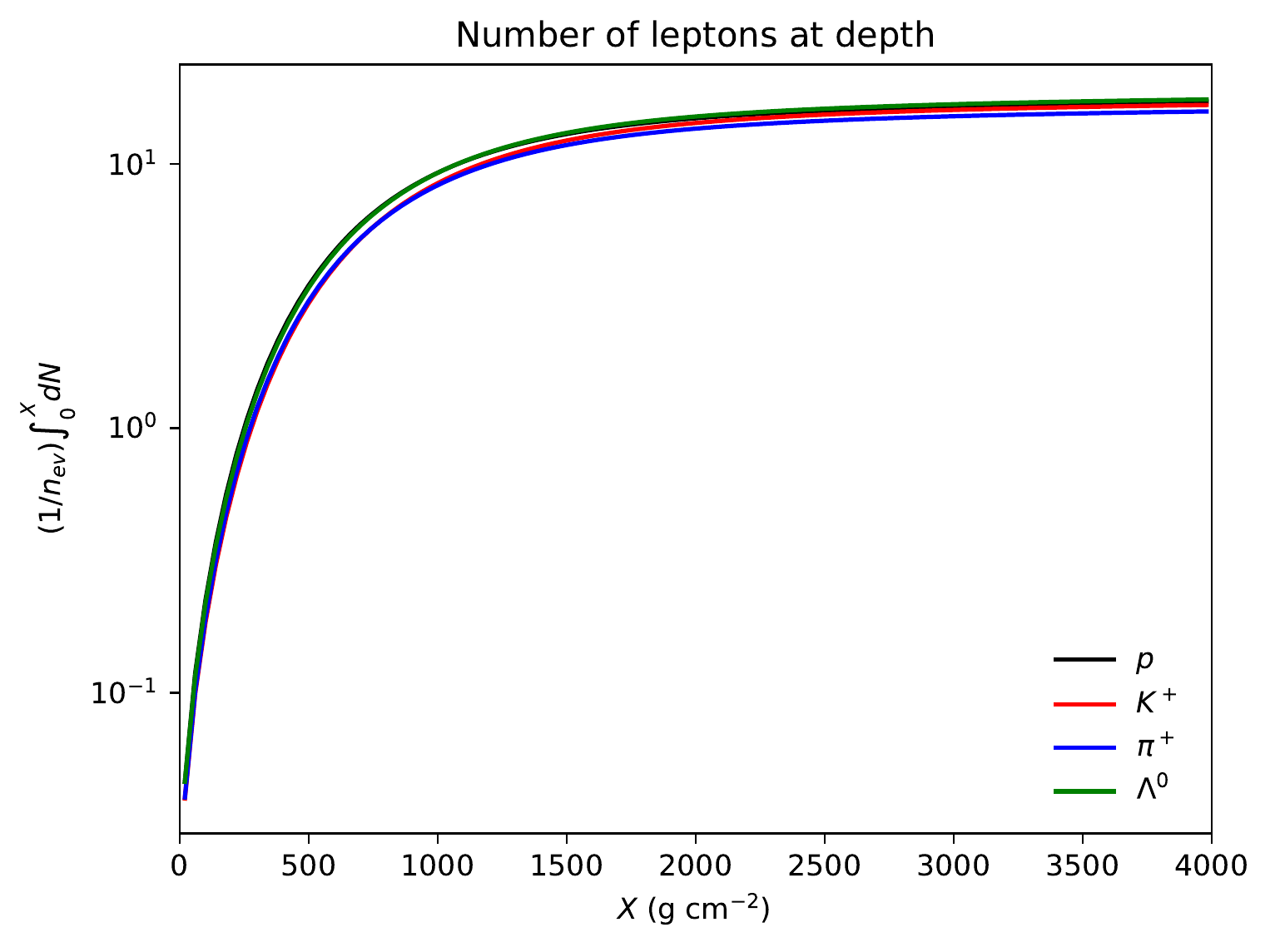}\\
(d)
\end{minipage}\\
\caption{Evolution of a cascade initiated by a 1000~GeV proton, $\pi^+$,
$\K^+$ or $\Lambda^0$ passing through a 3.5 m thick slab of lead.
Shown is the number of (a) interactions, (b) hadrons produced (full)
and decayed (dashed), (c) hadrons remaining and (d) muons and neutrinos
remaining. Hadrons that fall below the $E_{\mathrm{kin,min}}$ threshold
are removed from the numbers in (c), but have not been counted as decays
in (b).}
\label{fig:lead}
\end{figure}

The new code can also be used to track a cascade through a solid
material. We have taken lead as an example of a heavy element that is
used in some detectors, with rather different properties than the
light elements and low density of air. Here the decays of longer-lived
particles, such as  $\pi^{\pm}$,  $\K^{\pm}$, $\K_{\mathrm{L}}$ and
$\mu^{\pm}$, do not play as significant a role as in the atmosphere,
given the shorter distances a particle travels through a detector.
The maximal primary hadronic energy is also lower than for cosmic rays.
Taking LHC as example, the 7~TeV maximum translates into collision CM
energies below 115~GeV. When we now study the cascades in lead, only
hadronic interactions are considered, as before, \ie leptons and photons
are free-streaming. Some illustrative results are shown in \figref{fig:lead},
for a $p_z = 1$~TeV initial hadron of different kinds. The density of lead is
$\rho = 11.35$ g/cm$^3$, so an interaction depth of 4000 g/cm$^2$
corresponds to 3.5~m. Hadrons below $E_{\mathrm{kin,min}} = 0.2$~GeV are
assumed to stop in the matter and not interact any further. Thus the
number of hadrons vanishes after som depth.

The main conclusion of \figref{fig:lead} is that the different incoming
hadrons give rise to rather similar cascades. This is largely owing to
the rapid multiplication into a fairly similar set of secondary hadrons.
Baryons tend to have larger cross sections than mesons, and the proton
the largest of them all, so it is understandable why the proton cascade
starts somewhat earlier and also dies down earlier. Strange particles 
have somewhat lower cross sections than their non-strange counterparts,
which explains why the $\K^+$ curve starts slower than the $\pi^+$ one. 
But also other factors may be relevant, like how the leading-particle
spectrum of a collision affects the nature of subsequent collisions. Here
we expect a baryon beam to give a harder leading hadron than a meson,
and a strange hadron a harder spectrum than a non-strange one, within
the context of normal string fragmentation. This could partly compensate 
for the cross section differences. Further studies will be needed to
disentangle these and other factors that may contribute to the small
differences observed.

\section{Summary and outlook}

In this article we have extended the existing hadron--hadron
interaction framework of the \textsc{Pythia} event generator.
Traditionally it has been centered around $\p\p$ and $\p\pbar$
collisions. A few extensions to some meson--meson collision types
have been implemented as part of the Vector Meson Dominance scenario
of a photon fluctuating to and interacting like a flavour-diagonal
vector meson.

Now we have made a deeper study of almost all possible hadron--nucleon
collision types. This includes deriving new total and partial
cross sections at medium-to-high collision energies, based on the
DL and SaS ans\"atze, extended with the help of the Additive Quark Model
and Reggeon systematics where no data is available. It also includes
producing some twenty new PDF sets, here denoted SU21. One key
assumption has been that heavier valence quarks start out with a
larger fraction of the total hadron momentum, at the expense of
lighter quarks and gluons, so that all hadron constituents have
comparable average velocities. The same constituent-quark-mass ratios
as used in the AQM therefore come to characterize our new PDFs. 
A consistency check then is that the average number of multiparton
interactions is comparable in all collision types. This average is
the ratio of the integrated (mini)jet cross section, which directly
relates to the PDFs used, and the total (nondiffractive) cross section.
Both these numbers should reduce at comparable rates when light quarks
are replaced by heavier ones. 

Event properties nevertheless are not and should not be identical.
This is visible \eg in the rapidity distributions of charged particles,
which tend  to peak in the hemisphere of the heavier hadron, with its
(partly) harder PDFs, and for the same reason such hadronic collisions
tend to give somewhat harder $p_{\perp}$ spectra. Such differences should
be explored further and, to the extent data is or becomes available, it
would be interesting to compare.

It would also be interesting to explore the sensitivity of the cascade
to the different components of the full \textsc{Pythia} event simulation.
Considerable effort has gone into the separate modelling
of different hadron species, but how much of that actually affects the
end result? Is it important to use PDFs tailormade for each hadron,
or would one proton/baryon and one pion/meson PDF have been enough?
And what is the impact of minijets with its initial- and final-state
radiation? Jets are key features for LHC physics, where \textsc{Pythia}
likely is more developed and better tuned than many cosmic-ray generators,
but where effects may be overshadowed \eg by the beam-remnant description
in the forward direction. (The latter is the subject of a separate
ongoing study.)  If one wants to study how a charm or bottom hadron
interacts on its way through matter, on the other hand, a tailormade
description may be relevant. 

We do not claim any fundamentally new results in this article, but still
present some nice studies that point to the usefulness of the framework.
We show how hadronic cascades evolve in the atmosphere, spanning energy
scales from 10$^8$~GeV (or higher if wanted) to 0.2~GeV, how the energy
rapidly is spread among
many hadrons with low energy each, how hadron decays give muon and neutrino
fluxes, how the kinematics and dynamics leads to a wide spread of particles
that hit the ground, and more. Note that a complete record of all particles
is kept, so it is possible to ask rather specific questions, such as e.g.\
whether hard-jet production in the primary interaction correlate with
isolated energy/particle clusters on the ground. We also show, for the 
solid-target case, how hadrons with larger cross sections also begin their
cascades earlier, evolve faster and peter out sooner.

In the current article we have put emphasis on the applications to full
cascade evolution, in the atmosphere or in solid matter, rather than on
the single collision. One reason is that the full cascade offers
further technical challenges on top of modelling the individual collision,
which forces us to extend the capabilities of the \textsc{Pythia} code.
Previously it has not been feasible to switch collision energy or beam
type event by event, at least not without each time doing a complete
reinitialization, which then slows down event generation times by orders
of magnitude. The other reason is that we would like to be able to benefit
from and contribute to the understanding of hadronic collisions in
different environments. Currently there is one set of event generators
that is mainly used for LHC $\p\p$ physics, such as \textsc{Herwig}
\cite{Bellm:2019zci}, \textsc{Sherpa} \cite{Sherpa:2019gpd} or
\textsc{Pythia}, and another one for cosmic rays, see the Introduction,
with only EPOS as an example of a code used in both environments. 

Nevertheless, we are aware that we have not presented a full framework
for hadronic cascades. One would need to extend the \textsc{Angantyr}
framework for nuclear collisions so that it could also switch between
different collision beams and energies within a manageable time.
Ideally it would be validated at lower energies and, for the handling of
iron and other heavy cosmic rays, include a model of the nuclear breakup
region. This is a tall order, that is beyond our control. In the current
study we have instead introduced a quick-and-dirty fix, tuned to
reproduce some of the simpler \textsc{Angantyr} phenomenology, to handle
hadron--nucleus but not nucleus--nucleus collisions.

Furthermore, hadronic cascades is not the end of the story, but must be
part of a larger framework that encompasses all relevant processes, and
provides a more detailed modelling of the atmosphere.
The hope is that the code will find use in larger frameworks, such as
CORSIKA 8 for cosmic rays and GEANT4 for detector simulation. At the
very least, we offer a far more powerful replacement to the older
\textsc{Pythia}~6 code currently used in some such frameworks.
In the future we could also take on some other related tasks, such as
photoproduction in the cascades.

The \textsc{Pythia} generator is under active development in a number of
directions. This article should not be viewed as an endpoint but 
hopefully as a step on the way towards making \textsc{Pythia} even
more useful for a number of physics studies.

\section*{Acknowledgements}

Thanks to Christian Bierlich for useful discussions on \Angantyr.
Work supported in part by the Swedish Research Council, contract number
2016-05996, and in part by the MCnetITN3 H2020 Marie Curie Innovative 
Training Network, grant agreement 722104.

\bibliographystyle{utphys}
\bibliography{bibliography}

\begin{thebibliography}{100}
\bibitem{Zweig:1981pd}
G.~Zweig, ``{An SU(3) model for strong interaction symmetry and its breaking.
  Version 1},''.
\end{thebibliography}

\end{document}